\documentclass[a4paper,aps,prd,showpacs,nofootinbib,preprintnumbers,amsmath,amssymb,mcite,superscriptaddress,12pt]{revtex4-1}

\usepackage[table]{xcolor}
\usepackage{graphicx}
\usepackage{dcolumn}
\usepackage{bm}
\usepackage{esvect}
\usepackage{accents}
\usepackage{soul}

\usepackage{hyperref}
\hypersetup{colorlinks=true,linkcolor=magenta,anchorcolor=green,citecolor=cyan,filecolor=black,menucolor=black,urlcolor=brown}
\usepackage{slashed}
\newcommand{\be}{\begin{equation}}
\newcommand{\ee}{\end{equation}}
\newcommand{\ba}{\begin{eqnarray}}
\newcommand{\ea}{\end{eqnarray}}
\newcommand{\Lp}{\left (}
\newcommand{\Rp}{\right )}

\newcommand{\V}[1]{\vec{#1}}

\newcommand{\f}[2]{\frac{#1}{#2}}

\newcommand{\munu}{{\mu\nu}}

\newcommand{\h}{\f{1}{2}}

\newcommand{\Bmat}{\begin{matrix}}
\newcommand{\Emat}{\end{matrix}}
\newcommand{\Barr}{\begin{array}}
\newcommand{\Earr}{\end{array}}

\begin{document}


\title{ 
Casimir pressure with dissipation and quantum corrections from axion dark matter 
}

\author{Philippe Brax}
\affiliation{Institut de Physique Th\'eorique, Universit\'e  Paris-Saclay, CEA, CNRS, F-91191 Gif-sur-Yvette Cedex, France}
\author{Pierre Brun}
\affiliation{D\'epartement de Physique des Particules, CEA  Universit\'e  Paris-Saclay, F-91191 Gif-sur-Yvette Cedex, France}

\begin{abstract}
We study the effects of the oscillating axion field present in our environment on the Casimir pressure between two metallic plates. We take into account the finite conductivity of the boundary plates and model the interactions between matter and  photons in the Schwinger-Keldysh formalism. This allows us to take into account dissipation in the quantum field  description of this open quantum system and retrieve the Lifschitz results for the Casimir interaction between two metallic plates. We then compute the leading correction  to the Lifschitz theory  in inverse powers of the axion suppression scale and show that the Casimir pressure receives oscillating corrections depending on the product of the axion mass and the distance between the plates. This contribution is repulsive at large distance compared to the axion Compton wavelength as a consequence of the breaking of parity invariance by the axion dark matter background. 

\end{abstract}

\maketitle

\section{Introduction}

We consider the effects of an axion-like oscillating pseudo-scalar field coupled to photons on  the Casimir pressure \cite{Casimir_original,milton2001casimir,Klimchitskaya:2009cw,Bimonte:2017bir,Bimonte2013,Bimonte:2021maf,Bimonte:2022een,bordag2009advances,Milton:2004ya,Jaffe:2005vp,Graham:2002fw,Mobassem:2014jma,Favitta:2023hlx}.
The axion field is taken to be  oscillating in our environment as expected from a field whose fast oscillatory behaviour reproduces the effects of dark matter on large scales and inside the Milky Way. The quantum modes of the photon
are affected by the coupling to the axion. This leads to corrections to the Casimir pressure between the plates.

In the ideal case of infinitely reflexive metallic plates, the electromagnetic field in the cavity shows  resonances which are perfect and imply an infinite series of Dirac peaks for the induced electromagnetic field. This behaviour is regularised physically by the finite permittivity of the metal, which in the simple Drude approximation depends on the plasma frequency $\omega_{\rm pl}$ and the conductivity $\sigma$ \cite{Brax:2024cur}\footnote{For a recent description of the issues of the Drude model when comparing with data, see \cite{Klimchitskaya:2023ebx}}. In this case the number of resonances  becomes finite and their frequency values  are all bounded by the plasma frequency $\omega_{\rm pl}$. . The {non-zero value of} the resistivity moves the pole of the Green's functions away from the real axis. As a result, the modes of the photon field between the boundary plates penetrate inside them over a finite length. The resonances are widened by the dissipation induced by the metallic plates. {These effects are thoroughly discussed in~\cite{Brax:2024cur} and constitute the classical treatment of this effect. In the present paper we present the quantum description of the Casimir effect in the presence of dissipation and axion dark matter.}

{
The description of the quantum fluctuations of the photons when coupled to
a dissipative metal goes beyond a usual quantum field theory treatment\footnote{By this we mean the $in-out$ formalism of scattering amplitudes or the $in-in$ description of cosmological expectation values for instance. In both cases unitarity plays a crucial role and is lost here. See appendix \ref{app:in} for details.}. Indeed dissipation breaks unitarity and the Hamiltonian of the system becomes non-Hermitian\footnote{ Unitarity can be reintroduced by considering the complete system involving the photons and all the matter degrees of freedom. This approach to the Casimir effect with dissipation can be found in \cite{Guerout:2018mse}.}. The core of this work corresponds to  modifying  the usual QFT methods to compute quantum averages in this dissipative  case and then apply these results in the case of axions\footnote{{For a recent description of electromagnetism in matter in the Schwinger-Keldysh formalism where gauge invariance is considered, see \cite{Salcedo:2024nex}}.}.

At the classical level, dissipation can be treated in a Lagrangian framework by doubling the numbers of degrees of freedom, i.e. by introducing a "shadow" field corresponding to the time-reversed evolution.
The shadow field brings  energy from late times backwards so that the system comprised of  the real field and  its shadow becomes a conservative system~\cite{Galley:2012hxx}. In this paper we use this formal method, at the expense of a doubling of number of the degrees of freedom. In the end the shadow field is disregarded and the original field evolves in a dissipative system with increasing times.

The goal of this work is to use a path integral formulation  to calculate vacuum expectation values of the energy momentum tensor of axion electrodynamics. This eventually leads to the computation of the Casimir pressure between two metallic plates.  The usual way of doing this would be via  the Gellman-Low formula in the $in-out$ formalism 
\footnote{For the link between the $in-in$ and the $in-out$ formalisms in the case of a closed and non-dissipative quantum system, see \cite{Donath:2024utn} for instance.}.
Here this connection is lost because in the zero-temperature treatment we pursue, the evolution operator is not unitary.
The path integral formulation only stands
for the vacuum expectation of operators ${\cal O}(t)$ evaluated at a single time $\langle 0\vert {\cal O}(t)\vert 0\rangle$ 
where the vacuum is taken to be the one of the free theory when no dissipation is taken into account. In this case, a Schwinger-Keldysh representation of these expectation values is obtained.
The action of each field is integrated along the Schwinger-Keldysh path going from $t=-\infty$ to $t=+\infty$ and back. One can introduce the Schwinger-Keldysh representation of the action as involving two copies of each field, one on the lower branch of the Schwinger-Keldysh path from $t=-\infty$ to $t=+\infty$ and another one on the upper branch going from $t=+\infty$ to $t=-\infty$. This back-and-forth path is unrelated to the existence of the shadow field and shall be used for both, leading to another doubling of number of the fields\footnote{A simplified formalism where the shadow fields and the Keldysh copies are identified may be obtained using the classical action presented in the appendix \ref{app:diss}. We do not pursue this approach here and leave it to future work.}.

In addition, in the Coulomb gauge,  the two electromagnetic polarisations are described by two scalar fields. All in all, the two original polarisations of the photons are then described by eight fields, two of which are physical : the two polarisations going forward in time.
}

Using this formulation, we retrieve the usual Casimir pressure in the absence of dissipation and its Liftschitz generalisation when dissipation in the plates is taken into account \cite{lif1,lif2}. The Schwinger-Keldysh field theoretic treatment allows one to extend our results to the presence of an axion coupling to photons. The Casimir pressure is now simply corrected as the
two-point correlation function of the two photon polarisations is modified by the axion interaction. This can be calculated using the usual rules of perturbative quantum field theory in the Schwinger-Keldysh formulation. The pressure due to the coupling to the axion field vanishes with the mass of the axion and grows with the distance until it reaches the Compton length of the axion field. For larger distances, the pressure oscillates eventually and changes sign when $md={\cal O}(1)$.
{This is similar to  the results in \cite{Jiang:2018ivv} where the breaking of parity invariance, e.g.  by the background value of the axion field, allows for a change of the sign of the Casimir effect. Here the amplitude of the oscillating part due to the axion is not large enough to induce a change of sign and the breaking of the Kenneth-Klich theorem on the attractiveness of the Casimir interaction \cite{Kenneth:2006vr}. It simply reduces the magnitude of the Casimir interaction on large distances $d\gg m^{-1}$.}

The Casimir system with two metallic plates where dissipation and dispersion takes place is a testbed for dissipative open quantum system.   Indeed, as soon as the plates are not perfect, the reflection coefficients of the walls differ from unity and therefore energy flows from the vacuum cavity to the plates implying that the cavity is not a closed quantum system anymore. There is an exchange of energy with the plates which is reinforced in the dissipative case where currents inside the plaques can be excited and the Joule effect prevents the existence of thermal equilibrium if the transfer of heat into radiation is not taken into account.

Difficulties in the description of the Casimir effect in the presence of dissipation from first principle have already been noticed \cite{Bordag:2011zz,Bordag:2017qnh}. In particular the analytic continuation from the sum over the resonant frequencies in the cavity to the continuous integral used in the Lifschitz theory can be problematic. In this paper, we use the $in-in$ formalism for which the definition of the Feynman propagator with its $T$-product decomposition is crucial. As we will see, the retarded propagator is naturally defined with its resonances below the real axis in the complex plane of pulsations. The Feynman propagator requires to shift the resonances from below the negative real axis to above the negative real axis. {This can be achieved using a modification of the Schwarz reflection principle. 
In the limit of weak dissipation $\gamma \to 0$, where $\gamma$ is a damping time related to the conductivity and the plasma frequency via  $\sigma\gamma = \omega_{\rm pl}^2$}, the Feynman propagator is well defined apart from the usual poles at the resonances and the branch cuts for the continuum modes with a pulsation above the plasma frequency. For a plasma with $\gamma=0$, one can analytically continue the integral along the real axis defining the  propagator to the imaginary axis, i.e. a {Wick's} rotation. For finite values of the conductivity when $\gamma \ne 0$, the definition of the Feynman propagator is obstructed by a branch cut along the imaginary axis in the complex plane of pulsations. Using contour integrals, this induces a new contribution to the Feynman propagator  from the finite discontinuity across the imaginary axis. The analytic continuation to the imaginary axis of observables such as the Casimir pressure can still be performed and is well-defined, i.e. one can  Wick-rotate and  retrieve the Lifchitz theory \cite{lif1,lif2}. Then one can perform a standard perturbative calculation to take into account the effects of the axion coupling to the photons.

The effects that axions can have on the Casimir pressure have been already discussed in the literature. First of all, the coupling to matter of axions leads to quantum effects, such as Casimir-Polder forces, induced by the presence of light axions \cite{Klimchitskaya:2015kxa}. We focus {here} on a different aspect, when the coupling of axions to photons leads to a modification of the mode eigenfrequencies and a direct effect on the Casimir pressure. This has been discussed in the context of linear time or spatial variations of the axion field \cite{Kharlanov:2009pv,Ema:2023kvw,Fukushima:2019sjn,Brevik:2021ivj}. Some results concerning Green's functions and energy densities are discussed in the presence of an oscillating axion are presented in \cite{Favitta:2023hlx}. In this paper, we  consider the regime where the axion time variation is very rapid and  cannot be linearised. We average over the fast axion variations and deduce the effects of the axion on very large time scales compared to the inverse axion mass. We find that the correction to the Casimir pressure  oscillates in space with a period proportional to the Compton wavelength $1/m$. For large values of $md$, the pressure becomes repulsive and vanishes as $m^2d^2$ at short distance \cite{Kharlanov:2009pv}, {$d$ being the size of the cavity}.

The paper is arranged as follows. In section \ref{sec:matter} we describe the coupling between photons and the metallic plates, and the existence of narrow resonances. Then in section \ref{sec:path}, we consider the path integral associated to correlators in the $in-in$ formalism. In section \ref{sec:lif}, we recover the classic Lifschitz results {without axions} when the plates have a dissipative behaviour. Finally in section \ref{sec:axion} we calculate the leading correction to the Casimir pressure coming from the axion  contribution to the photon vacuum energy  in an idealised experimental setup. A few technical appendices complete the paper.
\section{Photons in matter coupled to axions}
\label{sec:matter}

\subsection{The action}

As we are interested in the axion in the presence of  metallic plates, we will calculate the quantum Casimir effect using the vacuum Lagrangian and its modified version taking into account the interaction of photons with matter. In vacuum we have
{\be
\mathcal{L} = -\frac{1}{4} F_\munu F^\munu - \frac{\phi}{4 M } F^{\munu}\tilde F_{\mu\nu}
\ee
}
where $\tilde F_{\mu\nu}=  \h \epsilon_{\mu\nu\rho\sigma}F^{\rho\sigma}$ is the dual field strength { and $M$ is an energy scale of the order of the Peccei-Quinn symmetry breaking scale}. We define $\epsilon^{0123}=1$. The pseudo-scalar field $\phi$  will be assumed to be time dependent, as representing the oscillations of the dark matter field in our environment. Notice that when the axion field becomes constant, the coupling between the axion and photons becomes irrelevant as $F\tilde F$ is a derivative term and we do not consider non-trivial topological effects.  We work in the Coulomb gauge where $A^0=0$ and $\partial_i A^i=0$ which selects two transverse polarisations. We are interested in situations where there is a background magnetic field parallel to the plates \cite{Brax:2024cur}. In terms of the fluctuation $a^\mu$ of the potential vector where
$A^\mu=\bar A^\mu +a^\mu$ and $\bar A^\mu$ corresponds to a constant magnetic field parallel to the boundary plates, we have for the action governing the dynamics of the fluctuating field \cite{Brax:2024cur}
{
\be
S= \int d^4 x \left (\frac{1}{2}(\vec e^2 -\vec b^2)+ \frac{\phi}{M} \vec e. \vec b - \vec {\cal J}.\vec  a\right )
\ee
}
where the have introduced the current
\be
{\cal J}^\mu= \frac{1}{2M} \partial_\nu \phi \epsilon^{\mu\nu\rho\sigma} \bar F_{\rho\sigma}.
\ee
In a time-dependent axionic background this corresponds to the current induced by the axion and the external background field
\be
\vec {\cal J}=-\frac{\dot \phi}{M} \vec B.
\ee
This action reproduces the equations of motion for the photon field in the presence of the  axion field.

When coupled to matter, in particular a metal where dissipation takes place, we would like  to find an action which reproduces the phenomenological field equations
of motion
\be
\partial_0 \vec d - \vec \nabla \wedge \vec b = - \frac{\dot \phi}{M} \vec b + \vec {\cal J}.
\ee
which corresponds to
\be
-\partial_0 ( \epsilon \star_t \partial_0 \vec a) + \Delta \vec a=- \frac{\dot \phi}{M} \vec b + \vec {\cal J}.
\ee
where we work in the Coulomb gauge with $a^0=0$ and $\vec e= -\partial_0 \vec a$. Here the displacement field is represented by the convolution
\be
\vec d= \vec e + \vec p = \epsilon \star_t \vec e
\ee
where $\epsilon(t)$ is the permittivity of the medium, $\vec p$ its polarisation and $\star_t$ represents the convolution in the time domain. This phenomenological equation is postulated and  leads to the wave equation in the medium
\be
(\Delta +\epsilon (\omega) \omega^2) \vec a=0
\ee
in the absence of axion source.
The propagation speed is now $1/\sqrt{\epsilon(\omega)}$. In real space, this change of speed becomes the convolution in time with the  permittivity function. The source term is not modified as this depends only on the external magnetic field and the axion background.

One way of finding the appropriate action governing the interaction of photons with matter and the axion field would be to go back to a microscopic model where atoms interact with the local electric and magnetic fields \cite{Bordag:2017qnh}. Here we will use a more effective  description where we introduce a second gauge field $\tilde a ^i$. This doubling of degrees of freedom can be seen as equivalent to introducing the time-reversed process in order to restore the Lagrangian structure of the theory despite dissipation \cite{Galley:2012hxx}.  In this approach we introduce the new action
\be
S= \int d^4 x (\vec{\tilde e}.\vec d -\vec{\tilde b}.\vec b+ \frac{\phi}{M} \vec {\tilde e}. \vec b+ \frac{\tilde \phi}{M} \vec { e}. \vec{\tilde b}  - \vec {\cal J} .\vec{\tilde  a} -\vec {\tilde{\cal J}}.\vec a).
\label{actfull}
\ee
where we have coupled the axion to the ``shadow" gauge field $\tilde a$. We have introduced the shadow axion field and source
\be
\tilde \phi(t)\equiv \phi(-t),\ \ \vec{\tilde{\cal J}}(t)\equiv \vec{\cal J}(-t).
\ee
{The shadow source involves a "shadow" background magnetic field which is identical to the original one in practice as the external magnetic field is time independent. The shadow axion couples to $\vec{\tilde b}$ in the same manner as it appears together with the shadow background magnetic field in $\vec{\tilde {\cal J}}$. This gives the correct sources in the Euler-Lagrange equations.}
The Euler-Lagrange equation for $\tilde a$ reduces to the phenomenological equation of motion
\be
-\partial_0 ( \epsilon \star_t \partial_0 \vec a) + \Delta \vec a=- \frac{\dot \phi}{M} \vec b + \vec {\cal J}.
\label{dis1}
\ee
whilst the equation for $a$ satisfies  the time reversed equation
\be
-\partial_0 ( \hat \epsilon \star_t \partial_0 \vec {\tilde a}) + \Delta \vec {\tilde a}=- \frac{\dot {\tilde \phi}}{M} \vec{\tilde b}+ \vec {\tilde {\cal J}}.
\ee
A classical solution of the two equations is given by $\tilde a(t)= a(-t)$ where $a(t)$ satisfies (\ref{dis1}) as $\phi(t)= \phi_0 \cos mt$ satisfies $\tilde \phi(t)= \phi(t)$  in the equations of motion. This justifies the identification of $\tilde a$ as a shadow field whose behaviour sees time in reverse.

Quantisation is made easier by separating the classical effects from the quantum effects.
We first separate  $a^\mu= a^\mu_{\rm cl}+ a^{\mu}_{\rm qu}$ where $a^\mu_{\rm cl}$ is the classical field sourced by ${\cal J}^\mu$ {and obeying the usual Maxwell equations}. We apply the same decomposition to $\tilde a$.
The resulting action can be written as
\be
S= S_{\rm  qu} + S_{\rm cl}
\ee
where the classical action is
\be
S_{\rm cl}= \int d^4 x (\vec{\tilde e}_{\rm cl}.\vec d_{\rm cl} -\vec{\tilde b}_{\rm cl}.\vec b_{\rm cl}+ \frac{\phi}{M} \vec {\tilde e}_{\rm cl}. \vec b_{\rm cl}+ \frac{\tilde \phi}{M}\vec { e}_{\rm cl}. \vec{\tilde b}_{\rm cl}  - \vec {\cal J} .\vec{\tilde  a}_{\rm cl}- \vec {\tilde{\cal J}}.\vec a_{\rm cl})
\ee
and the quantum action contains only quadratic terms in the quantum fields
\be
S_{\rm  qu}=\int d^4 x (\vec{\tilde e}_{\rm qu}.\vec d_{\rm qu} -\vec{\tilde b}_{\rm qu}.\vec b_{\rm qu}+ \frac{\phi}{M} \vec {\tilde e}_{\rm qu}. \vec b_{\rm qu}+ \frac{\tilde \phi}{M}\vec { e}_{\rm qu}. \vec{\tilde b}_{\rm qu})
\ee
where the classical source term has decoupled.
It is convenient to separate the quantum action into {two parts :} the free theory with the action
\be
S_{0}=\int d^4 x (\vec{\tilde e}_{\rm qu}.\vec d_{\rm qu} -\vec{\tilde b}_{\rm qu}.\vec b_{\rm qu}),
\ee
and the interaction terms
\be
S_I=\int d^4 x \frac{\phi}{M} (\vec {\tilde e}_{\rm qu}. \vec b_{\rm qu}+ \vec { e}_{\rm qu}. \vec{\tilde b}_{\rm qu}).
\ee
The free quantum modes satisfy
\be
-\partial_0( \epsilon\star_t \partial_0 \vec a_{\rm qu}) + \Delta \vec a_{\rm qu}= 0.
\ee
and
\be
-\partial_0( \hat \epsilon\star_t \partial_0 \vec {\tilde a}_{\rm qu}) + \Delta \vec {\tilde a}_{\rm qu}= 0.
\ee
which correspond in Fourier space to
\be
\epsilon(\omega) \omega^2 \vec a_{\rm qu} + \Delta \vec a_{\rm qu}= 0.
\ee
and
\be
\epsilon(-\omega) \omega^2 \vec{\tilde a}_{\rm qu} + \Delta \vec{\tilde a}_{\rm qu}= 0.
\ee
Notice that the modes for $\tilde a$ are obtained from the ones of $a$ by the exchange $\omega \to -\omega$. The modes taking into account the interactions will be obtained {later within} perturbation theory.

\subsection{Resonances}

We are interested in the Casimir situation where {boundary conditions are imposed in the $z$ direction. We consider metallic plates along the $(x,y)$ plane as sketched in Fig.~\ref{fig:setup}.
\begin{figure}[h]
\includegraphics[width=.7\textwidth]{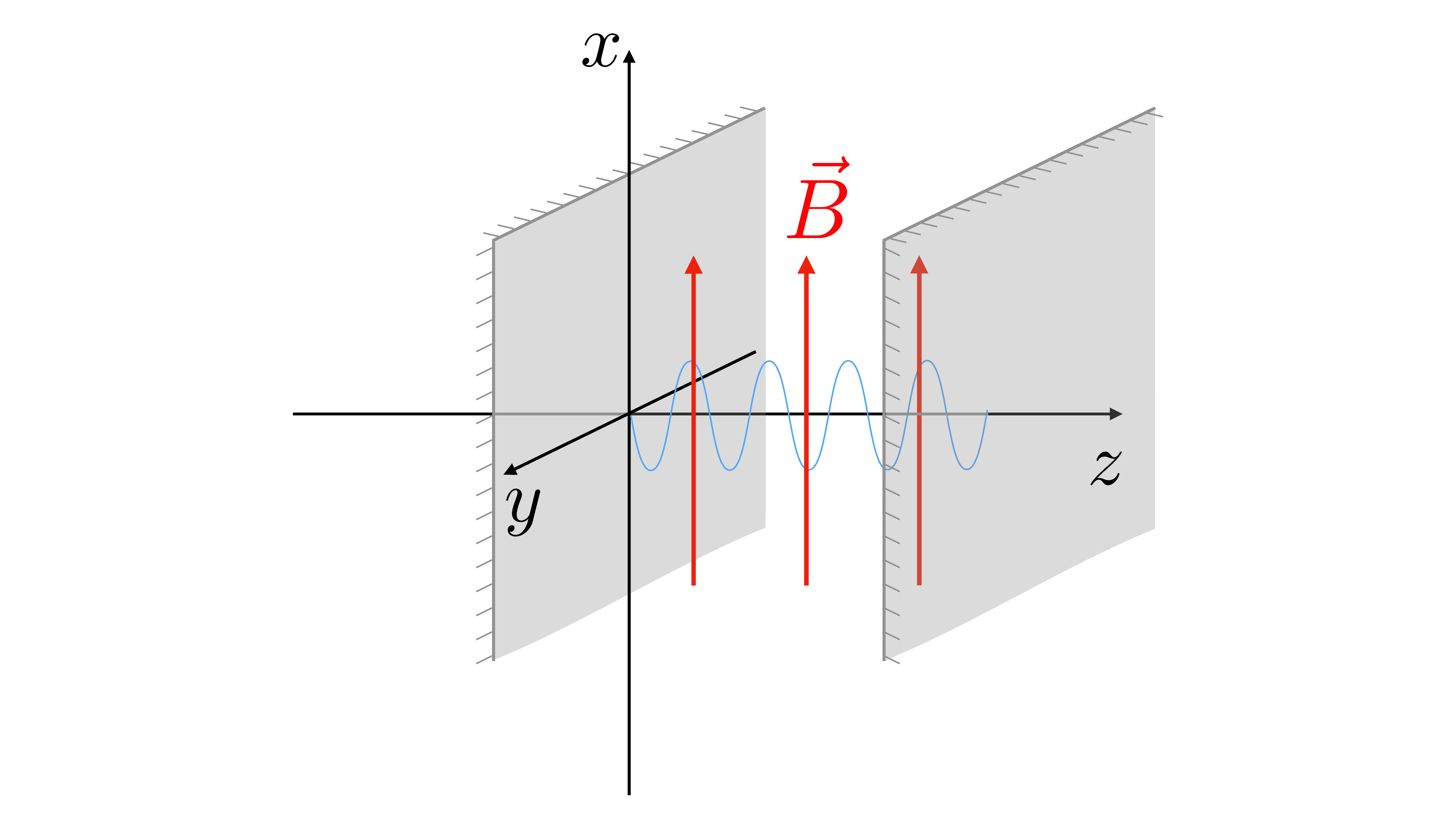}
\caption{A typical Casimir experimental setup. We add a constant magnetic field parallel to the plates. This leads to a classical Casimir pressure due to the presence of the axion which is studied in \cite{Brax:2024cur}. The quantum case studied here involves the modes in the presence of the background axion field. }
\label{fig:setup}
\end{figure}

The problem becomes (1+1)-dimensional and the modes
for $\vec a$ satisfy in momentum space
\be
 \vec a'' + (\omega^2 \epsilon(\omega)-p_\parallel^2) \vec a= 0,
\ee
where $\vec p_\parallel$ is the momentum parallel to the plates. Similarly we have
\be
 \vec{\tilde a}'' + (\omega^2 \bar \epsilon(\omega)-p_\parallel^2) \vec {\tilde a}= 0,
\ee
where we have used that $\bar \epsilon (\omega)= \epsilon (-\omega)$. This implies that the modes for $\vec{\tilde a}$ are simply the complex conjugate of the modes for $\vec a$. We can therefore restrict ourselves to finding the modes of $\vec a$.

We will consider a simple case and 
describe  the permittivity  in the material of the plates using  the Drude model \footnote{The constant magnetic field parallel to the plates does not alter the Lorenz force term in the equation of motion for the charge carriers and therefore does not change the permittivity \cite{Brax:2024cur}.}.
Inside the plates the permittivity is not equal to one and is given by
\be
\epsilon= 1- \frac{\omega_{\rm Pl}^2}{\omega^2 +i\gamma  \omega }
\ee
where $\omega_{\rm Pl}$ is the plasma frequency and $\gamma$ {is the inverse of a damping time related to the conductivity as
$
\sigma= \frac{\omega_{\rm Pl}^2}{\gamma}.
$}
Although this model is controversial when comparing to experimental data of the Casimir effect \cite{Klimchitskaya:2023ebx}, it is a good theoretical testbed to describe dissipation in a simple setting\footnote{For a modern perspective on the permittivity of matter, see \cite{Creminelli:2024lhd} for instance.}.

As the permittivity jumps across boundaries, we must impose conditions across these interfaces. There are two types of boundary conditions corresponding
to the two modes inside a cavity: the transverse electric (TE) and magnetic (TM) modes. We will give a full description of them below. In this
section, we first consider  the $TE$ modes where both $a$ and $\partial_z a$ are continuous across the boundaries. The $TM$ case is such that
$a$ and $\epsilon^{-1}(\omega)\partial_z a$ are continuous. The spectrum in this case is described in appendix \ref{sec:reso}.

We define in matter
\be
\xi= (p^2_\parallel- \omega ^2 \epsilon(\omega))^{1/2}
\ee
where {$p_\parallel$ is the modulus of the momentum in the $x-y$ plane, and} the square root is such that the real part of $\xi$ is always positive along the real axis in the $\omega$-complex plane.
In the  vacuum we define
\be
\Delta= (\omega^2 -p^2_\parallel)^{1/2}.
\ee
The modes are of the form $e^{-i\omega t + i\vec p_\parallel.\vec x_\parallel \pm  i\Delta z} $ in the vacuum,
$e^{-i\omega t + i\vec p_\parallel.\vec x_\parallel + \xi z} $ for $z<0$ and $e^{-i\omega t + i\vec p_\parallel.\vec x_\parallel - \xi z} $ for $z>d$. This choice guarantees the convergence of the modes at both $z=\pm \infty$. There are resonances in the cavity between the plates, and the resonance equation is given by
\be
\Theta (\omega, p_\parallel)=0
\ee
where, see appendix \ref{sec:ele},
\be
\Theta (\omega,p_\parallel)= \frac{1}{\Delta( (1- \frac{\xi^2}{\Delta^2}) \sin \Delta  d -2 \frac{\xi}{\Delta} \cos \Delta d)}.
\ee
This has solutions for
\be
\tan \Delta d= \frac{2\frac{\xi}{\Delta}}{1- \frac{\xi^2}{\Delta^2}}.
\ee
 In the case of a metal with dissipation,
we will only treat the case of weak dissipation implying that $\gamma$ is smaller than the frequencies of interest. {This is justified because in real situations, the plates are good conductors.} In particular as the resonances will be for $\omega_n \propto n/d$, this requires that $d\ll \gamma^{-1}$~\footnote{Another family of resonances exists when $d\gg \gamma^{-1}$ corresponding to $\omega_n \ll \gamma$ \cite{Brax:2024cur}. We will not consider this case and consider only $d\ll \gamma^{-1}$.}.  In this regime where  $\gamma \ll \vert \omega\vert \ll \omega_{\rm Pl}$ implying that $\Delta \ll \omega_{\rm Pl}$ we have
\be
\frac{\xi}{\Delta} \simeq \frac{\omega_{\rm Pl}}{\Delta}(1- i\frac{\gamma}{2\omega}),
\ee
which is large and therefore
\be
\tan \Delta d= - \frac{2\Delta}{\xi}.
\ee
The solutions are obtained by iteration and  are given by
\be
\Delta_n d= n\pi -2 \frac{n\pi}{\omega_{\rm Pl}d}(1+i\frac{\gamma}{2\omega_n^{(0)}}), \ n\ne 0
\label{res1}
\ee
where to  leading order
\be
\omega_n^{(0)}= {\rm sign}(n) \sqrt{ \frac{n^2\pi^2}{d^2}+p_\parallel^2}.
\ee
The imaginary part of the resonance frequency is obtained as
\be
\omega_n^2 =  \frac{n^2\pi^2}{d^2}+p_\parallel^2- i\,\frac{2n^2\pi^2}{\omega_{\rm Pl}d^2}\,\frac{\gamma}{\omega_n^{(0)}}.
\ee
 The solutions are
\be
\omega_n= \omega_n^{(0)}-i\,\frac{n^2\pi^2}{\omega_{\rm Pl}d^2}\,\frac{\gamma}{\frac{n^2\pi^2}{d^2}+p_\parallel^2}, \ n\ne 0
\label{reson}
\ee
corresponding to the retarded choice of poles below the real axis.
Here the fact that the poles are not on the real axis is due to the finite conductivity which induces the presence of finite-lifetime resonances. There are only a finite number of resonances of energy lower than the plasma frequency. 

 Above the plasma frequency, the plates become transparent and no resonances are present. Indeed the resonance equation becomes
 \be
 \tan \Delta d= 2i
 \ee
 or equivalently $\tanh \Delta d= 2$ which has no solution. Hence, there are no resonances above the plasma frequency. Of course we have only studied the two limiting cases of frequencies well below and well above the plasma frequencies. This is enough to state that no resonances are with a large frequency compared to $\omega_{\rm Pl}$ and that the spectrum contains only a finite number of resonances of finite width. Moreover, as we will see, the fact that the imaginary part of the eigen-frequencies along the positive real axis is negative follows from the fact that there is dissipation and the fields must decay at $t=+\infty$.

The case of the $TM$ polarisation is described in detail in appendix \ref{sec:reso}. The spectrum is still composed of resonances below the real axis. We will denote by $+$ the $TE$ polarisation and $-$ the $TM$ one.

\subsection{The retarded  and Feynman propagators}
\label{sec:ff}
\subsubsection{The retarded propagator}
The Casimir pressure is most easily calculated using a quantum formalism where the Feynman propagator in the presence of the non-trivial permittivity and the boundary plaques are present. Typically in the appendices (\ref{sec:ele}) and (\ref{sec:mag}), the Green's functions satisfying
\be
-\partial_0( \epsilon\star_t \partial_0 G) + \Delta G=  \delta^{(4)}(x^\mu-y^\mu)
\ee
are investigated. The electric and magnetic cases are simply distinguished by the type of boundary conditions that one must impose at the interface between the plates and the vacuum between the plaques.
In the following we will use time-translation invariance and space-translation invariance in the $(x,y)$ plane along the plaques.
In terms of Fourier decomposition the Green's functions satisfy
\be
(\partial_z^2 -p_\parallel^2 + \epsilon(\omega) \omega^2) G= \delta(z-z_0).
\ee
As well-known in the case of vacuum with no boundaries, i.e. quantum field theory in Minkowski space-time,  there are different Green's functions depending on prescriptions of contour integrals in the $\omega$-plane. The same cases can be obtained in the system with boundaries and a non-trivial permittivity. The solutions given in the appendices correspond to retarded propagators as we have shown in the previous section that the poles of the Green's functions are below the real axis. They are suited to the study of the radiation of the electric and magnetic fields when real currents and charge distributions are present inside the plates for instance.
This follows from the explicit representation
\be
G(z,\omega, p_\parallel;z_0)= \Theta(\omega, p_\parallel) H(z,\omega, p_\parallel;z_0)
\ee
where the functions $H(z,\omega,p_\parallel;z_0)$ can be found in the appendices.
We have seen that the poles of $\Theta$, giving rise to resonances, are in a  finite number and located at $\omega=\omega_n$ in the right half plane $\Re \omega >0$. The other resonances are situates at $-\bar\omega_n$ as we have the property $\bar \Theta (\omega, p_\parallel)= \Theta(-\bar\omega, p_\parallel)$ when $\omega$ is extended to the complex $\omega$-plane. This implies that the resonances appear in pairs $(\omega_n,-\bar\omega_n)$. As all the resonances $\omega_n$ in the right half plane have a negative imaginary part, their images $-\bar \omega_n$ have also a negative imaginary part too and are situated in the left half plane.
{This is illustrated in Fig.~\ref{fig:retarded_1}.}

\begin{figure}[h]
\centering
\includegraphics[width=.9\textwidth]{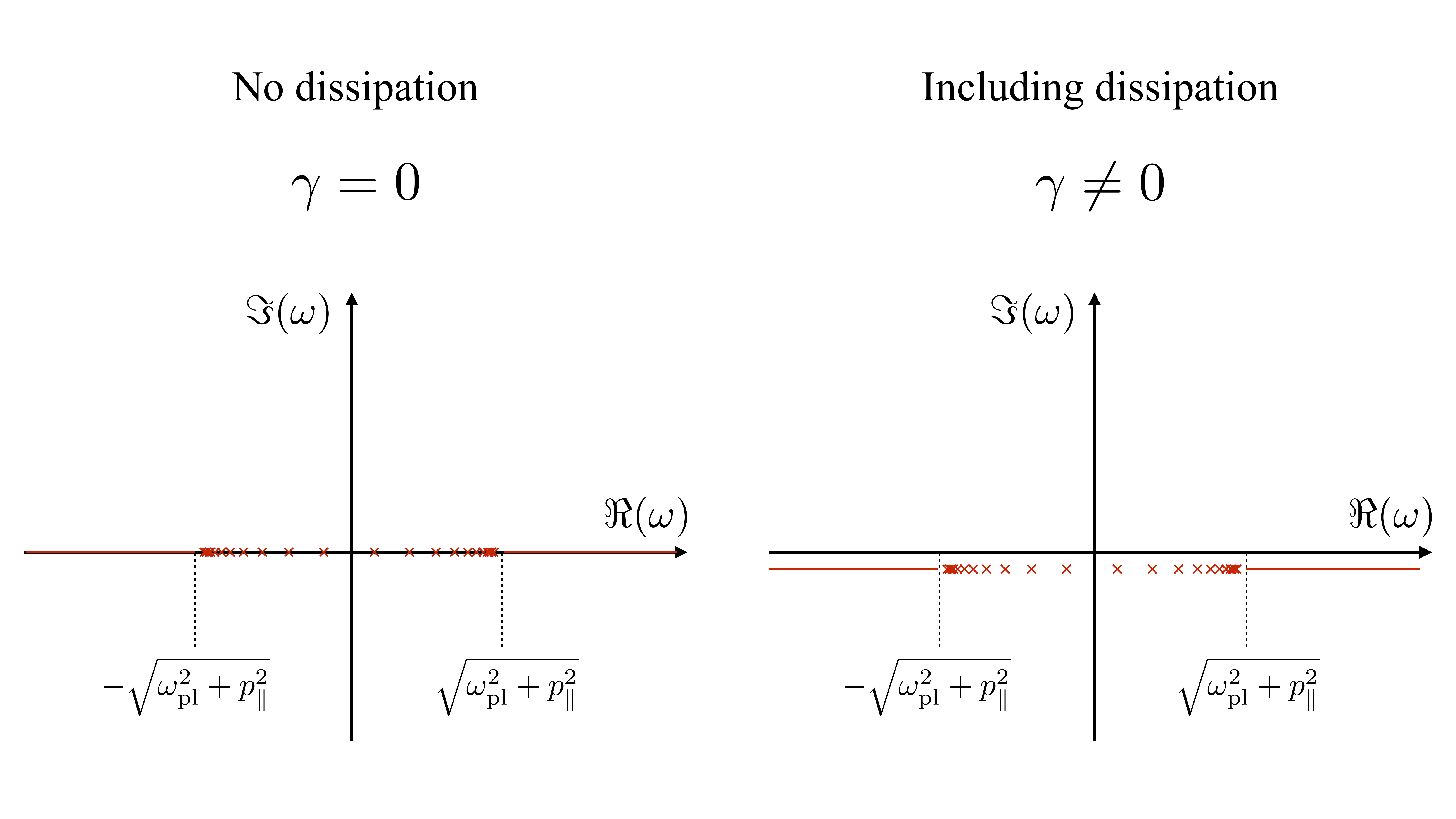}
\caption{{Location of the poles in the case without dissipation (left) and with dissipation (right). The cuts above the plasma frequency correspond to the transparency of the metal at high energy. \label{fig:retarded_1}}}
\end{figure}

We want to  reconstruct
\be
G(z, \vec x_\parallel, t;z_0,\vec x_{0\parallel},t_0)= \int \slashed{d} \omega \slashed d^2 {p_\parallel} e^{-i\omega (t-t_0)} e^{i\vec p_{\parallel}.(\vec x_\parallel-\vec x_{0\parallel})} G(z,\omega, p_\parallel;z_0).
\ee
The integrand  $G(z,\omega, p_\parallel;z_0)$ has poles in the complex $\omega$ plane which are located below the real  axis. When $t>t_0$, we can close the contour in the complex $\omega$ plane in the lower half-plane on a large circular contour. Inside the contour the poles are $\omega_n$ and $-\bar\omega_n$. The contributions from the $\omega_n$'s with residues $G_n(\omega_n, p_\parallel)$ given by
\be
G_n(\omega_n,p_\parallel)= \frac{1}{\frac{d\Theta^{-1}}{d\omega}\vert_{\omega=\omega_n}} H(z,\omega_n, p_\parallel)
\ee
lead to
\be
G_+(z, \vec x_\parallel, t;z_0,\vec x_0,t_0)\supset  -i\sum_n \int  \slashed d^2 {p_\parallel} e^{-i\omega_n (t-t_0)} e^{i\vec p_{\parallel}.(\vec x_\parallel-\vec x_{0\parallel})} G_n(\omega_n,p_\parallel) Y(t-t_0)
\ee
where the minus sign comes from the integral around each pole giving $-2\pi iG_n(p_\parallel)$. The function on the right-hand side is on-shell as the
frequencies are the eigen-frequencies of the system.

The Green's function depends on two complex functions $\Delta$ and $\xi$ which have branch cuts in the complex $\omega$-plane. The $\Delta$ function has a branch cut between $-p_\parallel$ and $p_\parallel$ where it takes opposite values across the cut. Now the Green's function can be explicitly seen to be an even function of $\Delta$, see  the appendices \ref{sec:ele} and \ref{sec:mag}. As a result, there is no branch cut due to $\Delta$ for the Green's functions. This is not the same for $\xi$ which has two branch cuts joining infinity from the two branch points
\be
\omega_{\pm}= \pm (\omega_{\rm Pl}^2 + p_\parallel^2)^{1/2}- \frac{i\gamma \omega^2_{\rm Pl}}{2(p^2_\parallel+\omega^2_{\rm Pl})}
\ee
below the real axis.
Above the branch cut starting at $\omega_+$, we have $\xi =-i \sqrt{\vert p^2_\parallel -\epsilon(\omega) \omega^2\vert}$ and the opposite value below the cut. The situation is reversed along the branch cut starting at $\omega_-$ with a value $i\sqrt{\vert p^2_\parallel -\epsilon(\omega) \omega^2\vert}$ above the cut and its opposite below the cut.
As a result there is a contribution from the continuum along the branch cut of $\xi$ given by
\be
G_+(z, \vec x_\parallel, t;z_0,\vec x_{0\parallel},t_0)\supset
\int_{\omega_+}^{\infty} \slashed{d} \omega \int  \slashed d^2 {p_\parallel} e^{-i\omega (t-t_0)} e^{i\vec p_{\parallel}.(\vec x_\parallel-\vec x_{0\parallel})}(G(-i\vert \xi\vert)- G(i \vert \xi\vert)) Y(t-t_0)
\ee
where we have denoted $\vert \xi\vert= \sqrt{\vert p^2_\parallel-\epsilon(\omega) \omega^2\vert}$ and simplified the notation to emphasize the dependence of $G$ only on $\xi$ above and below the cut. The contributions from the resonances and the continuum must be added to give the full expression of $G_+$.

Similarly the poles at $-\bar \omega_n$ have a residue
\be
G_n(-\bar\omega_n,p_\parallel)= -\frac{1}{\overline{\frac{d\Theta^{-1}}{d\omega}}\vert_{\omega=\omega_n}} \bar H(z,\omega_n, p_\parallel)
\ee
where $H(z,-\bar\omega_n, p_\parallel)=  \bar H(z,\omega_n, p_\parallel)$ leading to a contribution
\be
G_-(z, \vec x_\parallel, t;z_0,\vec x_0,t_0)\supset -i\sum_n \int  \slashed d^2 {p_\parallel} e^{i\bar\omega_n (t-t_0)} e^{i\vec p_{\parallel}.(\vec x_\parallel-\vec x_{0\parallel})} G_n(-\bar \omega_n,p_\parallel) Y(t-t_0)
\ee
which is also on-shell. The continuum gives rise to
\be
G_-(z, \vec x_\parallel, t;z_0,\vec x_{0\parallel},t_0)\supset
\int_{-\infty}^{\omega_-}\slashed{d} \omega \int  \slashed d^2 {p_\parallel} e^{-i\omega (t-t_0)} e^{i\vec p_{\parallel}.(\vec x_\parallel-\vec x_{0\parallel})}(G(i\vert \xi\vert)- G(-i \vert \xi\vert)) Y(t-t_0)
\ee
The two contributions to $G_\pm$ are distinguished by separating the frequencies into the ones with positive real parts and negative ones.
Now  when $t<t_0$, one must close the contour in the upper-half plane where there are no poles and the function vanishes. As a result we have
\be
G(z, \vec x_\parallel, t;z_0,\vec x_{0\parallel},t_0)= (G_+(z, \vec x_\parallel, t;z_0,\vec x_0,t_0) +G_-(z, \vec x_\parallel, t;z_0,\vec x_0,t_0)) Y(t-t_0)
\ee
This is a retarded Green's function where  the on-shell resonances and the continuum  enter in the evaluation.
Notice that the on-shell term  can be written as
\be
G_{\rm onshell}(z ,\vec x_\parallel, t;z_0,\vec x_{0\parallel},t_0) = -i\int \slashed{d} \omega \slashed d^2 {p_\parallel} e^{-i\omega (t-t_0)} e^{i\vec p_{\parallel}.(\vec x_\parallel-\vec x_{0\parallel})}  \slashed{\delta}( \Theta^{-1}(\omega,p_\parallel)) H(z,\omega, p_\parallel) Y(t-t_0)
\ee
where the on-shell condition is apparent. The integration contours mentioned here are illustrated in Fig.~\ref{fig:retarded_2}

\begin{figure}[h]
\centering
\includegraphics[width=.6\textwidth]{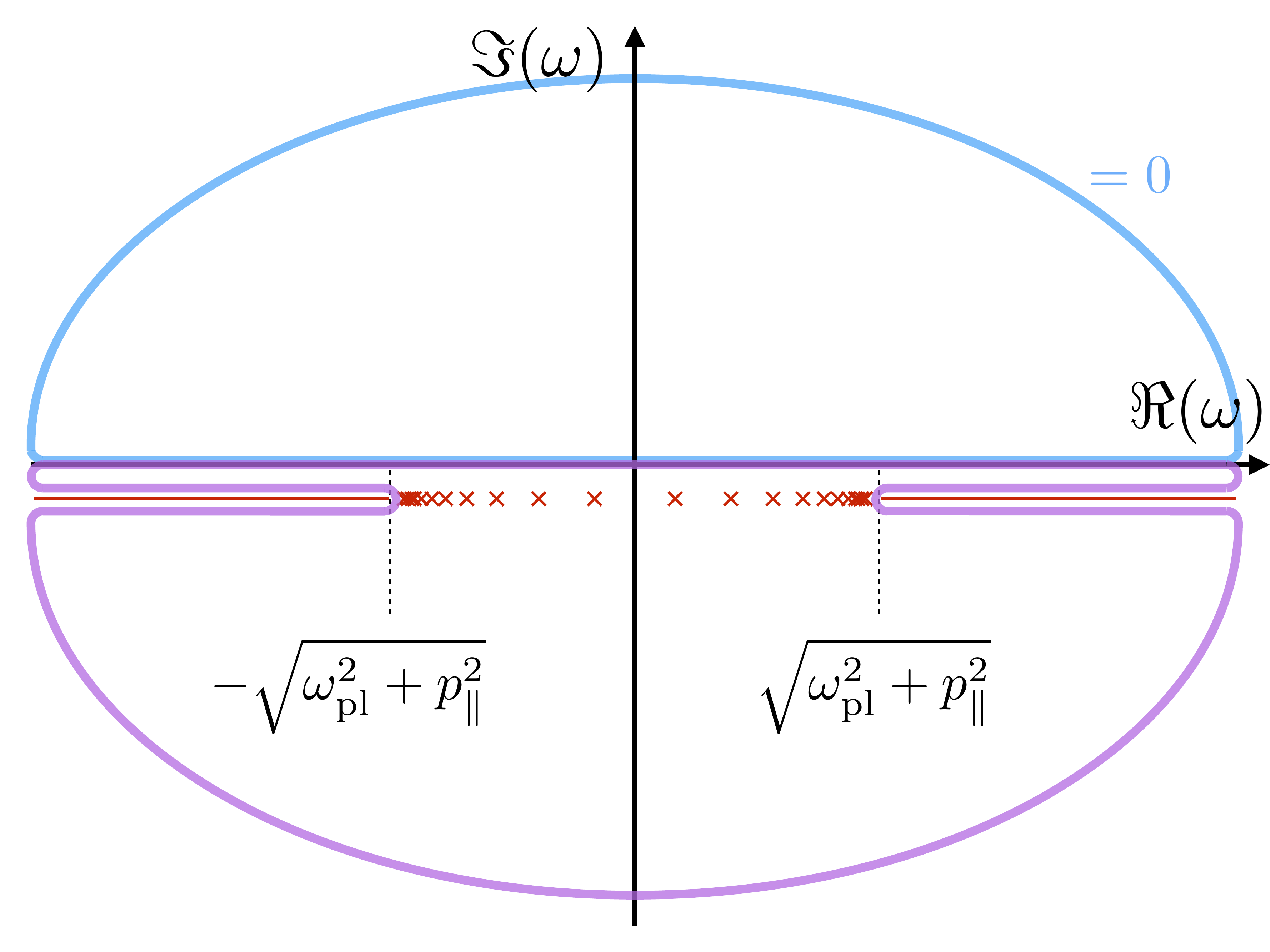}
\caption{{Integration contours for the retarded Green's function. In the upper half plane, the retarded Green's function vanishes. This is not the case when closing the contour below the axis dues to the poles and the transparency cuts. \label{fig:retarded_2}}}
\end{figure}

\subsubsection{The Feynman propagator}
\label{sec:feynman}
In the quantum field theoretic  treatment of the Casimir effect which will be expounded below, the Feynman propagator corresponding to the $T$-product of the field is required. The Feynman propagator can be obtained using an adaptation of  the Schwarz reflection principle where holomorphic functions are extended from the upper half plane to the lower one. In the right half place with $\Re \omega >0$, we take
\be
\Re \omega \ge 0: G_F(\omega)= G(\omega)
\ee
where we have omitted the dependence on the other variables to simplify the notation. In the left half plane we define
\be
\Re \omega \le 0: G_F(\omega)= \bar G(\bar \omega)= G(-\omega).
\ee
The Feynman propagator is obtained through a reflection with respect to the origin which sends the resonances in the right half plane to the left half plane above the negative real axis.
This defines the Feynman propagator in the whole $\omega$ complex plane.
{In terms of integrals, the Feynman propagator reads explicitly
\be
G_F(t,\vec x)= 2\int_0^\infty \slashed{d}\omega\int  \slashed{d}^2 p_\parallel \cos \omega t ~e^{i\vec p_\parallel. \vec x_\parallel} G(\omega, \vec p_\parallel, z)
\ee
which satisfies the same partial derivative equation as the retarded propagator\footnote{We use
$
\int_0^\infty d\omega \cos \omega t = \pi \delta(t)
$ implying that $-\partial_t(\epsilon(t) \star_t \partial_t )G_F + \Delta G_F= \delta^{(4)}(x^\mu) $.}. }
The locations of the poles for the definition of the Feynman propagator are shown in Fig.~\ref{fig:feynman_1}. There is a discontinuity on the imaginary axis if $\gamma\neq 0$, except at the origin.

\begin{figure}[h]
    \centering
    \includegraphics[width=0.6\textwidth]{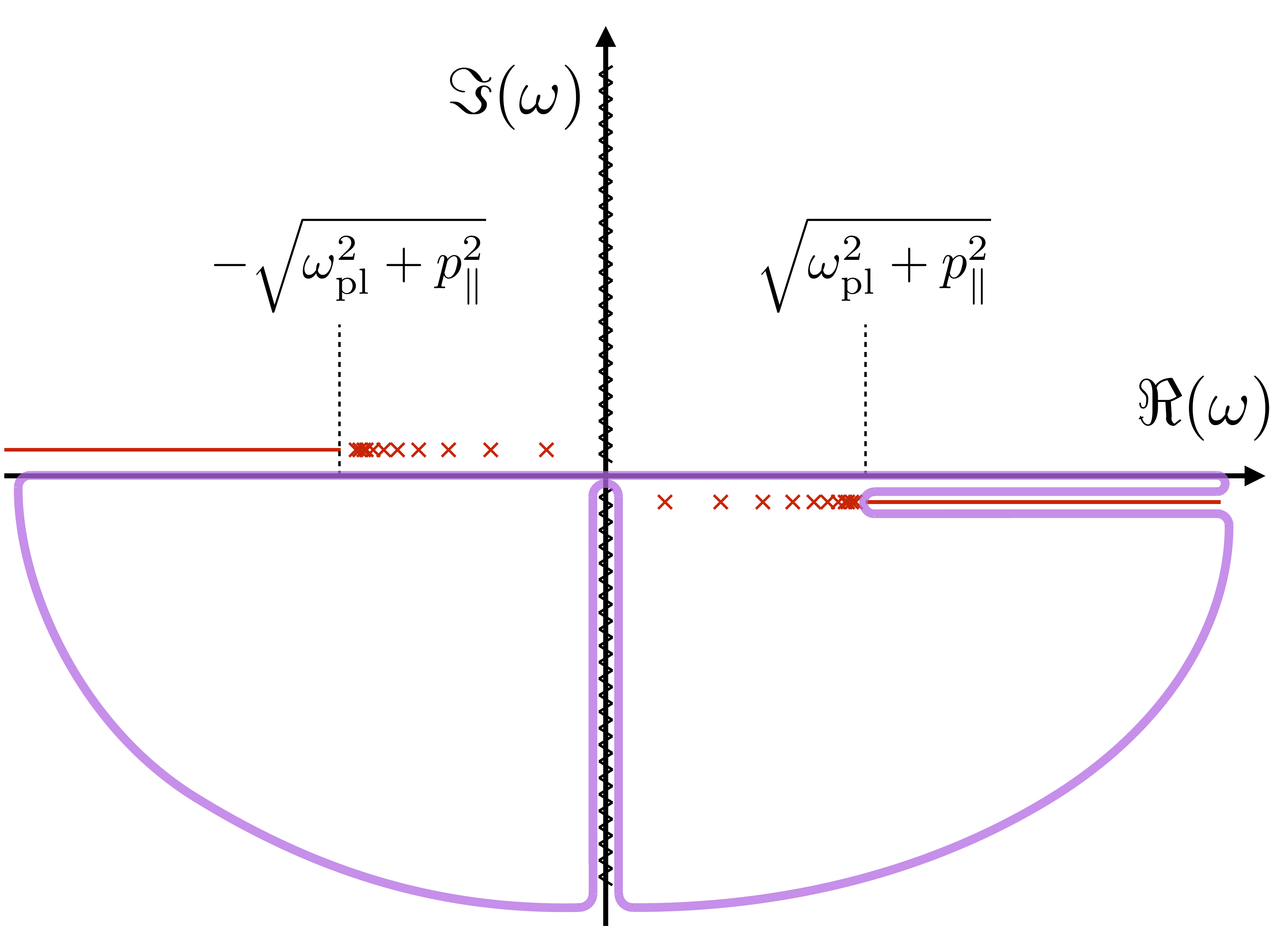}
    \caption{Location of the poles and cuts for the definition of the Feynman propagator and relevant integration contours.}
    \label{fig:feynman_1}
\end{figure}

Now the poles of the Feynman propagator are located at $\omega_n$ and $-\omega_n$ below and above the real axis. The residues of the poles at $-\omega_n$ are given by   $-G_n(\omega_n,p_\parallel)$.
Closing the contour of the inverse Fourier integral in the lower half $\omega$-plane, we have
\be
G_F(z, \vec x_\parallel, t;z_0,\vec x_{0\parallel},t_0)\supset- i\int  \slashed d^2 {p_\parallel} e^{-i\omega_n (t-t_0)} e^{i\vec p_{\parallel}.(\vec x_\parallel-\vec x_{0\parallel})}G_n(\omega_n,p_\parallel) Y(t-t_0)
\ee
whereas when closing in the upper half plane we have
\be
G_F(z, \vec x_\parallel, t;z_0,\vec x_{0\parallel},t_0)\supset - i\int  \slashed d^2 {p_\parallel} e^{i\omega_n (t-t_0)} e^{i\vec p_{\parallel}.(\vec x_\parallel-\vec x_{0\parallel})}G_n(\omega_n,p_\parallel) Y(t_0-t).
\ee
The contribution from the continuum along the branch cut of $\xi$ is given by
\be
G_F(z, \vec x_\parallel, t;z_0,\vec x_{0\parallel},t_0)\supset
\int_{\omega_+}^{\infty} \slashed{d} \omega \int  \slashed d^2 {p_\parallel} e^{-i\omega (t-t_0)} e^{i\vec p_{\parallel}.(\vec x_\parallel-\vec x_{0\parallel})}(G(-i\vert \xi\vert)- G(i \vert \xi\vert)) Y(t-t_0)
\ee
where we have denoted $\vert \xi\vert= \sqrt{\vert p^2_\parallel-\epsilon(\omega) \omega^2\vert}$ and simplified the notation to emphasize the dependence of $G$ only on $\xi$ above and below the cut. All in all we have
\begin{eqnarray}
&&G_F^+(z, \vec x_\parallel, t;z_0,\vec x_{0\parallel},t_0)=
(- i\int  \slashed d^2 {p_\parallel} e^{-i\omega_n (t-t_0)} e^{i\vec p_{\parallel}.(\vec x_\parallel-\vec x_{0\parallel})}G_n(\omega_n,p_\parallel)\nonumber \\ &&+\int_{\omega_+}^{\infty} \slashed{d} \omega \int  \slashed d^2 {p_\parallel} e^{-i\omega (t-t_0)} e^{i\vec p_{\parallel}.(\vec x_\parallel-\vec x_{0\parallel})}(G(-i\vert \xi\vert)- G(i \vert \xi\vert)))Y(t-t_0)\nonumber \\
\end{eqnarray}
Similarly, the closing of the contour in the upper half plane gives a contribution along the cut between $-\infty$ and $-\omega_+$
implying that the Feynman propagator is decomposed as
\be
G_F(z, \vec x_\parallel, t;z_0,\vec x_{0\parallel},t_0)= G_F^+(z, \vec x_\parallel, t;z_0,\vec x_{0\parallel},t_0)Y(t-t_0) + G_F^+(z_0, \vec x_{0\parallel}, t_0;z,\vec x_{\parallel},t)Y(t_0-t)
\ee
where $G_F^+$ is the positive frequency propagator comprising both the on-shell and the continuum contributions.
This is a  $T$-product and coincides with what is expected for a  Feynman propagator.
Similar properties are valid for the magnetic Green's function upon substituting $G\to G_M$.

This definition of the Feynman propagator is such that  the function $G_F(\omega)$ is not continuous across the imaginary axis. The Feynman propagator is only continuous on the imaginary axis provided
\be
G(i\omega)= G(-i\omega)
\ee
which is only valid  when $\gamma= 0$ and approximately in  the weak limit of dissipation $\gamma \to 0$. In this case the resonances are displaced from the real axis by an infinitesimal amount. When this is not the case, the Feynman propagator has a branch cut on the imaginary axis apart from the origin where it is continuous. This is discussed in the next section.
\subsubsection{The Feynman propagator for   finite dissipation}

When $\gamma \ne 0$ and is not taken to be infinitesimal, the Feynman propagator defined in the complex $\omega$-plane by a reflection across the origin is not continuous along the imaginary axis apart from the origin. When calculating the Fourier transform as an integral along the real line $\omega \in \mathbb{R}$, this is not an issue. When going to the complex plane and closing the contour in the lower half plane for instance, this requires to modify the integration path by including a contribution going from $-i\infty$ to $0$ and back to $-i\infty$ along both sides of the branch cut which are joined at the origin where the propagator is continuous. This gives a new contribution to $G_F^+$ which reads
\be
\delta G_F^+= \int_{-i\infty}^{0} \slashed{d} \omega \int  \slashed d^2 {p_\parallel} e^{-i\omega (t-t_0)} e^{i\vec p_{\parallel}.(\vec x_\parallel-\vec x_{0\parallel})}(G(\omega)- G(-\omega))Y(t-t_0)
\ee
This is necessary  to lift the obstruction to constructing the Feynman propagator as a continuous function across the imaginary axis. As expected, this new contribution preserves the structure of the Feynman propagator as a T-product. 

In practice, we will see that it is easier to calculate the Feynman propagator at equal times by first performing a Wick's rotation sending the real axis to the imaginary axis via a rotation of a $\pi/2$ angle. As the axis is rotated, it does not cross poles and branch cuts. The equality of the two integrals over $\omega$ is guaranteed as long the integrand converges to zero on a large circular quadrant joining the two axes. For very large $\vert\omega\vert$ the permittivity is close to unity and the Feynman propagator behaves like the one in empty space in $1/\vert \omega \vert $. This justifies the validity of Wick's rotation. {Instead of using the contour drawn in Fig.~\ref{fig:feynman_1}, the integral can be performed very close to the positive part of the imaginary axis after Wick's rotation, as  schematically shown in Fig.~\ref{fig:feynman_2}. Because of the parity on the imaginary axis, this gives half the desired result}.

\begin{figure}[h]
    \centering
    \includegraphics[width=0.6\textwidth]{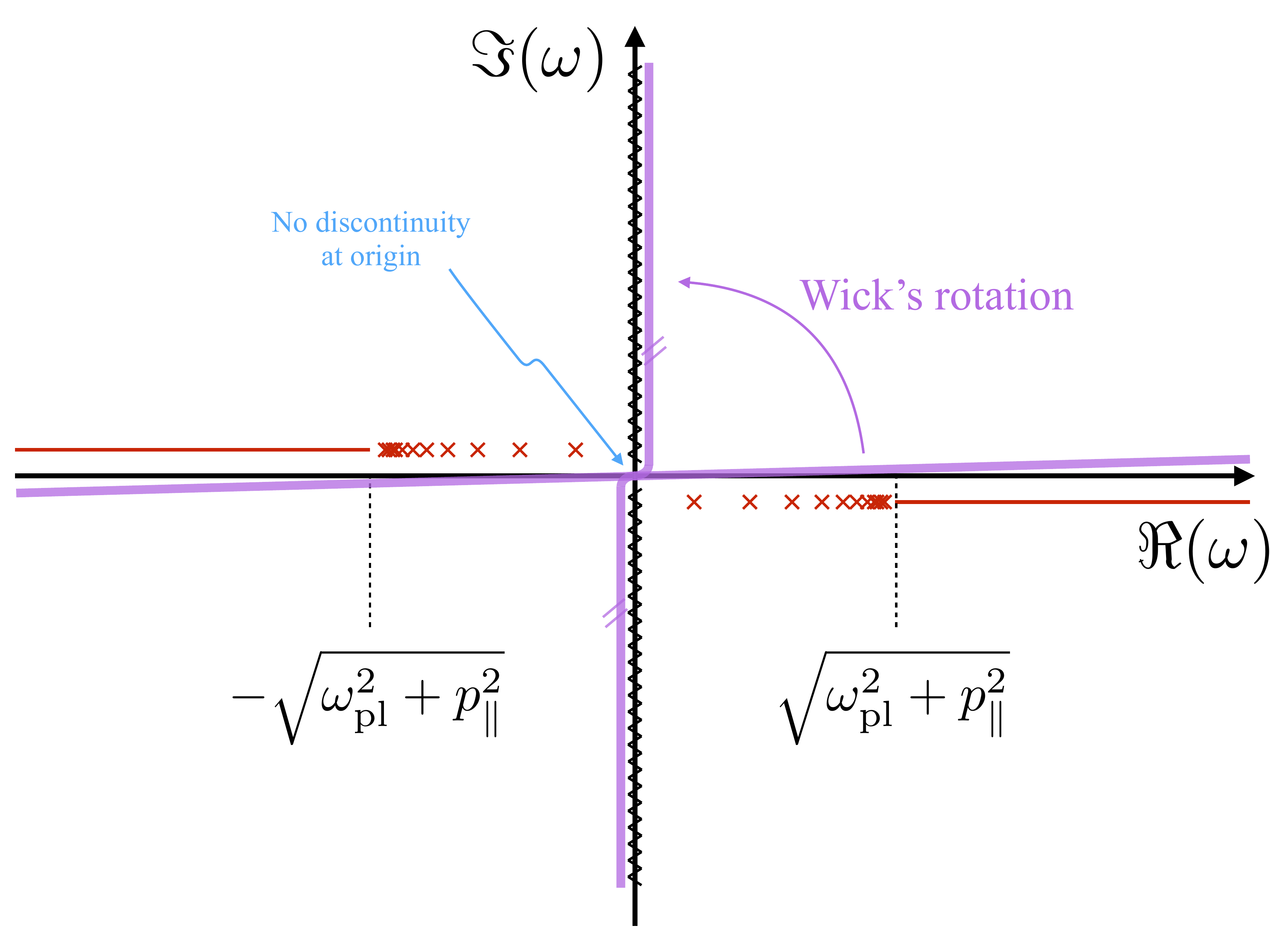}
    \caption{The Feynman propagator can be Wick-rotated to the vicinity of the imaginary axis. After this  Wick's rotation the integral is computed on a line infinitely close of the positive part of the imaginary axis.}
    \label{fig:feynman_2}
\end{figure}

After the rotation the integral for the Feynman propagator lies to the right of the imaginary axis in the upper half plane and to its left in the lower half. Analytically we have $G_F(i\omega)= G(i\omega)$ along this path. As a result, we can use the expressions for the retarded Green's function to obtain the Casimir pressure for instance as we shall see below.

\section{Path integral quantisation}
\label{sec:path}
\subsection{The two polarisations}
A  direct approach to quantisation bypasses the interpretation of the field $a_i$ in terms of particles, which is flawed as Poincar\'e invariance is broken and there is dissipation in the system.
We first decompose the vectors $a_i,\ i=(x,y,z),$ as 
\be
a_i(\vec x,t)= \sum_{\alpha=\pm}\int \slashed{d}^3 k e_{\alpha i}(k) a_\alpha (k,t) e^{i\vec k.\vec x}= \sum_{\alpha} e_{\alpha i}\star_x a_\alpha
\ee
where $\star_x$ is the convolution operator in space.  The polarisation vectors are complex such that $\bar e_{i\alpha}(k)= e_{i\alpha}(-k)$ and satisfy the familiar orthogonality relations
\be
\sum_{\alpha}  \bar e_{\alpha i}(k)  e_{\alpha j}(k)=\delta_{ij}- \frac{ k_i k_j}{k}, \ \ k^i e_{\alpha i}(k)=0,\ \ \bar e_{\alpha i}(k) e^i_{\beta}(k)=\delta_{\alpha \beta}
\ee
where $k^2=  k_i k^i$. The momenta are real. The vector field is Hermitian as long as the two scalar polarisation fields satisfy $a_\alpha^\dagger= a_\alpha$, i.e. they are real scalar fields. Similarly we introduce two scalar field $\tilde a _{\alpha}$ for the shadow vector field $\vec{\tilde a}$.
We define the Fourier tranform for each polarisation scalar as
\be
a_\alpha (\vec x,t)= \int \slashed{d}^3 k   a_\alpha (k,t) e^{i\vec k.\vec x}.
\ee
As a consequences we have for the displacement vector
\be
d_i=-\sum_{\alpha=\pm}\int \slashed{d}^3 k e_{\alpha i}(k) (\epsilon\star_t\dot a_\alpha) (k,t) e^{i\vec k.\vec x}= \sum_{\alpha=\pm } e_{\alpha i}\star_x \dot d_\alpha
\ee
and we have used the property $\partial_0 (\epsilon \star_t a_\alpha)(k,t)= (\epsilon \star_t \dot a_\alpha)(k,t)$.

\subsection{The Hamiltonian and dissipation}

Let us now come back to the quantisation procedure.
We can also calculate the associated Hamiltonian using the conjugate momenta
\be
\vec \pi_a= \hat \epsilon\star_t \partial_0 \vec{\tilde a} -\frac{\phi}{M} \vec {\tilde b}, \ \ \vec \pi_{\tilde a}=  \epsilon\star_t \partial_0 \vec{ a}-\frac{\tilde \phi}{M} \vec { b}
\ee
and then obtain
\be
H= \int d^3 x ( \partial_0 \vec{\tilde a} (\epsilon \star_t \partial_0 \vec a) + \partial_i \vec a.\partial_i \vec{\tilde a} + \vec {\cal J} .\vec{\tilde  a}+ \vec{\tilde{\cal J}}.\vec a))
\ee
where the  dependence on the axion field only appears in the source terms.
From the quantum action $S_{\rm qu}$ we get the Hamiltonian
\be
H_{\rm qu}= \int d^3x ( \partial_0 \tilde a^i (\epsilon\star_t \partial_0 a^i) + \partial_i \tilde a _j \partial_i a_j)
\ee
where we will suppress the ${\rm qu}$ index for convenience below as we are only dealing with the quantum properties from now on.
Notice that the explicit axion interactions do not enter the Hamiltonian. They appear in the definition of the canonical momenta.

We will separate the Hamiltonian corresponding to the free theory in vacuum and the part containing the matter permittivity. We have
\be
H_{0}=  \int d^3x ( \partial_0 \tilde a_i \partial_0 a_i  +  \partial_i \tilde a _j \partial_i a_j) =\sum_{\alpha=\pm}\int d^3 x( \partial _0 \tilde a_\alpha \partial_0 a_\alpha + \partial_i \tilde a _\alpha \partial_i a_\alpha)
\ee
corresponding to two systems of independent free scalar fields.
The free Hamiltonian is corrected by interaction terms
\be
H_I=  \sum_{\alpha=\pm} \int d^3x ( (\partial_0 \tilde  a_\alpha)  \chi\star_t \partial_0 a_\alpha)
\ee
where we have defined
\be
\chi(\omega)= \epsilon (\omega)-1.
\ee
Associated to the free Hamiltonian are the free conjugate momenta
\be
\vec \pi_a^0= \partial_0 \vec{\tilde a}, \ \ \vec \pi_{\tilde a}^0=   \partial_0 \vec{ a}
\ee
such that
\be
H_{0}=  \int d^3x ( \vec\pi^0_{a}.\vec \pi^0_{\tilde a} +  \partial_i \tilde a _j \partial_i a_j)
\ee
and the interaction Hamiltonian
\be
H_I=   \int d^3x ( \vec \pi^0_{\tilde a} .  \chi\star_t \vec \pi_a^0).
\ee
In the schr\"odinger picture, the free conjugate momenta are associated to the functional derivatives $\vec \pi_a^0=\frac{1}{i} \frac{\delta}{\delta \vec a}$ and similarly for $\vec \pi^0_{\tilde a}=\frac{1}{i} \frac{\delta}{\delta \vec{\tilde a} }$. As operators the two momenta commute. Now
let us take the Hermitian conjugate of the interaction Hamiltonian density
\be
 [(\partial_0 \tilde a_\alpha)  (\chi\star_t \partial_0 a_\alpha)]^\dagger = (\hat \chi\star_t\partial_0 a_\alpha)  (\partial_0 \tilde a_\alpha)
\ee
where $\hat \chi(t)= \chi(-t)$.
We see that
\be
H_I^\dagger\ne H_I
\ee
as soon as $\chi(-\omega)\ne \chi(\omega)$,
i.e. the interaction Hamiltonian is not Hermitian. This is a consequence of dissipation.

\subsection{Schwinger-Keldysh formalism}

We are interested in vacuum expectation values in this theory with dissipation. In the Heisenberg representation of quantum mechanics where
operators evolve and states are fixed, we shall quantise by taking as states (and the full Hilbert space of states) the one of the free theory. The operators evolve in time and the generator of the time evolution is the Hamiltonian. The free evolution operator  satisfies
\be
\partial_t U_{0}= -i H_{0} U_{0}
\ee
whose solution is taken as the time ordered product
\be
U_{0}(t)= T( e^{-i\int_{-\infty}^t H_{0}(t') dt'}).
\ee
It is convenient to define the operators in the interaction representation where this free evolution is taken into account
\be
a_{\alpha I}(t,\vec x)= U_{0}^\dagger(t) a_{\alpha I}(-\infty,\vec x)U_{0}(t).
\ee
The free evolution operator is unitary as $H_0$ is Hermitian.

The time evolution of the operators involves the total Hamiltonian which is not Hermitian hence the evolution operator is not unitary. It satisfies
\be
\partial_t U= -i H U
\ee
whose solution is again the time ordered product, i.e. operators are ordered from left to right from the greatest time to the lowest
\be
U(t)= T\Lp e^{-i\int_{-\infty}^t H (t') dt'}\Rp.
\ee
In the interaction picture we introduce
\be
U_I(t)= U_{0}^\dagger (t) U(t)
\ee
which satisfies the equation
\be
\partial_t U_{I}= -i  U_{0}^\dagger H_I(a_\alpha,t) U_{0} U_I.
\ee
We define
\be
{\cal H}_I(a_{I\alpha})=U_{0}^\dagger H_I(a_\alpha,t) U_{0}
\ee
as the interaction Hamiltonian in the interaction picture, i.e.
\be
{\cal H}_I(a_{\alpha I})=  \sum_{\alpha=\pm} \int d^3x  (\partial_0 \tilde  a_{\alpha I})  (\chi\star_t \partial_0 a_{\alpha I}) .
\ee
As result we have the T-product exponentiation
\be
U_I=  T\Lp e^{-i\int_{-\infty}^t {\cal H}_I (t') dt'}\Rp
\ee
which satisfies $U_I(-\infty)={\cal I}$ where ${\cal I}$ is the identity operator.
Similarly,  we have the time evolution
\be
a_{\alpha }(t)=  U_{I}^\dagger(t) a_{\alpha I}(t) U_{I}(t)
\ee
between the operators in the interaction picture $a_{\alpha I}$ and the quantum field operators $a_\alpha$.
The same is true for any operator
\be
{\cal O}(t)= U_{I}^\dagger(t) {\cal O}_{ I}(t) U_{I}(t)
\ee
where ${\cal O}_I= U_{0}^\dagger {\cal O} U_{0}$.
Notice that we have
\be
\partial_t U_{I}^\dagger = i  U_{I}^\dagger  {\cal H}_I^\dagger (a_\alpha,t).
\ee
whose solution is
\be
U_I^\dagger =  \bar T\Lp e^{i\int_{0}^t {\cal H}_I^\dagger (t') dt'}\Rp
\ee
where the $\bar T$ time ordering puts all the fields from left to right in increasing time order.

We will be interested in simple observables involving only one time and defined by the vacuum expectation value of operators ${\cal O}(t)$. For instance we will take
$ {\cal O}(t)= a_{\alpha}(\vec x, t) \tilde a_{\beta}(\vec y, t)$. We will consider the steady state situation reached after bringing the two plates together and waiting for a long time. In this case, we will have access to
\be
\langle 0\vert {\cal O}(+\infty) \vert 0 \rangle= \langle 0\vert T_C\left [ {\cal O}_I(t) U_I(-\infty,-\infty)\right ]\vert 0 \rangle
\ee
where we have identified the time-ordered product along the double line from $-\infty$ to $+\infty$ and back, {see appendix \ref{app:in} for details}. 
\be
\langle 0\vert T_C\left [ {\cal O}_I(+\infty) U_I(-\infty,-\infty)\right ]\vert 0 \rangle\equiv \langle 0\vert U_I^\dagger(+\infty) {\cal O}_I(+\infty) U_I(+\infty)\vert 0 \rangle.
\ee
Along the upper part of the path $C$ the evolution operator involves ${\cal H}_I$ and on the lowest part ${\cal H}_I^\dagger$.
This representation allows one to write this vacuum expectation value as the path integral
\be
\langle 0\vert {\cal O}(+\infty) \vert 0 \rangle= \int {\cal D} a^1_{\alpha}  {\cal D} a^2_{\alpha}{\cal D} \tilde a^1_{\alpha}  {\cal D} \tilde a^2_{\alpha} {\cal O}_I e^{iS(a_\alpha^1,\tilde a_\alpha^1) - iS^\dagger(a_\alpha^2,\tilde a_\alpha^2)}
\ee
involving two copies of the two polarisation scalars $a^a_{\alpha}, \tilde a^a_\alpha \ a=1,2$ labelling the upper and lower parts of the Schwinger-Keldysh contour.
The operator ${\cal O}_I$ is inserted on the upper path.
Of course as a boundary condition we must impose that $a_\alpha^1(\infty)=a_\alpha^2(\infty)$ and similarly $\tilde a_\alpha^1(\infty)=\tilde a_\alpha^2(\infty)$ which imposes constraints on the Green's functions.

The action in the Schwinger-Keldysh formalism is defined by
\be
S(a^a_\alpha,\tilde a^a_\alpha)= S(a_\alpha^1,\tilde a^1_\alpha) - S^\dagger(a_\alpha^2,\tilde a^2_\alpha).
\ee
The quantum action can be written explicitly as
\be
S_{\rm}\Lp a^1_{\alpha},\tilde a^1_\alpha\Rp=\int d^4x \Lp \partial_0 \tilde a_{\alpha }^1 \epsilon\star_t \partial_0  a_{\alpha }^1- \partial_i \tilde a_{\alpha}^1 \partial^i  a_{\alpha }^1 +\frac{\phi}{M} \partial_0 \tilde a_{\alpha }^1 P_{i\alpha\beta}\star_x \partial^i a_{\beta}^1+\frac{\tilde \phi}{M}\partial_0 a_{\alpha }^1 P_{i\alpha\beta}\star_x \partial^i \tilde a_{\beta}^1 \Rp
\ee
and similarly on the lower branch where we have defined the coupling tensor
\be
 P_{i\alpha\beta}= \int \slashed{d} k  P_{i\alpha\beta}(k) e^{i\vec k.\vec x}
\ee
where
\be
 P_{i\alpha\beta}(k)= \epsilon_{ijk} \bar e_{j\alpha} e_{k\beta}
\ee
is the vector product of the polarisation vectors. This is obviously zero if $\alpha=\beta$ by antisymmetry and orthogonal to the two polarisation vectors.
This coupling introduces a mixing between the two polarisation scalars. We will separate the quantum action into the free part
\be
S_{\rm free}(a^a_{\alpha},\tilde a^a_{\alpha})= \int d^4x ( \partial_0 \tilde a_{\alpha }^a \epsilon\star_t \partial_0  a_{\alpha }^a- \partial_i \tilde a_{\alpha }^a \partial^i  a_{\alpha }^a)
\ee
with
\be
S_{\rm free}= S_{\rm free}(a^1_{\alpha},\tilde a^1_{\alpha})-S^\dagger_{\rm free}(a^2_{\alpha},\tilde a^2_{\alpha})
\ee
and the interaction part {with the axion}
\be
S_{\rm int}(a^a_{\alpha},\tilde a^a_{\alpha})= \int d^4 x\Lp \frac{\phi}{M} \partial_0 \tilde a_{\alpha }^a P_{i\alpha\beta}\star_x \partial^i a_{\beta}^a+\frac{\tilde\phi}{M}\partial_0 a_{\alpha }^a P_{i\alpha\beta}\star_x \partial^i \tilde a_{\beta}^a\Rp
\ee
such that
\be
S_{\rm int}= S_{\rm int}(a^1_{\alpha},\tilde a^1_{\alpha})-S^\dagger_{\rm int}(a^2_{\alpha},\tilde a^2_{\alpha})
\ee
This allows one to use perturbation theory techniques when calculating vacuum expectation values of operators depending  on a single time.

\subsection{The propagators and the two-point functions}

\subsubsection{Polarisations and boundary conditions}

So far we have not distinguished the two polarisations and the associated scalar fields $a_\alpha$. This is valid in empty space when the permittivity is
uniformly equal to unity in space. When the permittivity is not a constant like in the case of the Casimir experiment set-up, we must specify the boundary conditions at the interface between two media. To simplify the discussion, let us consider a planar interface at $z=0$ between two media of permittivities $\epsilon_1=1$ and $\epsilon_2= \epsilon(\omega)$ in Fourier space. There are two types of electromagnetic waves which can be transmitted across the surface. The first one is the TE, transverse electric, polarisation when the electric field is parallel to the surface, say in the $x$ direction. The other polarisation is the TM, transverse magnetic, polarisation where the magnetic field is parallel to the interface, again say in the $x$ direction. {The two modes correspond to the two panels of Fig.~\ref{fig:modes}.}

\begin{figure}[h]
    \centering
    \includegraphics[width=0.7\textwidth]{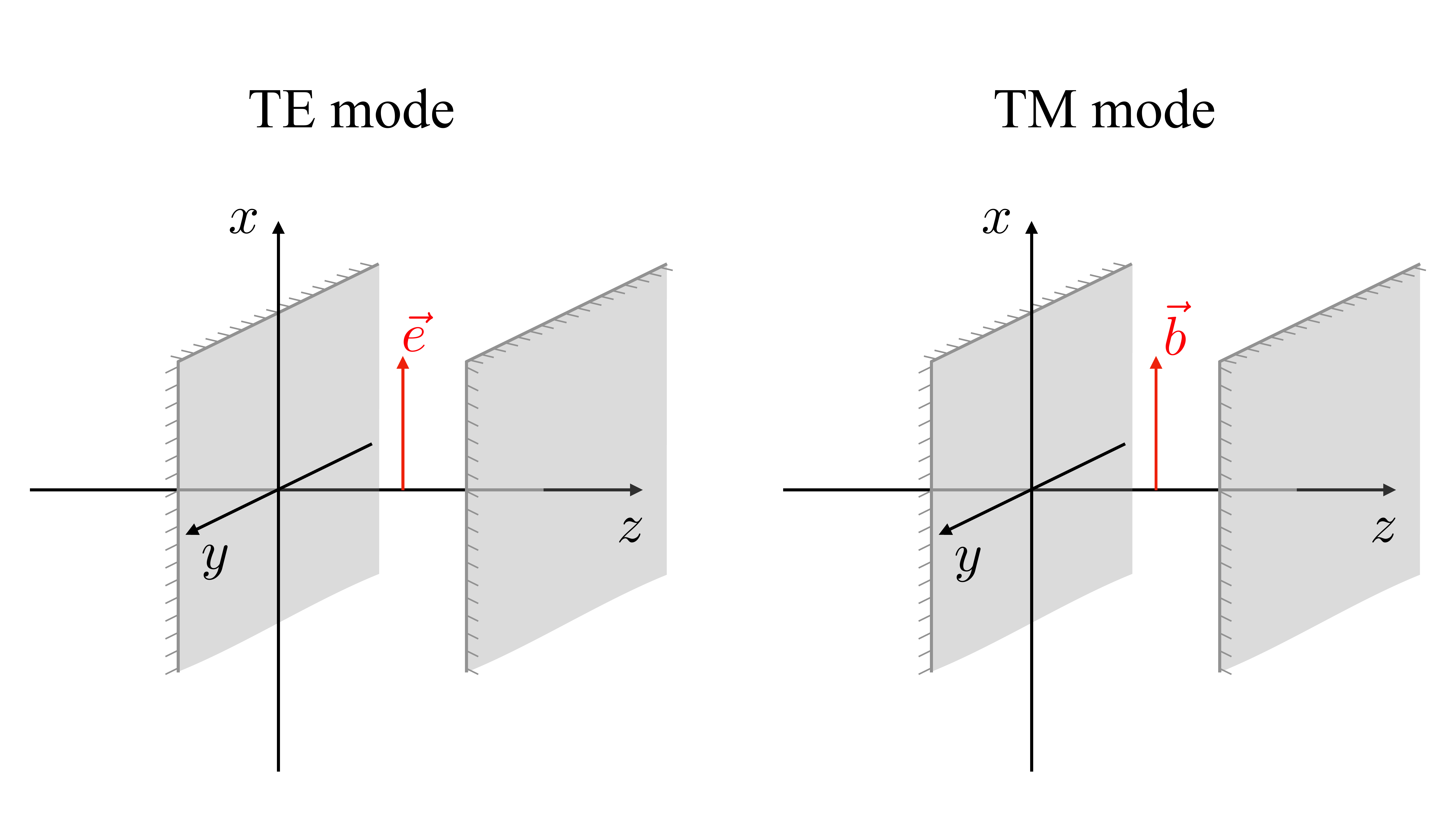}
    \caption{Illustration of the field's vectors in the TE and TM modes.}
    \label{fig:modes}
\end{figure}

Across the surface, both the transverse electric and magnetic fields must be continuous.
Let us first consider the TE case. We can write
\be
e_x= -\partial_0 a_x
\ee
implying that $a_x$ must be continuous across the interface. Maxwell's equation reads
\be
\partial_0 \vec b = -\vec \nabla \wedge \vec e
\ee
and implies that the parallel component of $\vec b$ is $b_y$ which is proportional to $\partial_z a_x$. The boundary conditions imply that for the TE polarisation we must impose
\be
[a_x]=0, \ \ [\partial_z a_x]=0
\ee
where $[f]=f_2-f_1$ is the jump of $f$ across the interface.
Considering the case of an incident wave with momentum $\Delta$ along the $z$ direction being transmitted to a wave of momentum $\xi$, we have the relations between the amplitudes of the reflected and transmitted waves
\begin{eqnarray}
&& A_++ A_-= B_+\nonumber \\
&& \Delta (A_+-A_-)= \xi B_+\nonumber \\
\end{eqnarray}
where $A_{\pm}$ are the amplitudes of the incident and reflected waves, and $B_+$ the transmitted one. We get that the reflection coefficient $r_{TE}= \frac{A_-}{A_+}$ is
\be
r_{\rm TE}= \frac{ 1-\frac{\xi}{\Delta}}{1+\frac{\xi}{\Delta}}.
\ee
Let us now turn to the case where the magnetic field is along the $x$ direction. This implies that $b_x=-\partial_z a_y$ and
the Maxwell equation
\be
\partial_0 \vec d= \vec \nabla \wedge \vec b
\ee
implies that in Fourier modes
\be
e_y=-i \frac{\partial^2_z a_y}{\epsilon \omega}.
\ee
Now integrating this equation between $z=-\delta$ and $z=\delta$, and upon using the continuity of $e_y$ across the interface implies that $\frac{1}{\epsilon} \partial_z a_y$ must be continuous. As a result we find that the TM polarisation boundary condition is
\be
[a_y]=0, \ \ \left [\frac{1}{\epsilon}\partial_z a_y \right ]=0.
\ee
For incident waves we find that amplitudes of the transmitted and reflected waves must satisfy
\begin{eqnarray}
&& A_++ A_-= B_+\nonumber \\
&& \Delta (A_+-A_-)= \frac{\xi}{\epsilon} B_+\nonumber \\
\end{eqnarray}
implying that the reflection coefficient for the TM modes is
\be
r_{\rm TM}= \frac{ 1-\frac{\xi}{\epsilon\Delta}}{1+\frac{\xi}{\epsilon\Delta}}.
\ee
for a wave of energy $\omega$. We will see that these boundary conditions follow from the action and influence the calculations of the propagators.
In the following we will denote by $+$ the TE polarisation and $-$ the TM polarisation.

\subsubsection{The TE propagators}

The two polarisations with their specific boundary conditions can be related to the operators defining the action. This  leads to the appropriate propagators with the TE and TM boundary conditions.
In order to simplify the notation we define the electric doublet
\be
A_+^a=\left (\begin{array} c
a_+^a \\
\tilde a^a_+\\
\end{array} \right)
\ee
associated with the TE polarisation.
The free action for the TE polarisation is obtained after integration by parts
\be
S_{\rm free}= \frac{1}{2} \int d^4 x (A_{+}^a)^T {\bf { D^E}}_{ab} A^b_{+}.
\ee
The differential operator is off diagonal
\be
{\bf{ D}}^E_{ab}=  \left ( \begin{array} {cc}
0& \hat {\cal D}^E_{ab}\\
{\cal D}^E_{ab}& 0 \\
\end{array}
\right )
\ee
where $a=1,2$ labels the two Keldysh branches of the ${\cal C}$ path. We have introduced the differential operators
\begin{eqnarray}
&&{\cal D}^E_{11}= {\cal D}_E, {\cal D}^E_{22}= - \tilde {\cal D}_E\nonumber \\
&& \hat {\cal D}^E_{11}= \tilde {\cal D}_E, \hat{\cal D}^E_{22}= -  {\cal D}_E\nonumber \\
\end{eqnarray}
and all the other entries of the matrix vanish, i.e. there is no coupling between the upper and lower branches of the Schwinger-Keldysh  contour.
The minus signs originate from the complex conjugation of $e^{iS}$  on the lower contour.
More precisely the pair $({\cal D},\tilde{\cal D})$ is such that
\be
{\cal D}_E= -\partial_0 (\epsilon\star_t \partial_0) + \Delta, \ \tilde {\cal D}_E= -\partial_0 (\hat \epsilon\star_t \partial_0)+ \Delta,
\ee
or in Fourier space (for the time part only)
\be
{\cal D_E}= \epsilon(\omega) \omega^2  + \Delta, \ \tilde {\cal D}_E= \epsilon(-\omega) \omega^2 + \Delta,
\ee
where we can now see that $\hat {\cal D}_E= {\cal D}_E^\dagger$ upon using $\bar\epsilon(\omega)= \epsilon(-\omega)$. As a result the differential operator ${\bf D}^E$ is Hermitian. We also define the contraction $x^ay_a= c_{ab} x^a y^a$ where $c_{11}=1, \ c_{22}=-1$ and zero otherwise. This is the "metric" in the Schwinger-Keldysh formalism allowing one to raise and lower indices. This also allows one to treat the fields on the two parts of the Schwinger-Keldysh path simultaneously.
We can invert the operator ${\bf D}^E$ by defining
\be
{\bf D}^E_{ab} {\bf \Delta}^{bc}_E(x;y)= i \delta^{(4)}(x^\mu-y^\mu)\delta^c_a
\ee
or more explicitly as
\be
{\bf{ \Delta}}^{ab}_E=  \left ( \begin{array} {cc}
0&  {\Delta }^{ab}_E\\
{\hat \Delta}^{ab}_E& 0 \\
\end{array}
\right ).
\ee
Let us define the two propagators
\be
{\cal D}_E \Delta_E(x;y)= i \delta^{(4)}(x^\mu-y^\mu)
\ee
and
\be
\tilde {\cal D}_E \tilde \Delta_E(x;y)= -i \delta^{(4)}(x^\mu-y^\mu).
\ee
As a result we get the set of propagators
\begin{eqnarray}
&&\Delta^{11}_E\equiv \Delta_E,\ \ \Delta^{22}_E\equiv \tilde \Delta_E\nonumber \\
&& \hat \Delta^{11}_E= \tilde \Delta_E,\ \ \hat \Delta^{22}_E=  \Delta_E\nonumber \\
\end{eqnarray}
corresponding to the fact that the propagators on the second branch of the Schwinger-Keldysh path are the time reversed of the propagators on the first branch.
In the following we will use
\be
\Delta_E= iG_{EF}
\ee
where $G_{EF}$ satisfies
\be
-\partial_0( \epsilon\star_t \partial_0 G_{EF}) + \Delta G_{EF}=  \delta^{(4)}(x^\mu-y^\mu)
\ee
This is the  Feynman propagator where we will retrieve its   T-product decomposition, see below.
Finally notice that across a boundary where $\epsilon$ jumps, like in the case of the transition between the vacuum and the plates in Casimir experiments, we can integrate both sides between $z=-\delta$ and $z=\delta$ where $z=0$ is the interface and get that
\be
\left [\frac{dG_{EF}}{dz}\right ]_{z=0}=0
\ee
i.e. the first derivative of $G_F$ does not jump across the interface. This is the TE boundary condition applied to the Green's function.

Let us connect these propagators with correlation functions. As the path integral of the free theory is Gaussian, we can immediately obtain
\be
e^{W(J)}= \int {\cal D} A^1_{+}  {\cal D} A^2_{+}  e^{iS_{\rm free}+\int d^4x A_{+}^{T a} {\bf J}_{a+}}
\ee
where ${\cal D} A^a_+= {\cal D} a^a_+ {\cal D}\tilde a ^a_+$ and
\be
{\bf J}_{a+}= \left ( \begin{array} c
J_{a+}\\
\tilde J_{a+} \\
\end{array}
\right )
\ee
Here $W(J)$ is the generating functional of connected correlation functions.
The Gaussian integral is obtained by solving
\be
{\bf D}^E_{ab} A_{+ }^b= i {\bf J}_{a+}
\ee
corresponding to
\be
A^a_+= {\bf \Delta}_E^{ab}{\bf J}_{b+}
\ee
and substituting back in the action
\be
W(J)= \frac{1}{2} {\bf J}_{a+}^T ({\bf\Delta}^{ab}_E)^T {\bf J}_{b+}.
\ee
Here the transpose operators acts on  the discrete matrix indices.
As a result we get for the two-point functions obtained  by Gaussian integration
\be
\langle A_{+}^a(x) A_{+}^b (y)\rangle= {\bf \Delta}_E^{ab}(x,y)
\ee
 and we denote by
\be
\langle A_{+}^a(x) A_{+}^b (y)\rangle= \int {\cal D} A^1_{+}  {\cal D} A^2_{+}  \;A_{+}^a(x) A_{+}^b (y) \; e^{iS_{\rm free}}.
\ee
These two-point functions are the TE two-point functions.

In a more explicit fashion, only the two-point functions between $a^a_\alpha$ and $\tilde a^a_\alpha$ are non-vanishing with
\be
\langle a^1_+(x) \tilde a^1_+ (y)\rangle=  \Delta_E (x;y), \ \ \langle \tilde a^1_+(x)  a^1_+ (y)\rangle=  \tilde \Delta_E (x;y).
\ee
We can now decompose the two-point functions as T-products.   For instance
the boundary condition at infinity $a^1(\infty)=a^2(\infty)$ implies that
\be
\langle a_{+}^1(x) \tilde a_{+}^1 (y)\rangle = Y(x^0-y^0)\langle a_{+}^2(x) \tilde a_{+}^1 (y)\rangle + Y(y^0-x^0) \langle a_{+}^1(x) \tilde a_{+}^2 (y)\rangle
\ee
where $Y$ is the Heaviside distribution.
For instance as $x^0\to \infty$, the field located at $x^\mu$ on the first branch reaches infinity where it becomes the field on the second branch. This implies that
the propagator for $x^0>y^0$ should  be identified with $\langle a_{\alpha}^2(x) a_{\beta}^1 (y)\rangle= \Delta_E^{21}\delta_{\alpha\beta}$ by continuity. This defines $\Delta_E^{21}$.
We have also the identification between the path integration and the quantum vacuum expectation values
\be
\langle 0\vert  T\left [a_{+}^a(x) \tilde a_{+}^b (x)\right ]\vert 0\rangle = \lim_{x^\mu \to y^\mu} \langle a_{+}^a(x) \tilde a_{+}^b (y)\rangle
\ee
where the limit has to be understood after the appropriate renormalisation procedure. 

\subsubsection{The TM propagator}

We now define the magnetic  doublet as
\be
A_-^a=\left (\begin{array} c
a_-^a \\
\hat \epsilon \star_t \tilde a^a_-\\
\end{array} \right)
\ee
associated with the TM polarisation. Notice that the second component is convolved with the time-reversed permittivity function.
The free action for the TM polarisation is obtained after integration by parts
\be
S_{\rm free}= \frac{1}{2} \int d^4 x (A_{-}^a)^T {\bf { D}}_{ab}^M A^b_{-}
\ee
and involves
the off-diagonal magnetic differential operator
\be
{\bf{ D}}_{ab}^M=  \left ( \begin{array} {cc}
0& \hat {\cal D}_{ab}^M\\
{\cal D}_{ab}^M& 0 \\
\end{array}
\right )
\ee
where we have introduced the differential operators
\begin{eqnarray}
&&{\cal D}_{11}^M= {\cal D}_M, {\cal D}_{22}^M= - \tilde {\cal D}_M\nonumber \\
&& \hat {\cal D}_{11}^M= \tilde {\cal D}_M, \hat{\cal D}_{22}^M= -  {\cal D}_M.\nonumber \\
\end{eqnarray}
The pair of magnetic operators $({\cal D}_M,\tilde{\cal D}_M)$ is such that
\be
{\cal D}_M= -\partial_0^2 + \epsilon^{-1} \star_t \Delta, \ \tilde {\cal D}_M= -\partial_0^2 + \hat \epsilon^{-1}\star_t\Delta,
\ee
or in Fourier space (for the time part only)
\be
{\cal D}_M= \omega^2  + \frac{\Delta}{\epsilon(\omega)}, \ \tilde {\cal D}_M=  \omega^2 + \frac{\Delta}{\epsilon(-\omega)},
\ee
i.e. the operator $\epsilon^{-1}\star_t$ is the convolution with the function whose Fourier transform is $1/\epsilon(\omega)$, respectively $\hat \epsilon^{-1}\star_t $ the convolution with the function of Fourier transform $1/\epsilon(-\omega)$.
We  see that $\hat {\cal D}_M= {\cal D}^\dagger_M$ upon using $\bar\epsilon(\omega)= \epsilon(-\omega)$. As a result the differential operator ${\bf D}_M$ is Hermitian.

We can invert the operator ${\bf D}^M$ by defining the propagators
\be
{\bf D}_{ab}^M {\bf \Delta}^{bc}_M (x;y)= i \delta^{(4)}(x^\mu-y^\mu)\delta^c_a
\ee
or more explicitly as
\be
{\bf{ \Delta}}^{ab}_M=  \left ( \begin{array} {cc}
0&  {\Delta }^{ab}_M\\
{\hat \Delta}^{ab}_M& 0 \\
\end{array}
\right ).
\ee
This requires the two magnetic propagators
\be
{\cal D}_M \Delta_M(x;y)= i \delta^{(4)}(x^\mu-y^\mu)
\ee
and
\be
\tilde {\cal D}_M \tilde \Delta_M(x;y)= -i \delta^{(4)}(x^\mu-y^\mu).
\ee
Hence we get the set of propagators
\begin{eqnarray}
&&\Delta^{11}_M\equiv \Delta_M,\ \ \Delta^{22}_M\equiv \tilde \Delta_M\nonumber \\
&& \hat \Delta^{11}_M= \tilde \Delta_M,\ \ \hat \Delta^{22}_M=  \Delta_M\nonumber \\
\end{eqnarray}
We will use in the following the magnetic Green's function
\be
\Delta_M= iG_{MF}
\ee
where $G_{MF}$ satisfies
\be
-\partial_0^2 G_{MF} + \epsilon^{-1}\star_t \Delta G_{MF}=  \delta^{(4)}(x^\mu-y^\mu)
\ee
or in Fourier space
\be
\omega^2 G_{MF} + \frac{\Delta}{\epsilon (\omega)} G_{MF}=  \delta^{(3)}(\vec x -\vec y).
\ee
Again across an interface where $\epsilon$ jumps, like at the interface between the vacuum and the plates in a Casimir experiment,  we have
\be
\left [\frac{1}{\epsilon (\omega)} \frac{dG_{MF}}{dz}\right ]_{z=0}= 0
\ee
which is the TM boundary condition applies to the Green's function.

Just like in the TE case, we can  connect these propagators with the correlation functions. The same reasoning leads to
\be
\langle A_{-}^a(x) A_{-}^b (y)\rangle= {\bf \Delta}_M^{ab}(x,y)\ee
where we denote by
\be
\langle A_{-}^a(x) A_{-}^b (y)\rangle= \int {\cal D} A^1_{-}  {\cal D} A^2_{-}  A_{-}^a(x) A_{-}^b (y)  e^{iS_{\rm free}}.
\ee
These two-point functions are the TM  two-point functions.
More explicitly, the only non-vanishing two-point functions between $a^a_-$ and $\tilde a^a_-$ are
\be
\langle a^1_-(x) \hat \epsilon \star_t\tilde a^1_\beta (y)\rangle=  \Delta_M (x;y), \ \ \langle \hat \epsilon \star_t \tilde a^1_-(x)  a^1_- (y)\rangle=  \tilde \Delta_M (x;y).
\ee
with the T-product decomposition
\be
\langle a_{-}^1(x) \hat \epsilon\star_t \tilde a_{-}^1 (y)\rangle = Y(x^0-y^0)\langle a_{-}^2(x)  \hat \epsilon\star_t\tilde a_{-}^1 (y)\rangle + Y(y^0-x^0) \langle a_{-}^1(x) \hat \epsilon\star_t\tilde a_{-}^2 (y)\rangle
\ee
where  $\langle a_{-}^2(x) \hat \epsilon \star_t \tilde a_{-}^1 (y)\rangle= \Delta^{21}_M$.
Finally we identify
\be
\langle 0\vert  T\left [ a_{-}^a(x) \hat \epsilon \star_t \tilde a_{-}^b (x) \right ]\vert 0\rangle = \lim_{x^\mu \to y^\mu} \langle a_{-}^a(x) \hat \epsilon\star _t \tilde a_{-}^b (y)\rangle
\ee
with  the appropriate renormalisation procedure.

\subsection{The link with electrodynamics}

We will elaborate further the link between the action involving the two fields $a_\alpha^1$ and $\tilde a_\alpha^1$ and classical electrodynamics. We focus on the part of the action which does not contain the axion field as the axion coupling to photon does not contribute to the energy-momentum tensor.
{All in all we consider eight fields. The two physical fields $a_+$ and $a_-$ are copied and time reversed for accounting for dissipation, and a second doubling of the field is necessary for integrating along the Schwinger-Keldysh contour, as shown in Fig.~\ref{fig:8fields} where the example of the TE field is considered.}

\begin{figure}
    \centering
    \includegraphics[width=0.5\textwidth]{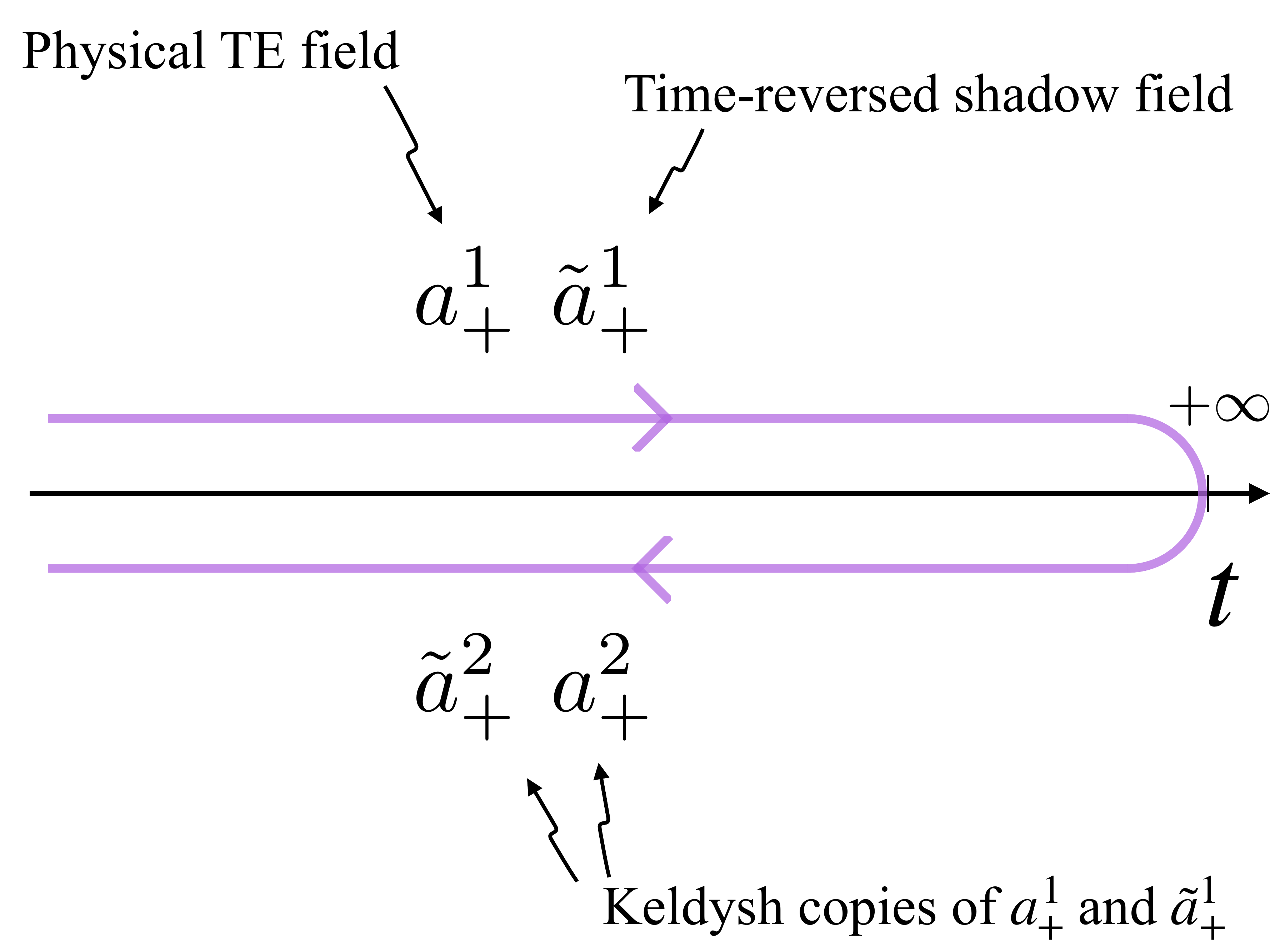}
    \caption{Illustration of the Keldysh contour in the case of the TE field.  }
    \label{fig:8fields}
\end{figure}

We have  seen that classically the shadow field is nothing but the time-reversed electromagnetic field. We first decompose  the electromagnetic energy momentum tensor obtained from the action
\be
S(a_\alpha^1,\tilde a_\alpha^1)=\int d^4x \; {\cal L}
\ee
where we only focus on the fields on the first branch on the Keldysh-Schwinger contour.
This leads to the decomposition into two separate parts
\be
{\cal T}^\alpha_{\mu\nu}= T_{\mu\nu}^\alpha+ \hat T_{\mu\nu}^\alpha
\ee
for each polarisation
where we have
\be
T_{\mu\nu}^\alpha = \partial_\nu a_\alpha^1\frac{\partial {\cal L} }{\partial \partial^\mu a^1_\alpha } -\frac{\eta_{\mu\nu}}{2} {\cal L},\  \hat T_{\mu\nu}^\alpha =  \partial_\nu \tilde a_\alpha^1 \frac{\partial {\cal L}}{\partial \partial^\mu \tilde a^1_\alpha } - \frac{\eta_{\mu\nu}}{2} {\cal L}.
\ee
Notice that this energy-momentum tensor is not symmetric and is  conserved.
More explicitly we have
\begin{eqnarray}
&& T_{00}^\alpha =  \frac{1}{2}\Lp \partial_0 a_{\alpha }^1(\hat \epsilon\star_t \partial_0 \tilde a^1_\alpha)   +  \partial_i  a_{\alpha}^1 \partial^i  \tilde a_{\alpha }^1\Rp\nonumber\\
&&\ T_{0i}^\alpha=   \partial_i  a_{\alpha }^1 (\hat \epsilon \star_t\partial_0\tilde a_{\alpha }^1), \   T_{i0}= \partial_0  a_{\alpha}^1 \partial_i \tilde a_\alpha^1 \nonumber \\
&&\ T_{ij}^\alpha= \partial_j   a_{\alpha }^1\partial_i   \tilde a_{\alpha }^1 - \frac{\delta_{ij}}{2}\Lp-\partial_0 a_{\alpha }^1 (\hat \epsilon\star_t \partial_0  \tilde a_{\alpha }^1)+  \partial_i  a_{\alpha}^1 \partial^i  \tilde a_{\alpha }^1\Rp\nonumber \\
\end{eqnarray}
and
\begin{eqnarray}
&&\hat T_{00}^\alpha=  \frac{1}{2}\Lp \partial_0  \tilde  a_{\alpha }^1( \epsilon\star_t \partial_0 a_\alpha^1) +  \partial_i  \tilde a_{\alpha}^1 \partial^i   a_{\alpha }^1\Rp\nonumber\\
&&\hat T_{0i}^\alpha=   \partial_i\tilde a_{\alpha }^1 (\epsilon\star_t \partial_0  a_{\alpha }^1)   ,\  \hat T^\alpha_{i0}= \partial_0 \tilde a_\alpha^1 \partial_i a_\alpha^1\nonumber \\
&&\hat T_{ij}^\alpha= \partial_j  \tilde a_{\alpha }^1\partial_i   a_{\alpha }^1 - \frac{\delta_{ij}}{2}\Lp-\partial_0 \tilde a_{\alpha }^1 ( \epsilon\star_t \partial_0   a_{\alpha }^1)+  \partial_i \tilde  a_{\alpha}^1 \partial^i   a_{\alpha }^1\Rp\nonumber \\
\end{eqnarray}
The
axion does not play a role in the energy momentum tensor as the original $F\tilde F$ term in the action of electrodynamics coupled to axions is topological, i.e. does not depend on the metric.

{This energy-momentum tensor has interesting properties in the classical case and in the absence of dissipation $\gamma=0$. There is no need for the doubling of the electromagnetic fields and in fact one can take $\tilde a(t)= a(t)$ as they both satisfy the same Maxwell equation. This implies that  we have the relation
\be 
T_{\mu\nu}= T^{\rm mink}_{\mu\nu}+ \partial_\lambda \Theta_{\mu\nu}^\lambda
\label{coho}
\ee
where the Minkowski energy-momentum tensor of electromagnetism in matter \cite{Medina:2017mcd} is given by
\begin{eqnarray}
&&T^{\rm mink}_{00}=  \frac{1}{2}\Lp\vec d.\vec e + \vec b.\vec b\Rp \nonumber \\
&& T_{0i}^{\rm mink}=  \Lp \vec d \wedge  \vec b \Rp_i, \ T_{i0}^{\rm mink}=\Lp \vec e \wedge \vec b\Rp_i\nonumber \\
&& T_{ij}^{\rm mink}=  -e_i d_j - b_i b_j +\frac{\delta_{ij}}{2}\Lp\vec e.\vec d + \vec b^2\Rp \nonumber \\
\end{eqnarray}
The correction tensor is given explicitly by
\begin{eqnarray}
    && \Theta^\lambda_{0i}=-\epsilon\star_t F^\lambda_0 a_i, \ \Theta^\lambda_{i0}= -F^\lambda_0 a_i\nonumber \\
    && \Theta^0_{ij}=-\epsilon\star_t F^0_i a_j,\ \Theta^k_{ij}=-F^k_i a_j.\nonumber \\
\end{eqnarray}
When $\epsilon=1$, i.e. in the case of electrodynamics in vacuum, this transformation is similar to the one between the Noether energy momentum tensor and the symmetric energy momentum tensor \cite{itzykson2012quantum}. The two energy momentum tensors $T^{\mu\nu}$  and $T^{\mu\nu}_{\rm mink}$ define the same conserved charge, i.e. the electromagnetic momentum
\be 
{\cal P}^\mu= \int d^3 x \;T^{0\mu}.
\ee
Indeed using the explicit antisymmetry of $F^{\mu\nu}$ we have
$\Theta^{00\mu}=0$ and therefore
\be
\int d^3 x\; \partial_\lambda \Theta^{\lambda 0 \mu}= \int d^3 x \;\partial_i \Theta^{i0\mu}=0
\ee
In this sense,  the two energy momentum tensors are equivalent
and we can identity $T^{\mu\nu}$ as the energy momentum tensor of electrodynamics in matter. 
Of course, the total derivatives in (\ref{coho}) would play a role when boundaries are present like in the Casimir setup. 
In the following section, we will show that the tensor $T^{\mu\nu}$ leads to the correct description of the Casimir effect in the presence of matter.}

{In the dissipative case when $\gamma\ne0$ we have classically  $\tilde a (t)= a(-t)$ and we find
\be
\hat T_{\mu\nu}(t)= T_{\mu\nu}(-t),
\ee
i.e. the total energy-momentum tensor is the sum of the energy-momentum tensor $T_{\mu\nu} $ and its time-reversed  $\hat T_{\mu\nu}$.} In particular the energy density of the system is the sum of the energy density of the forward and backward energy densities
\be
{\cal T}_{00}^\alpha(t)= T_{00}^\alpha(t)+ T_{00}^\alpha(-t).
\ee
This energy density is conserved which confirms that the dissipated energy by the forward process is brought back from infinity by the time-reversed dynamics.
As a result, we identify $T_{\mu\nu}$ with the energy-momentum tensor of  electrodynamics in the presence of dissipation.
{It involves  the shadow field in a crucial way.}

In the following we will calculate the quantum vacuum expectation values of $T_{\mu\nu}$ and identify this as the quantum vacuum contribution to the energy and pressure {of electromagnetism in matter}. We will confirm this identification by calculating the Casimir pressure when the plates are not ideal and retrieve the Lifschitz theory, i.e. the well known results \cite{lif1,lif2}.
The vacuum energy is obtained as
\be
E= \int d^3 x \;\langle 0\vert  T_{00}\vert 0\rangle
\ee
or equivalently
\be
E= \frac{1}{2} \sum_{\alpha=\pm} \int d^3 x\; \langle 0\vert ( \partial_0  a _\alpha^1 \hat \epsilon\star _t \partial_0 \tilde a^1_\alpha+
\partial_i  a _\alpha^1 \partial_i \tilde a^1_\alpha) \vert 0\rangle
\ee
Similarly for the pressure we have
\be
P_z= \frac{1}{2} \sum_{\alpha=\pm}\langle 0\vert ( \partial_z  a_{\alpha }^1 \partial_z \tilde a_{\alpha }^1+  \partial_0  a _\alpha^1 \hat \epsilon\star _t \partial_0 \tilde a^1_\alpha-
\partial_x  a _\alpha^1 \partial_x \tilde a^1_\alpha-\partial_y  a _\alpha^1 \partial_y \tilde a^1_\alpha) \vert 0\rangle
\ee
coming from $T_{zz}$ and
involving the electric and magnetic two-point functions.
In the following section, we will show that this pressure coincides with the result from Lifschitz's theory in the absence of axionic field. When the axion is taken into account, we will calculate the first correction to the quantum pressure due to the coupling to axions.

\section{The Lifschitz theory}
\label{sec:lif}
Before calculating the effects of axions on the Casimir effect, let us revisit the case with non-trivial permittivities in the plaques and no axions. We will use the previous energy and pressure defined from the action involving the two polarisation scalars and their shadow fields. We will retrieve the results of the Lifschitz theory which are commonly derived using different approaches \cite{lif1,lif2}.

\subsection{The vacuum energy with no axions}

We will calculate the vacuum expectation value of $T^{00}$ as obtained in the previous section.
Let us evaluate the potential energy first
\be
E_P=\frac{1}{2} \sum_{\alpha=\pm}  \langle 0\vert \int d^3 x \;\partial_i a^1_\alpha(x) \partial_i \tilde a_\alpha^1(x)  \vert 0\rangle
\ee
where the energy is the vacuum expectation value defined in the previous section. Here we consider the fields as living on the upper path of the Schwinger-Keldysh contour.
The kinetic energy is given by
\be
E_K=\frac{1}{2}\lim_{x^\mu\to x_0^\mu} \int d^3x\; \langle 0\vert  \partial_t  a _\alpha^1 (x) (\hat \epsilon\star _{t_0} \partial_{t_0} \tilde a^1_\alpha(x_0)\vert 0\rangle.
\ee
As the expectation values are taken at coincident points, the product of distributions is ill-defined and is understood in a limiting sense
using the point-splitting method with $x^\mu \to x_0^\mu$.
After an integration by parts the potential energy becomes
\be
E_P= -\frac{1}{2}\sum_{\alpha=\pm}  \langle 0\vert \int d^3 x  \; \partial^2_x a^1_\alpha (x) \tilde a_\alpha^1(x)  \vert 0\rangle
\ee
which is understood as the limit
\be
E_P=  -\frac{1}{2}\sum_{\alpha=\pm} \lim_{x^\mu\to x_0^\mu} \int d^3 x \;\partial_x^2 \langle  a^1_{\alpha}(x) \tilde a^1_{\alpha}(x_0) \rangle.
\ee
We first focus on the TE polarisation. 
Now we use the identity
\be
\langle  a^1_{+ } (x) \tilde a^1_{+ }(x_0) \rangle = i  G_{EF}(x;x_0)
\ee
As $x\ne x_0$ in the point-splitting limit we have also the equality
\be
\lim_{x^\mu\to x_0^\mu} \int d^3 x \; \partial_x^2 G_{EF}(x;x_0)= \lim_{x^\mu\to x_0^\mu} \int d^3 x \;\epsilon\star_{t} \partial_t^2 G_{EF}(x;x_0).
\ee
 Now as the Green's function is time-translation invariant, i.e. it only depends on $t-t_0$,  we have  $\partial^2_t G_{EF}(x,x_0)= -\partial_t\partial_{t_0} G_{EF}(x;x_0)$.
This implies that the potential energy and the kinetic energy for the TE polarisation are equal
\be
E_P^+=E_K^+=  \frac{i}{2}\lim_{x^\mu\to x_0^\mu} \int d^3 x \;\partial_{t}\partial_{t_0}  (\epsilon\star_{t} G_{EF})(x;x_0).
\ee
 The total energy is then
\be
E_+=  i \lim_{x^\mu\to x_0^\mu} \int d^3 x \;\partial_{t}\partial_{t_0}  (\epsilon\star_{t} G_{EF})(x;x_0).
\ee
Translation invariance in the parallel direction to the plates and time translation invariance leads to
\be
G_{EF}(x;x_0)= G_{EF}( t-t_0,\vec x_{\parallel}- \vec x_{\parallel 0},z;0,\vec 0, z_0)
\ee
implying that the integral over the surface area of the plates factors out and we get finally
in terms of Fourier modes
\be
\frac {E_+}{A}= i  \lim_{x\to x_0}\int dz \;\slashed{d} \omega \;\slashed{d}^2 p_\parallel\; e^{-i\omega (t-t_0)} \epsilon (\omega) \omega^2 G_{EF}(z,\omega, p_\parallel;z_0).
\ee
where we have defined the Fourier transform
\be
G_{EF}(z,\omega, p_\parallel;z_0)= \int dt\; d^2x_\parallel\; e^{-i\omega t} e^{i\vec x_\parallel.\vec p_{\parallel}} G_{EF}( t,\vec x_{\parallel},z;0,\vec 0, z_0)
\ee
which is given the appendix \ref{sec:ele} for the retarded Green's function and by the extension procedure in section \ref{sec:feynman} for the Feynman propagator.

The TM polarisation can be dealt with in the same fashion, see the appendix \ref{sec:mag}, using
\be
\langle  a^1_{- } (x) \tilde a^1_{- }(x_0) \rangle = i  \epsilon^{-1} \star_t G_{MF}(x;x_0)
\ee
implying that the energy for the TM polarisation is given by
\be
E_+=  i \lim_{x^\mu\to x_0^\mu} \int d^3 x \;\partial_{t}\partial_{t_0}  G_{MF}(x;x_0)
\ee
or equivalently
\be
\frac {E_-}{A}= i  \lim_{x^\mu\to x_0^\mu}\int dz \;\slashed{d} \omega\; \slashed{d}^2 p_\parallel \;e^{-i\omega (t-t_0)}  \omega^2 G_{MF}(z,\omega, p_\parallel;z_0).
\ee
Hence the total energy can be obtained from the following the integrals
\be
\frac {E}{A}= i  \lim_{x^\mu\to x_0^\mu}\int dz\; \slashed{d} \omega \;\slashed{d}^2 p_\parallel \;e^{-i\omega (t-t_0)}  \omega^2\Lp\epsilon(\omega)G_{EF}(z,\omega, p_\parallel;z_0)+  G_{MF}(z,\omega, p_\parallel;z_0)\Rp
\ee
leading to the vacuum energy.

\subsection{The ideal case}

Before focusing on the case with metallic plates, let us retrieve the Casimir energy when the plates are ideal with a vanishing Green's function apart from
inside the vacuum. This corresponds to the limit $\epsilon \to \infty$, see appendix \ref{sec:ele} for instance.
In this case we have $G_F=0$ for $z<0$ and $z>d$ whilst in between the plates
\be
 G_{EF}( z;z_0, \omega, p_\parallel)= \frac{1}{\Delta} \frac{\sin \Delta (z_0-d) \sin \Delta z}{\sin \Delta d}, \  G_{MF}( z;z_0, \omega, p_\parallel)= \frac{1}{\Delta} \frac{\cos \Delta (z_0-d) \cos \Delta z}{\sin \Delta d}.
 \label{ideal1}
\ee
Explicitly we have
\be
\int_0^d dz \; \left (G_F( z, \omega, p_\parallel;z)+G_{MF}( z, \omega, p_\parallel;z)\right )=\frac{d\cos \Delta d}{ \Delta \sin \Delta d} .
\ee
We will change variable from $\omega$ to $\Delta$ which is a double cover of the real line into the positive real line. In order to take into account the double counting we extend $\Delta$ to the negative real line by taking $\Delta = {\rm sign}(\omega) \sqrt{\omega^2 -p^2_\parallel}$. Then we get
\be
\frac {E}{A}= i  \frac{d}{2\pi}\lim_{t\to t_0}\int_{-\infty}^\infty d\omega \int \slashed{d}^2 p_\parallel \;e^{-i\omega (t-t_0)}\; \frac{\omega^2\cos \Delta d}{\Delta \sin \Delta d}.
\ee
In the complex $\Delta$ plane, the integrand has poles for
\be
\Delta_n= \frac{n\pi}{d}\Rightarrow \omega_n=  {\rm sign  } (n) \left(\frac{\pi^2 n^2}{d^2} +p_\parallel^2\right )^{1/2}
\ee
for $n\ne 0$. As $t>t_0$, the integral can be performed by closing the integration contour in the lower half-plane where  $\omega$ has a negative imaginary parts guaranteeing that the exponential factor $e^{-i\omega (t-t_0)}$ decays at infinity. As a memory of the shift of the poles when dissipation is present, we take the Feynman prescription that the poles are shifted from the real axis by
\be
\omega_n=  {\rm sign  } (n) \left (\left(\frac{\pi^2 n^2}{d^2} +p_\parallel^2\right)^{1/2}-i\delta\right)
\ee
where $\delta$ is infinitesimal.
As shown in Fig.~\ref{fig:ideal} the poles are integrated over circles with residues ${\rm Res_n}= \frac{ \omega_n }{d}$
we find that
\be
\frac {E}{A}= i  \frac{d}{2\pi} \times \int \slashed{d}^2 p_\parallel \;(-2\pi i)\; \sum_{n>0} \frac{ \omega_n }{d}
\ee
leading to
\be
\frac {E}{A}=  \sum_{n>0}\int \slashed{d}^2 p_\parallel \; {\omega_n}.
\label{vacE}
\ee
\begin{figure}
    \centering
    \includegraphics[width=0.5\textwidth]{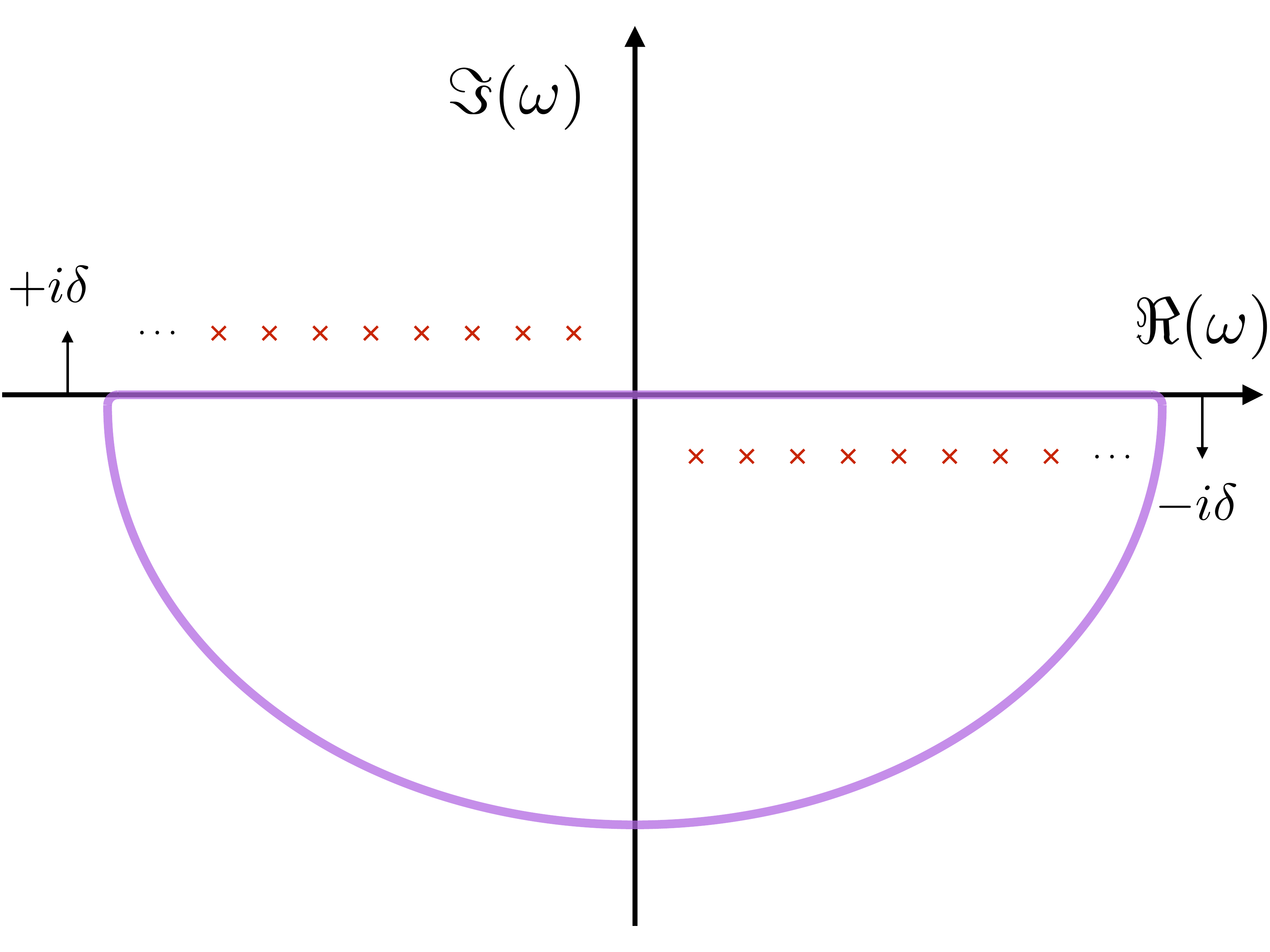}
    \caption{}
    \label{fig:ideal}
\end{figure}
This is the vacuum energy for the two polarisations, i.e. we retrieve that the vacuum energy is due to the presence of photon modes between the ideal plaques.
Notice the complementary roles that the $TE$ and $TM$ polarisations play, i.e. they contribute equally to the vacuum energy. Moreover if we had shifted the poles below the real axis as for retarded Green's functions, the energy would have vanished. Using the Feynman prescription is crucial.

\subsection{Metallic plates}

From now on and to simplify the notation, all the propagators are the Feynman propagators and we drop the $F$ index for simplicity.
In the case of metallic plates, the integrals of the Green's function inside the plates converges provided the two Green's functions $G_E(z,\omega, p_\parallel;z)$  and $G_M(z,\omega, p_\parallel;z)$ vanish when $z$ is very large. It turns out that $\lim_{z\to \infty} G_E(z,\omega, p_\parallel;z)= -\frac{1}{2 \xi}$ and $\lim_{z\to \infty} G_M(z,\omega, p_\parallel;z)= -\frac{\epsilon}{2 \xi}$. This means that one must renormalise the Green's functions according to
\be
G_{ER}(z,\omega, p_\parallel;z)= G_E(z,\omega, p_\parallel;z)+ \frac{1}{2\xi}, G_{MR}(z,\omega, p_\parallel;z)= G_M(z,\omega, p_\parallel;z)+ \frac{\epsilon}{2\xi}
\ee
Physically this amounts to subtracting the pressure exerted on the plate by the vacuum when one regularises the plate to be of finite and very large width $L$. The plate at $z=d+L$ compensates part of the pressure exerted at $z=d$. This is the origin of this renormalisation.
In practice this implies that 
\be
G_{ER}(z,\omega, p_\parallel;z)= \frac{\Theta_E(\omega,p_\parallel)}{2} \left(\frac{\xi}{\Delta}+ \frac{\Delta}{\xi}\right)\left( \sin \Delta d + \frac{\xi}{\Delta}\left( \cos \Delta (2z-d) -\cos \Delta d\right)\right ), \ 0\le z\le d
\ee
and
\be
G_{MR}(z,\omega, p_\parallel;z)= \frac{\Theta_M(\omega,p_\parallel)}{2} \left (\frac{\xi}{\epsilon \Delta}+ \frac{\epsilon \Delta}{\xi}\right )\left( \sin \Delta d + \frac{\xi}{\epsilon\Delta}\left( \cos \Delta (2z-d) -\cos \Delta d\right)\right ), \ 0\le z\le d
\ee
In the plates we find
\begin{eqnarray}
&&G_{ER}(z,\omega, p_\parallel;z)= \frac{\Theta_E(\omega,p_\parallel)}{2} \sin \Delta d \left(\frac{\xi}{\Delta}+ \frac{\Delta}{\xi}\right) e^{-2\xi (z-d)}\ \ z\ge d \nonumber \\
&&G_{ER}(z,\omega, p_\parallel;z)= \frac{\Theta_E(\omega,p_\parallel)}{2} \sin \Delta d \left(\frac{\xi}{\Delta}+ \frac{\Delta}{\xi}\right) e^{2\xi z)}\ \ z\le 0 \nonumber \\
\end{eqnarray}
and
\begin{eqnarray}
&&G_{MR}(z,\omega, p_\parallel;z)= \frac{\Theta_M(\omega,p_\parallel)}{2} \sin \Delta d \left(\frac{\xi}{\epsilon\Delta}+ \frac{\epsilon\Delta}{\xi}\right) e^{-2\xi (z-d)}\ \ z\ge d \nonumber \\
&&G_{MR}(z,\omega, p_\parallel;z)= \frac{\Theta_M(\omega,p_\parallel)}{2} \sin \Delta d \left(\frac{\xi}{\epsilon\Delta}+ \frac{\epsilon\Delta}{\xi}\right ) e^{2\xi z)}\ \ z\le 0 \nonumber \\
\end{eqnarray}
 As these expressions correspond to the renormalised Feynman propagators, they are defined in the right hand plane where $\Re \omega >0$. In the left half plane the  Feynman propagators are simply obtained by exchanging $\omega \to -\omega$. This implies that integrals involving the Feynman propagator along the real line are such that
\be
\int_{-\infty }^\infty  d\omega \;G_{EF} (\omega)= 2 \int_0^{\infty} d \omega \;G_{EF}(\omega)= 2 \int_0^{\infty} d \omega\; G_E(\omega)
\ee
{as $G_{EF}(\omega)= G_E(\omega)$
for $\omega >0$. We will use this below. }

\subsubsection{The electric energy}

The contribution from the plaques to the energy is given by
\begin{eqnarray}
&&\frac{E_{\rm plaque}^+}{A}= i \lim_{t\to t_0}\int_{0}^\infty \slashed{d}\omega \int \slashed{d}^2 p_\parallel\; e^{-i\omega (t-t_0)} \epsilon(\omega) \omega^2 \frac{\Theta_E(\omega,p_\parallel)}{2\xi} \left(\frac{\xi}{\Delta}+ \frac{\Delta}{\xi}\right)\sin \Delta d\nonumber\\
&&
+i \lim_{t\to t_0}\int_{-\infty}^0\slashed{d}\omega \int \slashed{d}^2 p_\parallel \;e^{-i\omega (t-t_0)} \epsilon(-\omega) \omega^2 \frac{\Theta_E(-\omega,p_\parallel)}{2\xi(-\omega)} \left(\frac{\xi(-\omega)}{\Delta}+ \frac{\Delta}{\xi(-\omega)}\right )\sin \Delta d
\end{eqnarray}
where on the second line all the contributions are evaluated at $-\omega$ and we do not specify $\Delta(-\omega)$ as $\Delta$ is an even function.
The contribution from the vacuum between the plaques is
\begin{eqnarray}
&&\frac{E_{\rm V}^+}{A}= i \lim_{t\to t_0}\int_0^\infty \slashed{d}\omega \int \slashed{d}^2 p_\parallel\; e^{-i\omega (t-t_0)} \omega^2 \frac{\Theta_E(\omega,p_\parallel)}{2\Delta} \xi d \left (\frac{\xi}{\Delta}+ \frac{\Delta}{\xi}\right ) \left(-\cos \Delta d +\left(\frac{\Delta}{\xi}-\frac{1}{\Delta d}\right ) \sin \Delta d\right )\nonumber \\
&& + i \lim_{t\to t_0}\int_{-\infty}^0 \slashed{d}\omega \int\slashed{d}^2 p_\parallel \; e^{-i\omega (t-t_0)} \omega^2 \frac{\Theta_E(-\omega,p_\parallel)}{2\Delta} \xi(-\omega) d \left(\frac{\xi(-\omega)}{\Delta}+ \frac{\Delta}{\xi(-\omega)}\right) \nonumber \\ && \times \left (-\cos \Delta d +\left(\frac{\Delta}{\xi(-\omega)}-\frac{1}{\Delta d}\right) \sin \Delta d\right)\nonumber \\ .
\end{eqnarray}
Adding the two integrals we get
\begin{eqnarray}
&&\frac{E_{\rm }^+}{A}= \frac{i}{2} \lim_{t\to t_0}\Big [\int_0^\infty \slashed{d}\omega\int   \slashed{d}^2 p_\parallel \;\Theta_E(\omega,p_\parallel) F(\omega,p_\parallel)e^{-i\omega (t-t_0)}\nonumber \\ && +\int_{-\infty}^0 \slashed{d}\omega\int \slashed{d}^2 p_\parallel\; \Theta_E(-\omega,p_\parallel) F(-\omega,p_\parallel)e^{-i\omega (t-t_0)}\Big ]\nonumber \\
\label{VV}
\end{eqnarray}
where we have introduced the even function of $\omega$ 
\be
 F_E(\omega,p_\parallel)= \omega^2\left(\frac{\xi}{\Delta}+ \frac{\Delta}{\xi}\right )\left[ \frac{\epsilon(\omega)}{\xi} \sin \Delta d +\frac{\xi d}{\Delta} \left(-\cos \Delta d +\left(\frac{\Delta}{\xi}-\frac{1}{\Delta d}\right) \sin \Delta d\right )\right ].
\ee
{In these integrals, the limit $t\to t_0$ is taken to be with $t\le t_0$ when $\omega \ge 0$ and $ t\ge t_0$ when $\omega \le 0$. This allows one to reduce }
the electric energy to a single integral with $t\le t_0$
\be
\frac{E_{\rm }^+}{A}= i \lim_{t\to t_0}\int_0^\infty \slashed{d}\omega\int   \slashed{d}^2 p_\parallel \;\Theta_E(\omega,p_\parallel) F_E(\omega,p_\parallel)e^{-i\omega (t-t_0)}
\ee
which will be calculated after Wick's rotation.
\subsubsection{The magnetic energy}

The calculation runs similarly to the electric case
with  the magnetic energy in the plaques given by
\be
\frac{E_{\rm plaque}^-}{A}= i \lim_{t\to t_0}\int \slashed{d}\omega  \;\slashed{d}^2 p_\parallel\; e^{-i\omega (t-t_0)}  \omega^2 \frac{\Theta_M(\omega,p_\parallel)}{2\xi} \left(\frac{\xi}{\epsilon\Delta}+ \frac{\epsilon\Delta}{\xi}\right )\sin \Delta d
\ee
whereas the contribution from the vacuum between the plaques is
\be
\frac{E_{\rm V}^-}{A}= i \lim_{t\to t_0}\int \slashed{d}\omega \;\slashed{d}^2 p_\parallel\; e^{-i\omega (t-t_0)} \omega^2 \frac{\Theta_M(\omega,p_\parallel)}{2\Delta \epsilon } \xi d \left(\frac{\xi}{\epsilon\Delta}+ \frac{\epsilon \Delta}{\xi}\right) \left(-\cos \Delta d +\left(\frac{\epsilon\Delta}{\xi}-\frac{1}{ \Delta d}\right) \sin \Delta d\right).
\ee
Adding the two integrals we get, where here $t\le t_0$, 
\be
\frac{E_{\rm }^-}{A}= i \lim_{t\to t_0}\int_0^\infty \slashed{d}\omega \int \slashed{d}^2 p_\parallel\; \Theta_M(\omega,p_\parallel) F_M(\omega,p_\parallel)e^{-i\omega (t-t_0)}
\label{VV1}
\ee
where we have introduced
\be
 F_M(\omega,p_\parallel)= \omega^2\left(\frac{\xi}{\epsilon\Delta}+ \frac{\epsilon\Delta}{\xi}\right)\left[ \frac{1}{\xi} \sin \Delta d +\frac{\xi d}{\epsilon\Delta} \left(-\cos \Delta d +\left(\frac{\epsilon\Delta}{\xi}-\frac{1}{\Delta d}\right) \sin \Delta d\right)\right]
\ee
The integrals giving the energy and  defined along the real axis such as (\ref{VV1}) are oscillatory and difficult to manipulate. It is more convenient to perform a Wick's rotation in the complex plane and integrate along the imaginary axis.

\subsection{Wick's rotation}

\subsubsection{The electric energy}

The oscillating integrals we found in the previous section can be evaluated after performing Wick's rotation. This is only possible as  the Feynman propagators does not have obstructions by poles and branch cuts when rotating the $\omega$-integral by an angle of $\pi/2$ in the complex plane.
{In taking the limit $t\to t_0$, we consider $t\le t_0$ implying that the integral along a large circle in the first quadrant is exponentially convergent to zero.}
Then, after Wick's rotation $\omega \to i\omega$ we have 
\be
\frac{E_{\rm }^+}{A}= -\frac{1}{2\pi} \int \slashed{d}^2 p_\parallel \int_{0}^\infty d\omega\; \Theta_E(i\omega,p_\parallel) F_E(i\omega,p_\parallel)
\ee
where now 
\be
\Delta= \left (\omega^2 + p_\parallel^2\right )^{1/2},\ \xi= \left (p_\parallel^2 + \omega^2 \epsilon (i\omega)\right )^{1/2}
\ee
where in the metallic case
\be
\epsilon (i\omega)= 1+ \frac{\omega_{\rm pl}^2}{\omega^2 + \gamma \omega}
\ee
which has no singularity on the positive imaginary axis where $\omega> 0$. This transformation implies that 
\be
\Theta_E(i\omega,p_\parallel)=-\frac{1}{\Delta\left( \left(1+ \frac{\xi^2}{\Delta^2}\right ) \sinh \Delta d + \frac{2\xi}{\Delta} \cosh \Delta d\right )}
\ee
and
\be
 F_E(i\omega,p_\parallel)= \omega^2\left(-\frac{\xi}{\Delta}+ \frac{\Delta}{\xi}\right)\left [ \frac{\epsilon(i\omega)}{\xi} \sinh \Delta d +\frac{\xi d}{\Delta} \left(\cosh \Delta d +\left(\frac{\Delta}{\xi}-\frac{1}{\Delta d}\right) \sinh \Delta d\right )\right ]
\ee
which is real.
Finally we obtain
\begin{eqnarray}
&&\frac{E_{\rm }^+}{A}= -\frac{1}{2\pi} \int \slashed{d}^2 p_\parallel \int_{0}^\infty d\omega  \frac{\omega^2}{\Delta\left( \left(1+ \frac{\xi^2}{\Delta^2}\right ) \sinh \Delta d + \frac{2\xi}{\Delta} \cosh \Delta d\right)} \left(\frac{\xi}{\Delta}- \frac{\Delta}{\xi}\right )\times \nonumber \\
&&\left [ \frac{\epsilon(i\omega)}{\xi} \sinh \Delta d +\frac{\xi d}{\Delta} \left(\cosh \Delta d +\left(\frac{\Delta}{\xi}-\frac{1}{\Delta d} \right ) \sinh \Delta d \right )\right].\nonumber\\
\end{eqnarray}
Let us take the large $d$ limit corresponding to the infinite space case. We have
\be
\frac{E_{\infty }^+}{A}= -\frac{1}{2\pi} \int \slashed{d}^2 p_\parallel \int_{0}^\infty d\omega  \frac{\omega^2}{\left (1  + \frac{\xi}{\Delta}\right)^2  }\left(\frac{\xi}{\Delta}- \frac{\Delta}{\xi}\right )\left[ \frac{\epsilon(i\omega)}{\xi}  +\frac{\xi d}{\Delta} \left(1 +\left(\frac{\Delta}{\xi}-\frac{1}{\Delta d}\right)\right)\right ].
\ee
The energy of empty space-time must be  renormalised to zero, as we assume that locally space-time is of the Minkowski type corresponding to a vanishing cosmological constant. As a result we define the renormalised energy
$
\frac{E_{\rm R}^+}{A}=\frac{E_{\rm }^+}{A}-\frac{E_{\infty }^+}{A}.
$
This turns out to be
\be
\frac{E_{\rm R}^+}{A}=  -\frac{d}{\pi} \int \slashed{d}^2 p_\parallel \int_{0}^\infty d\omega\; \frac{\omega^2}{\Delta} \frac{1}{\left(\frac{1+\frac{\xi}{\Delta}}{1-\frac{\xi}{\Delta}}\right )^2 e^{2\Delta d}-1} \left(1- \frac{2 p_\parallel^2}{\omega^2}\frac{1}{\xi d}\right )
\ee
where the factor $\left(1- \frac{2 p_\parallel^2}{\omega^2}\frac{1}{\xi d}\right )$ corresponds to the contribution from the boundary plates, i.e. it reduces to unity when concentrating on the cavity of size $d$ only.

\subsubsection{The magnetic energy}

After Wick's rotation $\omega \to i\omega$, the magnetic energy becomes
\be
\frac{E_{\rm }^-}{A}= -\frac{1}{2\pi} \int \slashed{d}^2 p_\parallel \int_{0}^\infty d\omega \; \Theta_M(i\omega,p_\parallel) F_M(i\omega,p_\parallel)
\ee
with
\be
\Theta_M(i\omega,p_\parallel)=-\frac{1}{\Delta \left ( \left(1+ \frac{\xi^2}{\Delta^2}\right) \sinh \Delta d + \frac{2\xi}{\Delta} \cosh \Delta d\right )}
\ee
and
\ba
F_M(i\omega,p_\parallel) &&= \omega^2\left (-\frac{\xi}{\epsilon(i\omega)\Delta}+ \frac{\epsilon(i\omega)\Delta}{\xi}\right )\times  \nonumber \\ &&
 \left [ \frac{1}{\xi} \sinh \Delta d +\frac{\xi d}{\epsilon(i\omega)\Delta} \left (\cosh \Delta d +\left (\frac{\epsilon(i\omega)\Delta}{\xi}-\frac{1}{\Delta d}\right ) \sinh \Delta d\right )\right]
\ea
which is real.
Finally we obtain
\begin{eqnarray}
\frac{E_{\rm }^-}{A} &&= -\frac{1}{2\pi} \int \slashed{d}^2 p_\parallel \int_{0}^\infty d\omega  \;\frac{\omega^2}{\Delta\left ( \left(1+ \frac{\xi^2}{\epsilon^2(i\omega)\Delta^2}\right) \sinh \Delta d + \frac{2\xi}{\epsilon(i\omega)\Delta} \cosh \Delta d \right )}
\times \nonumber \\
&& \left (\frac{\xi}{\epsilon(i\omega)\Delta}- \frac{\epsilon(i\omega)\Delta}{\xi}\right )
\left[ \frac{1}{\xi} \sinh \Delta d +\frac{\xi d}{\epsilon(i\omega)\Delta} \left (\cosh \Delta d +\left (\frac{\epsilon(i\omega)\Delta}{\xi}-\frac{1}{\Delta d}\right) \sinh \Delta d\right)\right ].
\nonumber \\
\end{eqnarray}
Let us take the large $d$ limit corresponding to the infinite space case again. We have
\ba
\frac{E_{\infty }^-}{A} &&= -\frac{1}{2\pi} \int \slashed{d}^2 p_\parallel \int_{0}^\infty d\omega\;  \frac{\omega^2}{\left(1  + \frac{\xi}{\epsilon^2(i\omega)\Delta}\right )^2  }\times \nonumber \\
&& \left(\frac{\xi}{\epsilon(i\omega)\Delta}- \frac{\epsilon(i\omega)\Delta}{\xi}\right )
\left [ \frac{1}{\xi}  +\frac{\xi d}{\epsilon(i\omega)\Delta} \left(1 +\left(\frac{\Delta}{\xi}-\frac{1}{\Delta d}\right )\right )\right ].
\ea
We define the renormalised energy
$
\frac{E_{\rm R}^-}{A}=\frac{E_{\rm }^-}{A}-\frac{E_{\infty }^-}{A}.
$
This turns out to be
\be
\frac{E_{\rm R}^-}{A}=  -\frac{d}{\pi} \int \slashed{d}^2 p_\parallel \int_{0}^\infty d\omega\; \frac{\omega^2}{\Delta} \frac{1}{\left (\frac{1+\frac{\xi}{\epsilon(i\omega)\Delta}}{1-\frac{\xi}{\epsilon(i\omega)\Delta}}\right)^2 e^{2\Delta d}-1} \left (1- \frac{2 p_\parallel^2}{\epsilon(i\omega)\omega^2}\frac{1}{\xi d}\right )
\ee
This can be explicitly evaluated in the ideal case.
\subsubsection{The ideal case}

The ideal case is easily retrieved by sending $\epsilon \to \infty$ and $\xi \to \infty$ implying that
\be
\frac{E_{\rm R}}{A}=  -\frac{2d}{\pi} \int \slashed{d}^2 p_\parallel \int_{0}^\infty d\omega\; \frac{\omega^2}{\Delta} \frac{1}{e^{2\Delta d}-1}
\ee
which can be written by parity as
\be
\frac{E_{\rm R}}{A}=  -\frac{d}{\pi} \int \slashed{d}^2 p_\parallel \int_{-\infty}^\infty d\omega \;\frac{\omega^2}{\Delta} \frac{1}{e^{2\Delta d}-1}
\ee
Notice that $\vec \Delta=(\omega, \vec p_\parallel)$ is a 3-vector whose norm is $\Delta$. Rotation invariance of the integral for the vector $\vec \Delta$ implies that
\be
\frac{E_{\rm R}}{A}=  -\frac{d}{3\pi(2\pi)^2} \int d^3 \Delta \frac{\Delta}{e^{2\Delta d}-1}= - \frac{4d}{3(2\pi)^2}\int_0^\infty d\Delta \frac{\Delta^3}{e^{2\Delta d}-1}=
-\frac{1}{12(2\pi)^2d^3}\Gamma(4) \zeta(4)
\ee
or more concisely
\be
\frac{E_{\rm R}}{A}=-\frac{\pi^2}{720\, d^3}
\ee
as $\Gamma (4)=6$ and $\zeta(4)= \frac{\pi^2}{90}$.
The pressure is then given by taking the derivative with respect to $d$
\be
P= -\frac{d{(E_R/A)}}{dd}= -\frac{\pi^2}{240 \,d^4}
\ee
as expected.

In the non-ideal case, the vacuum energy comprises the energy due to the fluctuations in the empty cavity and in matter. The fact that the reflection coefficients for waves coming from the cavity and leaving it into the plates are not strictly equal to unity implies that the cavity as seen as a closed system of dimension $d$ is in fact an open system where energy flows into the plate continually. As a result, the energy in the cavity and the pressure exerted by the quantum fluctuations on the interface betweeen the cavity and matter are not simply related by one being the derivative of the other. We must calculate the pressure directly. This is done in the following section.

\subsection{The vacuum pressure}
We will evaluate directly the pressure acting on the plate at $z=d$. For this we calculate the vacuum expectation value of the $T_{zz}$ part of the energy momentum tensor.
This is given by
\be
P_z= \frac{1}{2}\sum_{\alpha=\pm} \langle 0\vert \left ( \partial_z  a_{\alpha }^1 \partial_z \tilde a_{\alpha }^1+  \partial_0  a _\alpha^1 \hat \epsilon\star _t \partial_0 \tilde a^1_\alpha-
\partial_x  a _\alpha^1 \partial_x \tilde a^1_\alpha-\partial_y  a _\alpha^1 \partial_y \tilde a^1_\alpha\right ) \vert 0\rangle
\ee
which can be conveniently written in terms of the electric and magnetic contributions
\be
P_z^+= \frac{i}{2} \lim_{z\to z_0}\int \slashed{d}\omega \slashed{d}^2 p_\parallel \; \left(\partial_z\partial_{z_0}-p_\parallel^2+\omega^2\right) G_{EF}(z,\omega,p_{\parallel};z_0)
\ee
and
\be
P_z^-= \frac{i}{2} \lim_{z\to z_0}\int \slashed{d}\omega \slashed{d}^2 p_\parallel\;  \left (\partial_z\partial_{z_0}-p_\parallel^2+\omega^2\right ) G_{MF}(z,\omega,p_{\parallel};z_0)
\ee
evaluated at $z=z_0=d$ after the appropriate renormalisation.
As we are interested in the pressure due to the vacuum on the plate, the electric contribution is given by 
\be
P_z(d)^+=\frac{i}{2} \lim_{z\to d-, \ z_0\to d-}\int \slashed{d}\omega \slashed{d}^2 p_\parallel \;\left (\partial_z\partial_{z_0}+\omega^2 -p_\parallel^2\right) G_{EF}(z,\omega,p_{\parallel};z_0)
\ee
where the limit is taken with $z,z_0\le d$. A similar result holds for the magnetic pressure
\be
P_z(d)^-=\frac{i}{2} \lim_{z\to d-, \ z_0\to d-}\int \slashed{d}\omega \slashed{d}^2 p_\parallel \; \left (\partial_z\partial_{z_0}+\omega^2 -p_\parallel^2\right ) G_{MF}(z,\omega,p_{\parallel};z_0).
\ee
Let us evaluate these integrals.

\subsubsection{The electric pressure}
Let us start with the term involving the two derivatives in the electric case 
\be
\partial_z\partial_{z_0} G_{E}(z,\omega,p_{\parallel};z_0)\vert_{z=z_0=d}=\Theta(\omega,p_\parallel) \;\xi \;\Delta \left(\sin \Delta d -\frac{\xi}{\Delta} \cos \Delta \right)
\ee
where $G_E$ is the Fourier transform of the Green's function which can be then transformed into the Feynman propagator by the Schwarz reflection procedure.
The same derivative expression evaluated as $z\to \infty$ gives
\be
\partial_z\partial_{z_0} G_E(z,\omega,p_{\parallel};z_0)\vert_{z=z_0\to \infty}=\Theta(\omega,p_\parallel) \;\xi^2 \;\left (-\cos \Delta d + \frac{\sin \Delta d}{2}\left ( \frac{\Delta}{\xi}-\frac{\xi}{\Delta}\right )\right ).
\ee
Obviously this term corresponds to a contribution to the pressure exerted on the outer side of the plaque and we must subtract the two quantities to get
\be
\partial_z\partial_{z_0} G_{ER}(z,\omega,p_{\parallel};z_0)=\Theta(\omega,p_\parallel) \;\frac{\xi^2 }{2}\;\left ( \frac{\Delta}{\xi}+\frac{\xi}{\Delta}\right )\sin \Delta d.
\ee
where $R$ stands for renormalised.
Similarly the term in $\left (\omega^2-p^2_\parallel\right )  G_E(z,\omega,p_{\parallel};z)$ must be understood as $(\omega^2-p^2_\parallel)  G_{ER}(z,\omega,p_{\parallel};z)$
where $G_R$ vanishes as $z\to\infty$, i.e. the pressure coming from the outer edge of the plaque has been removed.
Here notice that we have taken the outer pressure as coming from vacuum, i.e. we have taken $\epsilon(\omega)=1$ for the evaluation of the Green's functions. We have then
\be
G_{ER}(d,\omega,p_{\parallel};d)=\frac{\Theta_E(\omega,p_\parallel)}{2} \left ( \frac{\Delta}{\xi}+\frac{\xi}{\Delta}\right )\sin \Delta d.
\ee
As a result the contribution to the pressure can be written as 
\be
P_{z}^+=  \frac{i}{4}\int \slashed{d}\omega \slashed{d}^2 p_\parallel \; \Theta_E(\omega,p_\parallel)\left( \frac{\Delta}{\xi}+\frac{\xi}{\Delta}\right )\left (\xi^2- p_\parallel^2 +\omega^2\right )\sin \Delta d.
\ee
This can be evaluated after Wick's rotation as
\be
P_z^+=  \frac{1}{4}\int \slashed{d}\omega\; \slashed{d}^2 p_\parallel \;\frac{\xi^2 -\Delta^2}{\Delta} \left (\frac{\xi}{\Delta}- \frac{\Delta}{\xi}\right )\frac{\sinh \Delta d}{\left (1+\frac{\xi^2}{\Delta^2}\right ) \sinh \Delta d + \frac{2\xi}{\Delta} \cosh \Delta d}
\ee
Again this result does not vanish when $d\to \infty$. This would correspond to the pressure in empty space-time which is taken to vanish as we assume that
locally space-time is Minkowski. After
removing the pressure in empty space-time  when $d\to \infty$ we get
\be
P_z^+=  -\int \slashed{d}\omega \;\slashed{d}^2 p_\parallel  \;\frac{\Delta}{{\left (\frac{1+\frac{\xi}{\Delta}}{1-\frac{\xi}{\Delta}}\right )^2 e^{2\Delta d}-1}}.
\ee
This is the Lifschitz's formula for the pressure coming from the $TE$ modes with two plates of the same permittivity $\epsilon$. Notice that this  integral can be written as
\be
P_z^+=  -\int \slashed{d}\omega \;\slashed{d}^2 p_\parallel \; \frac{\Delta}{{r_{TE}^{-2}\, e^{2\Delta d}-1}}.
\ee
where $r_{TE}$ is the reflection coefficient for the $TE$ modes evaluated for imaginary frequencies $r_{TE}(i\omega)= \frac{1-\frac{\xi}{\Delta}}{1+\frac{\xi}{\Delta}}$ where $\xi=\left (p_\parallel^2 +\omega^2 \epsilon(i\omega)\right )^{1/2}$ depends on the permittivity for imaginary frequencies.

\subsubsection{The magnetic  pressure}
The two derivative term of the magnetic Green's function is
\be
\partial_z\partial_{z_0} G_M(z,\omega,p_{\parallel};z_0)\vert_{z=z_0=d}=\Theta_M(\omega,p_\parallel)\;\frac{\xi \Delta}{\epsilon}\;\left (\sin \Delta d -\frac{\xi}{\epsilon\Delta} \cos \Delta d\right ).
\ee
The same derivative expression evaluated as $z\to \infty$ using the expressions in the plates gives
\be
\partial_z\partial_{z_0} G_M(z,\omega,p_{\parallel};z_0)\vert_{z=z_0\to \infty}=\Theta_M(\omega,p_\parallel) \xi^2  \left (-\cos \Delta d + \frac{\sin \Delta d}{2}\left ( \frac{\epsilon\Delta}{\xi}-\frac{\xi}{\epsilon\Delta}\right )\right ).
\ee
Now we must remove the pressure exerted by the vacuum outside the plates as the size of the plates goes to infinity. This is obtained by subtracting $\epsilon^{-2}\partial_z\partial_{z_0} G_M(z,\omega,p_{\parallel};z_0)\vert_{z=z_0\to \infty}$. Indeed we know that for the $TM$ polarisation $\lim_{\delta\to 0}\partial_z G_M\vert_{L+\delta}=\lim_{\delta \to 0}\epsilon^{-1}\partial_z G_M\vert_{L-\delta}$ for a plate of size $L$ and the derivatives are taken at the right boundary  of the plates.
The renormalised contribution to the pressure is then
\be
\partial_z\partial_{z_0} G_{MR}(z,\omega,p_{\parallel};z_0)=\Theta_M(\omega,p_\parallel)\; \frac{\xi^2 }{2\epsilon^2}\;\left ( \frac{\epsilon\Delta}{\xi}+\frac{\xi}{\epsilon\Delta}\right )\sin \Delta d.
\ee
where $R$ stands for renormalised.
Similarly the term in $\left (\omega^2-p^2_\parallel\right )  G_M(z,\omega,p_{\parallel};z)$ must be understood as $\left (\omega^2-p^2_\parallel\right )  G_{MR}(z,\omega,p_{\parallel};z)$
where $G_{MR}$ vanishes as $z\to\infty$, i.e. the pressure coming from the outer edge of the plaque has been removed. Hence we have
\be
G_{MR}(d,\omega,p_{\parallel};d)=\frac{\Theta_M(\omega,p_\parallel)}{2} \left ( \frac{\epsilon\Delta}{\xi}+\frac{\xi}{\epsilon\Delta}\right )\sin \Delta d.
\ee
As a result the contribution to the pressure can be written as
\be
P_{z}^-=  \frac{i}{4}\int \slashed{d}\omega\; \slashed{d}^2 p_\parallel\;  \Theta_M(\omega,p_\parallel)\left ( \frac{\epsilon \Delta}{\xi}+\frac{\xi}{\epsilon \Delta}\right)\left (\frac{\xi^2}{\epsilon^2}- p_\parallel^2 +\omega^2\right )\sin \Delta d.
\ee
This can be evaluated after Wick's rotation as
\be
P_z^-=  \frac{1}{4}\int \slashed{d}\omega\; \slashed{d}^2 p_\parallel \;\frac{\frac{\xi^2}{\epsilon^2(i\omega)} -\Delta^2}{\Delta} \left (\frac{\xi}{\epsilon(i\omega)\Delta}- \frac{\epsilon(i\omega)\Delta}{\xi}\right )\frac{\sinh \Delta d}{\left(1+\frac{\xi^2}{\epsilon(i\omega)\Delta^2}\right ) \sinh \Delta d + \frac{2\xi}{\epsilon(i\omega)\Delta} \cosh \Delta d}.
\ee
Again this result does not vanish when $d\to \infty$. This would correspond to the pressure in empty space-time which is taken to vanish as we assume that
locally space-time is Minkowski. After
removing the pressure in empty space-time  obtained when $d\to \infty$ we get
\be
P_z^-=  -\int \slashed{d}\omega\; \slashed{d}^2 p_\parallel  \;\frac{\Delta}{{\left (\frac{1+\frac{\xi}{\epsilon(i\omega)\Delta}}{1-\frac{\xi}{\epsilon(i\omega)\Delta}}\right )^2 e^{2\Delta d}-1}}.
\ee
This is Lifschitz formula for the pressure coming from the $TM$ modes with two plates with the same permittivity $\epsilon$. Notice that the same integral can be written as
\be
P_z^-=  -\int \slashed{d}\omega \;\slashed{d}^2 p_\parallel  \;\frac{\Delta}{{r_{TM}^{-2}\, e^{2\Delta d}-1}}.
\ee
where $r_{TM}$ is the reflection coefficient for the $TM$ modes evaluated for imaginary frequencies.

\subsubsection{The ideal case}
In the ideal case where $r_{TE}=r_{TM}=1$, the electric and magnetic contributions to the Casimir pressure are equal and the pressure  reduces to
\be
P_z= -2\int \slashed{d}\omega \;\slashed{d}^2 p_\parallel \; \frac{\Delta}{e^{2\Delta d}-1},
\ee
which can be evaluated as
\be
P_z= -\frac{2\times 4\pi}{16(2\pi)^3 d^4}\int_0^\infty dx \frac{x^3}{e^x-1}= - \frac{2\times 4\pi}{16(2\pi)^3 d^4}\Gamma(4) \zeta(4)= -\frac{\pi^2}{240\, d^4}.
\ee
This is the usual Casimir pressure in the ideal case.

Now that we have retrieved the classical Lifschitz theory in the Schwinger-Keldysh formalism, we can explore the effects of the axion coupling to photons on the Casimir pressure.

\section{Axion contribution to the Casimir effect}
\label{sec:axion}
\subsection{Perturbative expansion}
The Casimir pressure is corrected by the axion interaction to photons as the two point correlation function $\langle a_\alpha (x) \tilde a_\alpha (y)\rangle$, where $\alpha=\pm$ corresponds to the two polarisations, is corrected. This gives rise to corrections to $G_E(z,\omega, p_\parallel;z_0)$ and  $G_M(z,\omega, p_\parallel;z_0)$ which then induce a change in $P_z$. Let us start with the electric Green's function first.
As we must take the average over the axion oscillations, the first correction to $G_E$ reads at second order for the connected Green's function
\be
i\delta G_E (x;x_0)= -\frac{1}{2}\left \langle a^1_\gamma (x) \tilde a^1_\gamma(x_0) \left (S_{\rm int}(a^1_{\alpha},\tilde a^1_{\alpha})-S^\dagger_{\rm int}(a^2_{\alpha},\tilde a^2_{\alpha})\right )^2\right \rangle_c \label{eq:expansion},
\ee
as represented in Fig.~\ref{fig:expansion}.
This can be calculated using Wick's theorem} and contracting fields using the propagators. We are only interested in the corrections to the connected Green's function so we only keep contractions which do not contribute to disconnected pieces.
This involves three contributions. Let us start with the first one
\be
I= - \frac{1}{2}\left \langle a^1_\gamma (x) \tilde a^1_\gamma(x_0) S_{\rm int}^2(a^1_{\alpha},\tilde a^1_{\alpha})\right \rangle_c,
\ee
or more explicitly
\begin{eqnarray}
I &&=-\frac{1}{2}\int d^4x_1  d^4 x_2  \frac{\phi(t_1)}{M} \frac{\phi(t_2)}{M} \times \nonumber \\ 
&&\Big\langle a^1_\gamma (x) \tilde a^1_\gamma(x_0)
\left (\partial_0 \tilde a_{\alpha }^1 P_{i\alpha\beta}\star_{x_1} \partial^i a_{\beta}^1+\partial_0 a_{\alpha }^1 P_{i\alpha\beta}\star_{x_1} \partial^i \tilde a_{\beta}^1\right )\nonumber \\ 
&&\left (\partial_0 \tilde a_{\delta }^1 P_{i\delta\kappa}\star_{x_2} \partial^i a_{\kappa}^1+\partial_0 a_{\delta }^1 P_{i\delta\kappa}\star_{x_2} \partial^i \tilde a_{\kappa}^1\right )\Big\rangle_c.\nonumber \\
\end{eqnarray}

\begin{figure}[h]
    \centering
    \includegraphics[width=0.4\textwidth]{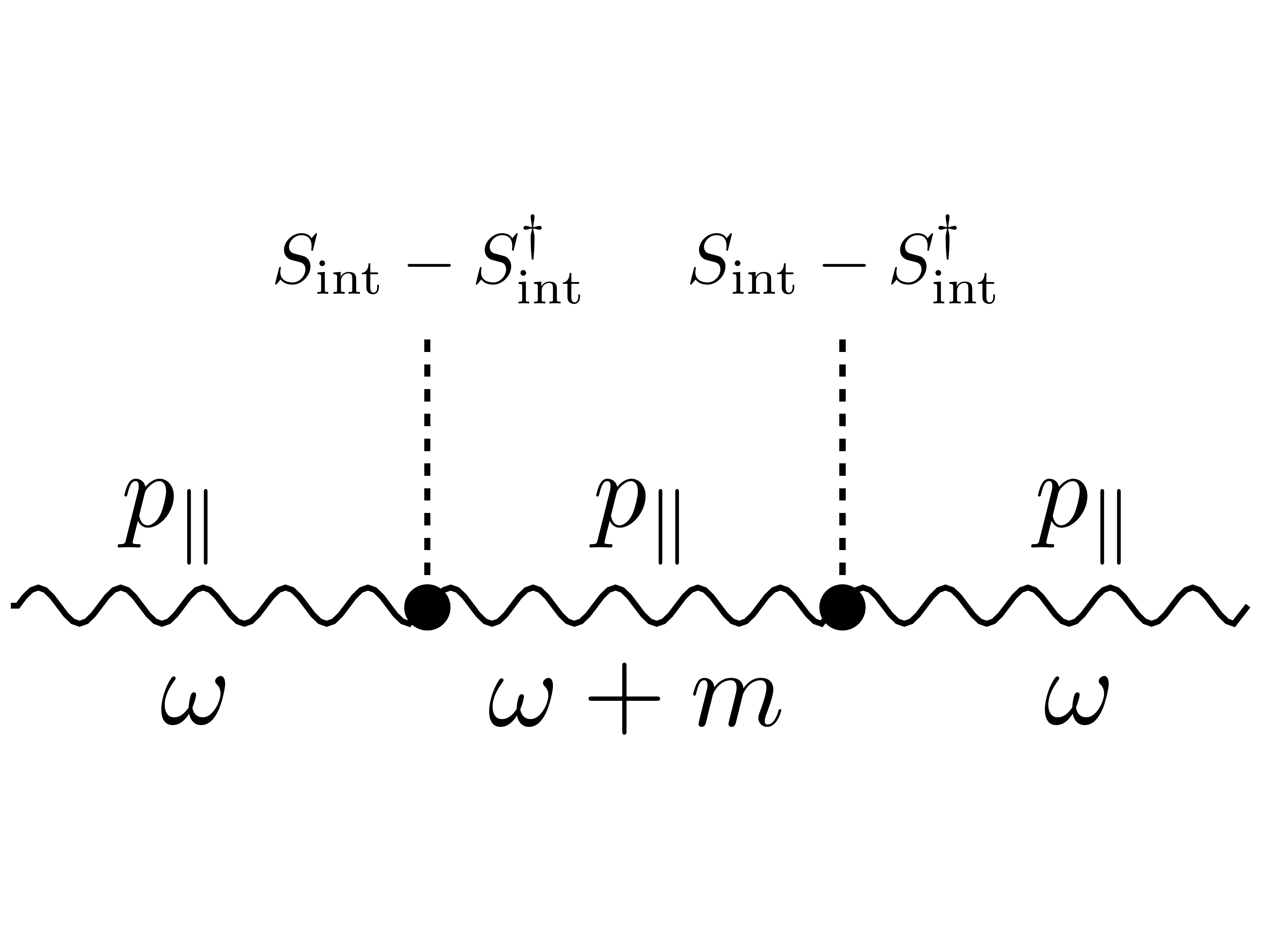}
    \caption{Diagrammatic representation of the insertions in Eq.~\ref{eq:expansion}. Notice that eventually the transverse momentum is conserved at each interaction vertex whilst the axion insertion leads to a change of the energy by $\pm m$. }
    \label{fig:expansion}
\end{figure}
Notice that we have explicitly used that $\tilde \phi\equiv \phi$ as the cosine function is even.
This can be evaluated using Wick's theorem and noticing that the coupling $P_{i\alpha \beta}$ mixes the polarisations, i.e. $\alpha\ne \beta$.
In the following, we denote by $\bar \alpha$ the complementary polarisation to $\alpha$, i.e. $\bar +\equiv -,\ \bar -\equiv +$. We also denote
\be
\left \langle a^1_+(x) \tilde a_+^1(y)\right \rangle = \Delta_+ (x;y), \quad\left \langle a^1_-(x) \tilde a_-^1(y)\right \rangle = \Delta_-(x;y), 
\ee
the latter being  related to the  magnetic Green's functions as
\be
\Delta_-(x;y)= \epsilon^{-1}\star_{t_x}\Delta_{MF}(x;y)
\ee
where
\be
\Delta_+(x;y)= iG_{MF}(x;y),\quad \ \Delta_M(x;y)= i G_{MF}(x;y).
\ee
Expanding the result we get
\begin{eqnarray}
I&& =-\frac{1}{2}\int d^4x_1  d^4 x_2  \frac{\phi(t_1)}{M} \frac{\phi(t_2)}{M}
\times\nonumber\\
&&\Big [ \partial_{t_1}\Delta_\gamma(x;x_1) P_{i\gamma \bar\gamma}(x_1)\star_{x_1} P_{j\bar\gamma \gamma}(x_2) \star_{x_2} \partial^i_{x_1}\partial_{t_2}
\Delta_{\bar\gamma}(x_1;x_2) \partial^j_{x_2}\tilde \Delta_\gamma (x_0;x_2)\nonumber \\
&& +
\partial_{t_1}\Delta_\gamma(x;x_1) P_{i\gamma \bar\gamma}(x_1)\star_{x_1} P_{j\gamma \bar \gamma}(x_2) \star_{x_2} \partial^i_{x_1}\partial_{x_2}^j
\Delta_{\bar\gamma}(x_1;x_2) \partial_{t_2}\tilde \Delta_\gamma (x_0;x_2)\nonumber \\
&&+
P_{i\bar \gamma \gamma}(x_1)\star_{x_1}\partial_{x_1}^i\Delta_\gamma(x;x_1)  P_{j\bar\gamma \gamma}(x_2) \star_{x_2} \partial_{t_1}\partial_{t_2}
\Delta_{\bar\gamma}(x_1;x_2) \partial^j_{x_2}\tilde \Delta_\gamma (x_0;x_2)\nonumber \\
&&+ P_{i\bar \gamma \gamma}(x_1)\star_{x_1}\partial_{x_1}^i\Delta_\gamma(x;x_1)  P_{j\gamma \bar\gamma}(x_2) \star_{x_2} \partial_{t_1}\partial^j_{x_2}
\Delta_{\bar\gamma}(x_1;x_2) \partial_{t_2}\tilde \Delta_\gamma (x_0;x_2)
\Big].\nonumber \\
\end{eqnarray}
The other correlation functions involve the propagators such as $\Delta^{12}$. The second term is
\begin{eqnarray}
II &&=-\frac{1}{2}\int d^4x_1  d^4 x_2  \frac{\phi(t_1)}{M} \frac{\phi(t_2)}{M} \times
\nonumber\\
&&\Big\langle a^1_\gamma (x) \tilde a^1_\gamma(x_0)
\left (\partial_0 \tilde a_{\alpha }^2 \bar P_{i\alpha\beta}\star_{x_1} \partial^i a_{\beta}^2+\partial_0 a_{\alpha }^2 \bar P_{i\alpha\beta}\star_{x_1} \partial^i \tilde a_{\beta}^2\right )\nonumber \\ &&\left (\partial_0 \tilde a_{\delta }^2 \bar P_{i\delta\kappa}\star_{x_2} \partial^i a_{\kappa}^2+\partial_0 a_{\delta }^2 \bar P_{i\delta\kappa}\star_{x_2} \partial^i \tilde a_{\kappa}^2\right )\Big\rangle_c.\nonumber \\
\end{eqnarray}
This is given explicitly by
\begin{eqnarray}
II &&=-\frac{1}{2}\int d^4x_1  d^4 x_2  \frac{\phi(t_1)}{M} \frac{\phi(t_2)}{M}\times
\nonumber \\
&&\Big [ \partial_{t_1}\Delta_\gamma^{12}(x;x_1) \bar P_{i\gamma \bar\gamma}(x_1)\star_{x_1} \bar P_{j\bar\gamma \gamma}(x_2) \star_{x_2} \partial^i_{x_1}\partial_{t_2}
\Delta_{\bar\gamma}^{22}(x_1;x_2) \partial^j_{x_2}\tilde \Delta_\gamma^{12} (x_0;x_2)\nonumber \\
&& +
\partial_{t_1}\Delta_\gamma^{12}(x;x_1) \bar P_{i\gamma \bar\gamma}(x_1)\star_{x_1} \bar P_{j\gamma \bar \gamma}(x_2) \star_{x_2} \partial^i_{x_1}\partial_{x_2}^j
\Delta_{\bar\gamma}^{22}(x_1;x_2) \partial_{t_2}\tilde \Delta_\gamma^{12} (x_0;x_2)\nonumber \\
&&+
\bar P_{i\bar \gamma \gamma}(x_1)\star_{x_1}\partial_{x_1}^i\Delta_\gamma^{12}(x;x_1)  \bar P_{j\bar\gamma \gamma}(x_2) \star_{x_2} \partial_{t_1}\partial_{t_2}
\Delta^{22}_{\bar\gamma}(x_1;x_2) \partial^j_{x_2}\tilde \Delta_\gamma^{12} (x_0;x_2)\nonumber \\
&&+ \bar P_{i\bar \gamma \gamma}(x_1)\star_{x_1}\partial_{x_1}^i\Delta^{12}_\gamma(x;x_1)  \bar P_{j\gamma \bar\gamma}(x_2) \star_{x_2} \partial_{t_1}\partial^j_{x_2}
\Delta_{\bar\gamma}^{22}(x_1;x_2) \partial_{t_2}\tilde \Delta_\gamma^{12} (x_0;x_2)
\Big].\nonumber \\
\end{eqnarray}
The third term is obtained from
\begin{eqnarray}
III_a &&=\frac{1}{2}\int d^4x_1  d^4 x_2  \frac{\phi(t_1)}{M} \frac{\phi(t_2)}{M}\times \nonumber\\ && \Big \langle a^1_\gamma (x) \tilde a^1_\gamma(x_0)
\left (\partial_0 \tilde a_{\alpha }^1  P_{i\alpha\beta}\star_{x_1} \partial^i a_{\beta}^1+\partial_0 a_{\alpha }^1 P_{i\alpha\beta}\star_{x_1} \partial^i \tilde a_{\beta}^1\right)\nonumber \\ && \left (\partial_0 \tilde a_{\delta }^2 \bar P_{i\delta\kappa}\star_{x_2} \partial^i a_{\kappa}^2+\partial_0 a_{\delta }^2 \bar P_{i\delta\kappa}\star_{x_2} \partial^i \tilde a_{\kappa}^2\right )\Big\rangle_c\nonumber \\
\end{eqnarray}
and the one where the two brackets are interverted.
This gives rise to eight terms in two separate contributions
\begin{eqnarray}
III_a &&=\frac{1}{2}\int d^4x_1  d^4 x_2  \frac{\phi(t_1)}{M} \frac{\phi(t_2)}{M}\times
\nonumber \\
&&\Big [ \partial_{t_1}\Delta_\gamma^{11}(x;x_1)  P_{i\gamma \bar\gamma}(x_1)\star_{x_1} \bar P_{j\bar\gamma \gamma}(x_2) \star_{x_2} \partial^i_{x_1}\partial_{t_2}
\Delta_{\bar\gamma}^{12}(x_1;x_2) \partial^j_{x_2}\tilde \Delta_\gamma^{12} (x_0;x_2)\nonumber \\
&& +
\partial_{t_1}\Delta_\gamma^{11}(x;x_1)  P_{i\gamma \bar\gamma}(x_1)\star_{x_1} \bar P_{j\gamma \bar \gamma}(x_2) \star_{x_2} \partial^i_{x_1}\partial_{x_2}^j
\Delta_{\bar\gamma}^{12}(x_1;x_2) \partial_{t_2}\tilde \Delta_\gamma^{12} (x_0;x_2)\nonumber \\
&&+
 P_{i\bar \gamma \gamma}(x_1)\star_{x_1}\partial_{x_1}^i\Delta_\gamma^{11}(x;x_1)  \bar P_{j\bar\gamma \gamma}(x_2) \star_{x_2} \partial_{t_1}\partial_{t_2}
\Delta^{12}_{\bar\gamma}(x_1;x_2) \partial^j_{x_2}\tilde \Delta_\gamma^{12} (x_0;x_2)\nonumber \\
&&+  P_{i\bar \gamma \gamma}(x_1)\star_{x_1}\partial_{x_1}^i\Delta^{11}_\gamma(x;x_1)  \bar P_{j\gamma \bar\gamma}(x_2) \star_{x_2} \partial_{t_1}\partial^j_{x_2}
\Delta_{\bar\gamma}^{12}(x_1;x_2) \partial_{t_2}\tilde \Delta_\gamma^{12} (x_0;x_2)
\Big]\nonumber \\
\end{eqnarray}
and
\begin{eqnarray}
III_b&& =\frac{1}{2}\int d^4x_1  d^4 x_2  \frac{\phi(t_1)}{M} \frac{\phi(t_2)}{M}\times\nonumber\\
&&\Big [ \partial_{t_1}\Delta_\gamma^{12}(x;x_1) \bar P_{i\gamma \bar\gamma}(x_1)\star_{x_1}  P_{j\bar\gamma \gamma}(x_2) \star_{x_2} \partial^i_{x_1}\partial_{t_2}
\Delta_{\bar\gamma}^{21}(x_1;x_2) \partial^j_{x_2}\tilde \Delta_\gamma^{11} (x_0;x_2)\nonumber \\
&& +
\partial_{t_1}\Delta_\gamma^{12}(x;x_1) \bar P_{i\gamma \bar\gamma}(x_1)\star_{x_1}  P_{j\gamma \bar \gamma}(x_2) \star_{x_2} \partial^i_{x_1}\partial_{x_2}^j
\Delta_{\bar\gamma}^{21}(x_1;x_2) \partial_{t_2}\tilde \Delta_\gamma^{11} (x_0;x_2),\nonumber \\
&&+
\bar P_{i\bar \gamma \gamma}(x_1)\star_{x_1}\partial_{x_1}^i\Delta_\gamma^{12}(x;x_1)   P_{j\bar\gamma \gamma}(x_2) \star_{x_2} \partial_{t_1}\partial_{t_2}
\Delta^{21}_{\bar\gamma}(x_1;x_2) \partial^j_{x_2}\tilde \Delta_\gamma^{11} (x_0;x_2)\nonumber \\
&&+ \bar P_{i\bar \gamma \gamma}(x_1)\star_{x_1}\partial_{x_1}^i\Delta^{12}_\gamma(x;x_1)   P_{j\gamma \bar\gamma}(x_2) \star_{x_2} \partial_{t_1}\partial^j_{x_2}
\Delta_{\bar\gamma}^{21}(x_1;x_2) \partial_{t_2}\tilde \Delta_\gamma^{11} (x_0;x_2)
\Big].\nonumber \\
\end{eqnarray}
All these contributions give rise to Feynman integrals corresponding to tree-level Feynman diagrams where two vertices containing one axion insertion each are drawn and joined by lines corresponding to the propagators linked to the external points see Fig.~\ref{fig:expansion}.

\subsection{Feynman integrals}

All the Feynman integrals  are constructed in the same fashion. Let us give an explicit example using the first term of $I$ that we denote by $I_1$. First of all writing $\phi(t)= \frac{\phi_0( e^{imt}+ e^{-imt})}{2}$, the source term contains terms in $\frac{\phi_0^2}{4} e^{i\epsilon_1 mt_1+i\epsilon_2 mt_2}$
where $\epsilon_{1,2}=\pm$. As we take the average over $t=t_0$ to obtain the correction to the Casimir pressure, the end result is non-vanishing for contributions with $\epsilon_2=-\epsilon_1$ only. If this is not the case then the diagram is associated to an oscillating integral whose average vanishes.

There are two contributions corresponding to the two possibilities for $\epsilon_1=\pm$.
The first one is given by
\begin{eqnarray}
I_1 &&\supset  -\frac{\phi_0^2}{8M^2} \int \slashed{d}^2 p_\parallel\; \slashed{d}\omega \; dz_1 \;dz_2 \;e^{ip_\parallel.(x-x_0)_\parallel- i\omega(t-t_0)}\omega (\omega+m) \times\nonumber \\
&&\Delta_\gamma(z;z_1,\omega,p_\parallel)\; \hat\Delta_{\bar \gamma}(z_1;z_2,\omega+m,p_\parallel)\; \check{\tilde \Delta}_\gamma(z_0;z_2;\omega,p_\parallel).\nonumber \\
\end{eqnarray}
We have used the relation
$
k^i P_{i\gamma \bar \gamma}(\vec k)= \vert \vec k \vert
$
as the triple $(\vec k, \vec e_\gamma, \vec  e_{\bar\gamma})$ is of volume $\vert \vec k \vert$. Notice that reversing the order of the vectors changes the overall sign leading to the changes of signs in the second and third contributions $I_2$ and $I_3$. We have introduced the modified propagator
\be
\check \Delta_\gamma(z_1,z_2;\omega, p_\parallel)= \int \slashed{d}p_z \;e^{ip_z z_1} \sqrt{p_z^2 +p_\parallel^2}  \;\Delta_\gamma(z_2;\omega, p_z, p_\parallel)
\ee
where we have Fourier-transformed the propagator with respect to the first argument only
\be
\Delta_\gamma (z_2;\omega, p_z, p_\parallel)= \int dz_1 \;e^{-ip_z z_1} \Delta_\gamma (z_1,z_2;\omega, p_\gamma).
\ee
Similarly we define $\check{\tilde \Delta}_\gamma$ by first taking the Fourier transform of $\tilde \Delta$ with respect to the second argument and then multiplying by $\sqrt{p_z^2 +p_\parallel^2}$ before taking the inverse Fourier transform.
The other contribution to $I_1$ is simply obtained by changing $m\to -m$.
This  corresponds to a contribution to the Fourier transform of the Green's function
\begin{eqnarray}
i\delta G_{I_1}(z;z_0,\omega,p_\parallel)&&=\frac{\phi_0^2}{4M^2}\int dz_1\; dz_2 \;\Delta_\gamma(z;z_1,\omega,p_\parallel)  \check{\tilde \Delta}_\gamma(z_0;z_2;\omega,p_\parallel)\omega
\times\nonumber \\
&&\big [  (\omega-m) \check\Delta_{\bar \gamma}(z_1;z_2,\omega-m,p_\parallel)+(\omega+m) \check\Delta_{\bar \gamma}(z_1;z_2,\omega+m,p_\parallel)
\big ].\nonumber \\
\end{eqnarray}
The second contribution in the $I$ list is given by
\begin{eqnarray}
 i\delta G_{I_2}(z;z_0,\omega,p_\parallel)&&=
-\frac{\phi_0^2}{8M^2}\int dz_1\; dz_2 \;\Delta_\gamma(z;z_1,\omega,p_\parallel)  \check{\tilde \Delta}_\gamma(z_0;z_2;\omega,p_\parallel)\omega^2
\times\nonumber \\
&&\big [  \check\Delta_{\bar \gamma}(z_1;z_2,\omega-m,p_\parallel)+ \check\Delta_{\bar \gamma}(z_1;z_2,\omega+m,p_\parallel)
\big ].\nonumber \\
\end{eqnarray}
The third contribution in the $I$ list is given by
\begin{eqnarray}
i\delta G_{I_3}(z;z_0,\omega,p_\parallel)&&\supset-\frac{\phi_0^2}{8M^2}\int dz_1\; dz_2\; \Delta_\gamma(z;z_1,\omega,p_\parallel)   \check{\tilde \Delta}_\gamma(z_0;z_2;\omega,p_\parallel)\times
\nonumber \\
&&\big [  (\omega-m)^2 \check\Delta_{\bar \gamma}(z_1;z_2,\omega-m,p_\parallel)+(\omega+m)^2 \check\Delta_{\bar \gamma}(z_1;z_2,\omega+m,p_\parallel)
\big ].\nonumber \\
\end{eqnarray}
Finally the last contribution reads
\be
i\delta G_{I_4}(z;z_0,\omega,p_\parallel)=i\delta G_{I_1}(z;z_0,\omega,p_\parallel).
\ee
We can now collect the four terms and obtain the simplified result. This gives
\begin{eqnarray}
i\delta G_{I}(z;z_0,\omega,p_\parallel)&&=-\frac{m^2\phi_0^2}{8M^2}\int dz_1\; dz_2 \;\Delta_\gamma(z;z_1,\omega,p_\parallel)  \check{\tilde \Delta}_\gamma(z_0;z_2;\omega,p_\parallel)\times
\nonumber \\
&&\big [  \check\Delta_{\bar \gamma}(z_1;z_2,\omega-m,p_\parallel)+ \check\Delta_{\bar \gamma}(z_1;z_2,\omega+m,p_\parallel)
\big ].\nonumber \\
\end{eqnarray}
We can comment on this result. First of all, this can be seen as the Feynman integral for the propagator of each polarisation where two insertions of the interaction vertex are added. As this is a tree level Feynman diagram, momentum is conserved and the three propagators have the same transverse momentum $p_\parallel$. At each vertex the pulsation $\omega$ is shifted by either $m$ or $-m$, see Fig. \ref{fig:expansion}. The time derivatives  at the vertex result in factors of  $m$ corresponding to a vertex in $\dot \phi$. In fact we know that the correction must vanish when $m$ vanishes as the term $F\tilde F$ in the Lagrangian is a total derivative.  In this correction to the propagator, only the propagator $\Delta\equiv \Delta^{11}$ is involved. We can now read off the other contributions
\begin{eqnarray}
i\delta G_{II}(z;z_0,\omega,p_\parallel)&&=-\frac{m^2\phi_0^2}{8M^2}\int dz_1 \;dz_2\; \Delta_\gamma^{12}(z;z_1,\omega,p_\parallel)  \check{\tilde \Delta}_\gamma^{12}(z_0;z_2;\omega,p_\parallel)\times
\nonumber \\
&&\big [  \check\Delta_{\bar \gamma}^{22}(z_1;z_2,\omega-m,p_\parallel)+ \check \Delta_{\bar \gamma}^{22}(z_1;z_2,\omega+m,p_\parallel)
\big ].\nonumber \\
\end{eqnarray}
The third contribution gives
\begin{eqnarray}
i\delta G_{III_a}(z;z_0,\omega,p_\parallel)&&=\frac{m^2\phi_0^2}{8M^2}\int dz_1 dz_2 \Delta_\gamma^{11}(z;z_1,\omega,p_\parallel)  \check{\tilde \Delta}_\gamma^{12}(z_0;z_2;\omega,p_\parallel)\times
\nonumber \\
&&\big [  \check\Delta_{\bar \gamma}^{12}(z_1;z_2,\omega-m,p_\parallel)+ \check\Delta_{\bar \gamma}^{12}(z_1;z_2,\omega+m,p_\parallel)
\big ],\nonumber \\
\end{eqnarray}
and the fourth one
\begin{eqnarray}
i\delta G_{III_b}(z;z_0,\omega,p_\parallel)&&=\frac{m^2\phi_0^2}{8M^2}\int dz_1 dz_2 \Delta_\gamma^{12}(z;z_1,\omega,p_\parallel)  \check{\tilde\Delta}_\gamma^{11}(z_0;z_2;\omega,p_\parallel)\times
\nonumber \\
&&\big [  \check \Delta_{\bar \gamma}^{21}(z_1;z_2,\omega-m,p_\parallel)+ \check\Delta_{\bar \gamma}^{21}(z_1;z_2,\omega+m,p_\parallel)
\big ].\nonumber \\
\end{eqnarray}
The structure of these corrections are similar to the ones found in \cite{Favitta:2023hlx} here extended to the Schwinger-Keldysh case relevant to the dissipative case.

\begin{figure}[h]
    \centering
    \includegraphics[width=0.6\textwidth]{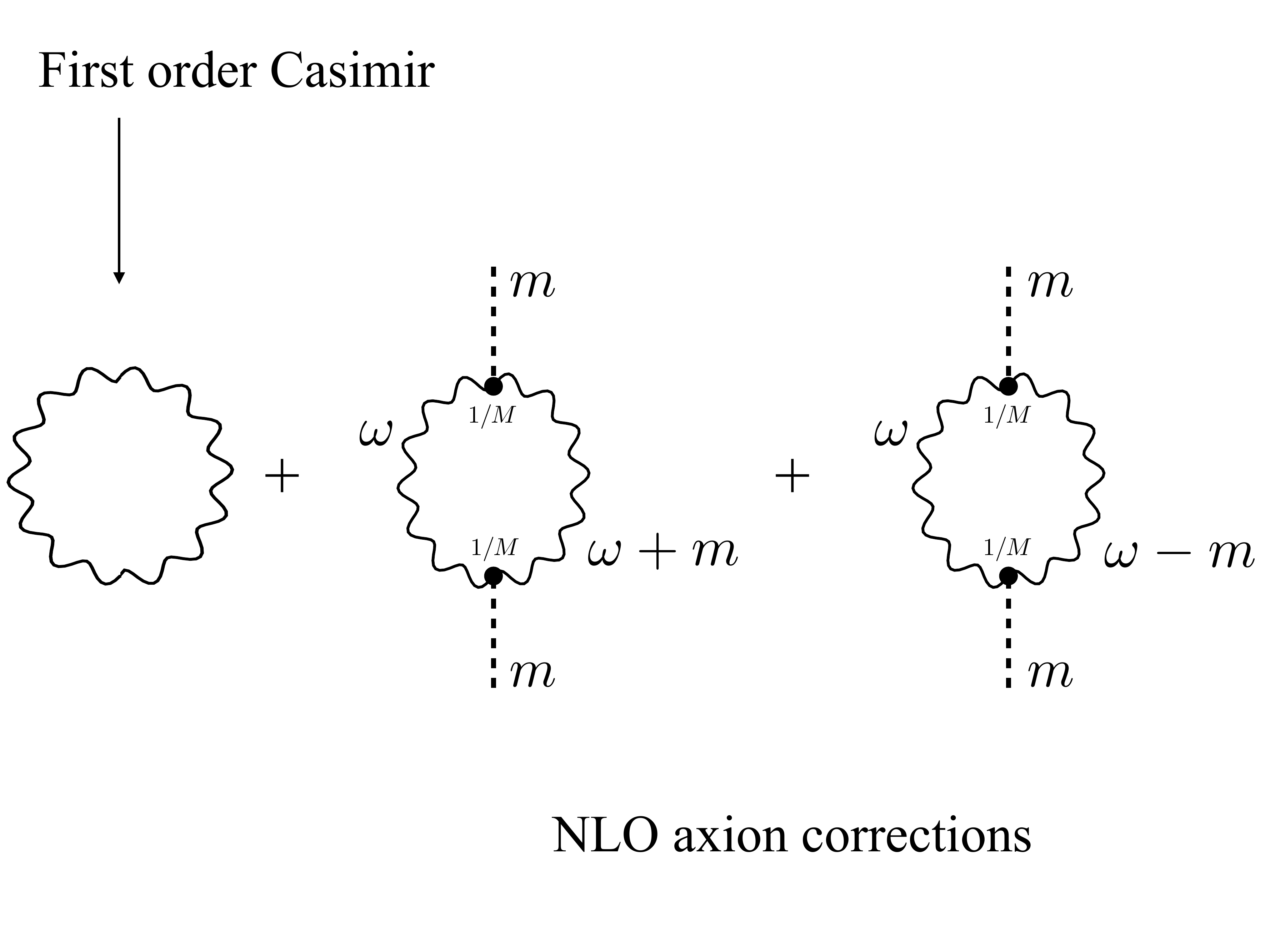}
    \caption{{Feynman diagrams corresponding to the quantum Casimir effect and the leading-order axion corrections. The first order casimir effect corresponds to the Lifschitz theory. }}
    \label{fig:diagrams}
\end{figure}

We can now concentrate on the corrections to the Green's function at coinciding points {as illustrated in Fig.~\ref{fig:diagrams}}. In principle all the integrals are given above and could be evaluated numerically after an appropriate renormalisation when necessary. In the following, we will focus on the case $\gamma \to 0$ as most of the calculations can be performed analytically and they illustrate the physics that one expects from the axion contribution to the Casimir pressure.

\subsection{The vanishing dissipation limit $\gamma \to 0$}
Let us recall some of the properties of the propagators before discussing the different  contributions to the axionic Casimir pressure. We will restrict ourselves to the weak dissipation limit $\gamma \to 0$ as this simplifies the calculation and as for the ideal situation in the electromagnetic  Casimir case, this provides a useful estimate of the effect without too many calculational complications.

For the diagonal propagators involving the first part of the Keldysh path we
can use the decomposition defined in section \ref{sec:ff}
\be
\Delta^{11}_\gamma(z;z_0,\omega, p_\parallel)= i \Theta_\gamma (\omega,p_\parallel) H_\gamma(z;z_0, \omega, p_\parallel)
\ee
when $\Re (\omega) >0$ and we replace $\omega \to -\omega$ when $\Re (\omega) <0$. This corresponds to the definition of the Feynman propagator.
Similarly we have
\be
\Delta^{21}_\gamma(z;z_0,\omega, p_\parallel)\supset iY(\Re(\omega))\;\slashed{\delta}\left(\Theta^{-1}_\gamma(\omega,p_\parallel)\right ) \; H_\gamma(z;z_0, \omega, p_\parallel)
\ee
corresponding to the positive energy resonances, i.e. the resonances whose real parts are positive.
This is complemented by a contribution coming from the continuum with pulsations above $\omega_{\rm Pl}$.
A similar result holds for $\Delta^{12}_\gamma$ with the negative frequency resonances.
In the Feynman integrals, we also need the modified propagators $\check \Delta_\gamma$. They involve the modified Fourier transform $\check{H}$ which is defined like $\check \Delta_\gamma$.
Finally, we notice that the Green's functions $\tilde \Delta$ exchange the positive and negative frequency parts with the ones of $\Delta$.
Using this,
we see that $\delta G_{II}$ vanishes as it is only non-vanishing when $\Re(\omega)=0$ and this does not contribute to the pressure which involves an integral over $\omega$. Similarly, the contribution $\delta G_{III_b}$ is only non-vanishing when $\Re(\omega) \in [-m,0]$ and $\delta G_{III_b}$ when $\Re(\omega) \in [0,m]$.

In the following, we will only be interested in axion masses $m\lesssim \omega_{\rm Pl}$ corresponding to resonances for distances larger than the London length. In this regime, we see that the $\delta G_{III}$ channels do not receive contributions from the continuum. Only the resonances can contribute. Collecting everything we have therefore 
\begin{eqnarray}
\delta G_{I}(z;z_0,\omega,p_\parallel)&&\supset \frac{m^2\phi_0^2}{8M^2}\;\Theta_\gamma(\omega,p_\parallel)\;\tilde \Theta_{ \gamma}(\omega, p_\parallel) 
\times \nonumber\\
&&\int dz_1 dz_2 H_\gamma(z;z_1,\omega,p_\parallel)  \check{\tilde H}_\gamma(z_0;z_2;\omega,p_\parallel)
\nonumber \\
&&\Big [  \Theta_{\bar \gamma}(\omega-m,p_\parallel)\check H_{\bar \gamma}(z_1;z_2,\omega-m,p_\parallel)+  \Theta_{\bar \gamma}(\omega+m,p_\parallel)\check H_{\bar \gamma}(z_1;z_2,\omega+m,p_\parallel)
\Big ].\nonumber \\
\end{eqnarray}
The third contributions give
\begin{eqnarray}
\delta G_{III_a}(z;z_0,\omega,p_\parallel)&&\supset-\frac{m^2\phi_0^2}{8M^2}\;Y\left (-\Re(\omega)\right ) \;Y\left(\Re(\omega)+m\right)\;\Theta_\gamma(\omega,p_\parallel)
\times \nonumber\\
&&\slashed{\delta}\left (\Theta^{-1}_{\bar \gamma}(\omega+m,p_\parallel)\right )\;\slashed{\delta}\left (\tilde \Theta^{-1}_{ \gamma}(\omega,p_\parallel)\right )\times \nonumber \\ 
&& \int dz_1 dz_2 H_\gamma(z;z_1,\omega,p_\parallel)  \check{\tilde  {H_\gamma}}(z_2;z_0;\omega,p_\parallel)
\check H_{\bar \gamma}(z_1;z_2,\omega+m,p_\parallel),
\nonumber \\
\end{eqnarray}
and similarly
\begin{eqnarray}
\delta G_{III_b}(z;z_0,\omega,p_\parallel)&&\supset-\frac{m^2\phi_0^2}{8M^2}\;Y\left(\Re(\omega)\right ) \;Y\left(m-\Re(\omega)\right )\;\Theta_\gamma(\omega,p_\parallel) \times \nonumber\\
&&\slashed{\delta}\left (\Theta^{-1}_{\bar \gamma}(\omega-m,p_\parallel)\right )\;\slashed{\delta}\left(\tilde \Theta^{-1}_{ \gamma}(\omega,p_\parallel)\right )\times \nonumber \\ 
&& \int dz_1 dz_2 H_\gamma(z;z_1,\omega,p_\parallel) \check{\tilde {H}}_\gamma(z_2;z_0;\omega,p_\parallel)
\check H_{\bar \gamma}(z_2;z_1,\omega-m,p_\parallel).
\nonumber \\
\end{eqnarray}
As can be seen, the last two terms are only non-vanishing when two conditions are satisfied, i.e. either
\be
\Theta^{-1}_{\bar \gamma}(\omega+m,p_\parallel)=0,\quad
\tilde \Theta^{-1}_{\gamma}(\omega,p_\parallel)=0
\label{firtc}
\ee
or
\be
\Theta^{-1}_{\bar \gamma}(\omega-m,p_\parallel)=0,\quad
\tilde \Theta^{-1}_{\gamma}(\omega,p_\parallel)=0.
\ee
These equations must have solutions for either $\Re(\omega)\in [-m,0]$ or $\Re(\omega)\in [0,m]$. When  solutions exist, they would be in a finite number.
The resonances are given by (\ref{reson}) with an infinitesimal $\gamma$. The condition (\ref{firtc}) must have solutions with the same imaginary parts. This implies that the integer numbers labeling the resonances must be equal. This is not compatible with the real parts being equal implying that there is no resonance when the system is weakly dissipative.
This implies that
\be
\delta G(z;z_0,\omega, p_\parallel)= \delta G_I(z;z_0,\omega, p_\parallel)
\ee
which can be explicitly calculated analytically.

\subsection{The axionic pressure in the ideal case}

We will simplify the calculation of the correction to the Green's function by approximating the propagators by their ideal expressions. This is a good approximation when the penetration length is much smaller than the distance between the plates in such a way that one can neglect the penetration of the modes into the plaques.
In this case the two Green's functions reduce to
\be
 G_E( z;z_0, \omega, p_\parallel)= \frac{1}{\Delta} \frac{\sin \Delta (z_0-d) \sin \Delta z}{\sin \Delta d}, \  G_M( z;z_0, \omega, p_\parallel)= \frac{1}{\Delta} \frac{\cos \Delta (z_0-d) \cos \Delta z}{\sin \Delta d}
 \label{ideal}
\ee
and vanish outside the interval $[0,d]$. As a result the quantum correction to the Green's functions becomes
\begin{eqnarray}
\delta G_{E}(z;z_0,\omega,p_\parallel)&&=\frac{m^2\phi_0^2}{8M^2}\int dz_1 \;dz_2\; G_E(z;z_1,\omega,p_\parallel)  \check G_E(z_2;z_0;\omega,p_\parallel)
\times \nonumber \\
&&\left [  \check G_{E}(z_1;z_2,\omega-m,p_\parallel)+ \check G_{E}(z_1;z_2,\omega+m,p_\parallel)
\right  ]\nonumber \\
\end{eqnarray}
for the electric case. We have used $\tilde G(x;y)= G(y,x)$. The correction to the magnetic Green's function is given by
\begin{eqnarray}
\delta G_{M}(z;z_0,\omega,p_\parallel) &&=\frac{m^2\phi_0^2}{8M^2}\int dz_1 \;dz_2 \;G_M(z;z_1,\omega,p_\parallel) \check G_M(z_2;z_0;\omega,p_\parallel)
\times \nonumber \\
&&\left [  \check G_M(z_1;z_2,\omega-m,p_\parallel)+ \check G_M(z_1;z_2,\omega+m,p_\parallel)
\right ].\nonumber \\
\end{eqnarray}
An easy calculation gives that for the modified propagators,
\be
 \check G_E( z;z_0, \omega, p_\parallel)=\omega  G_E( z;z_0, \omega, p_\parallel);\quad \check G_M( z;z_0, \omega, p_\parallel)=\omega  G_M( z;z_0, \omega, p_\parallel).
 \ee
The explicit evaluation of the integrals gives for the corrections
\begin{eqnarray}
\delta G_{E}(z;z_0,\omega,p_\parallel) &&=-\frac{m^2\omega\phi_0^2}{2M^2}\frac{\sin \Delta z \sin \Delta (z_0-d)}{\sin^2 \Delta d }
\times \nonumber \\
&&\left [\frac{ \left (\cos \Delta d -\cos \Delta^- d\right )^2}{\left (\Delta^2 -(\Delta^-)^2\right )^2 } \frac{\omega-m}{\Delta^- \sin \Delta^- d} + \frac{ \left (\cos \Delta d -\cos \Delta^+ d\right )^2}{\left(\Delta^2 -(\Delta^+)^2\right )^2 } \frac{\omega+m}{\Delta^+ \sin \Delta^+ d} \right ]
\nonumber \\
\end{eqnarray}
where $\Delta^\pm $ are associated to $\omega \pm m$. For the magnetic Green's function we have
 \begin{eqnarray}
\delta G_{M}(z;z_0,\omega,p_\parallel) &&=-\frac{m^2\omega\phi_0^2}{2M^2}\frac{\cos \Delta z \cos \Delta (z_0-d)}{\Delta^2\sin^2 \Delta d }
\times \nonumber \\
&&\left  [\frac{ \left (\cos \Delta d -\cos \Delta^- d\right )^2}{\left (\Delta^2 -(\Delta^-)^2\right )^2 } \frac{\Delta_-(\omega-m)}{ \sin \Delta^- d}+ \frac{ \left (\cos \Delta d -\cos \Delta^+ d\right )^2}{\left (\Delta^2 -(\Delta^+)^2\right )^2 } \frac{\Delta_+(\omega+m)}{ \sin \Delta^+ d} \right ] .
\nonumber \\
\end{eqnarray}
Notice that the Green's functions are simply corrected by momentum-dependent form factors.
Using these expressions we can get the pressure after summing over the two polarisations. Let us first consider the correction to the electric pressure
\begin{eqnarray}
\delta P_E(d)&&=-i \frac{m^2\phi_0^2}{4M^2}\int \slashed{d}\omega\; \slashed{d}^2 p_\parallel\;  \frac{\omega \Delta^2 \cos \Delta d}{\sin^2 \Delta d }
\times \nonumber \\
&&\left  [\frac{ \left (\cos \Delta d -\cos \Delta^- d\right )^2}{\left (\Delta^2 -(\Delta^-)^2\right )^2 } \frac{\omega -m }{\Delta^- \sin \Delta^- d}+ \frac{ \left (\cos \Delta d -\cos \Delta^+ d\right )^2}{\left (\Delta^2 -(\Delta^+)^2\right )^2 } \frac{\omega +m }{\Delta^+ \sin \Delta^+ d} \right ] .
\nonumber \\
\end{eqnarray}
This can be evaluated after Wick's rotation $\omega \to i\omega$ where we have
$
\Delta^\pm= \sqrt{ (\omega \mp im)^2 +p^2_\parallel}
$ leading to
\ba
\delta P_E(d) &&= \frac{m^2\phi_0^2}{2M^2}\times
\nonumber \\
&&\Re \left ( \int \slashed{d}\omega\; \slashed{d}^2 p_\parallel\; \omega(\omega+im)\frac{\Delta^2}{\Delta^-(\Delta^2 -(\Delta^-)^2)^2} \frac{\cosh \Delta d (\cosh \Delta d -\cosh \Delta^- d)^2}{\sinh^2 \Delta d  \sinh \Delta^- d}
\right  ).
\ea
This expression can be simplified as
\be
\delta P_E = \frac{\phi_0^2}{2M^2}\; \frac{1}{d^4} \;f_E(md),
\ee
where the form factor is given by the integral
\be
f_E(md)=
\Re \left ( \int \slashed{d}x \;\slashed{d}^2 \V y\; x(x+imd)  \frac{\hat \Delta^2}{\hat \Delta^-(2x+imd)^2}  \frac{\cosh \hat\Delta  \left (\cosh \hat \Delta  -\cosh \hat \Delta^- \right )^2}{\sinh^2 \hat \Delta  \sinh \hat \Delta^- }
\right  ),
\ee
where we have defined $x=\omega d$ and $\V y= \V p_\parallel d$ and obtain
\be
\hat \Delta= (x^2 +\V y^{\; 2})^{1/2}, \ \hat\Delta^-= ( (x+imd)^2 +\V y^{\;2})^{1/2}.
\ee
The same method yields for the magnetic contribution
\ba
\delta P_M(d) &&= \frac{m^2\phi_0^2}{2M^2} \times \nonumber\\
&&\Re \left ( \int \slashed{d}\omega \slashed{d}^2 \V p_\parallel \omega(\omega+im)\frac{\Delta_-}{\left (\Delta^2 -(\Delta^-)^2\right )^2} \frac{\cosh \Delta d \left(\cosh \Delta d -\cosh \Delta^- d\right)^2}{\sinh^2 \Delta d  \sinh \Delta^- d}
\right  ),
\ea
corresponding to
\be
\delta P_M = \frac{\phi_0^2}{2M^2} \;\frac{1}{d^4} \;f_M(md).
\ee
where the form factor is given by the integral
\be
f_M(md)=
\Re \left  ( \int \slashed{d}x \;\slashed{d}^2 \V y \; x(x+imd)  \frac{\hat \Delta_-}{(2x+imd)^2}  \frac{\cosh \hat\Delta  \left (\cosh \hat \Delta  -\cosh \hat \Delta^- \right )^2}{\sinh^2 \hat \Delta  \sinh \hat \Delta^- }
\right  ).
\ee
Notice that the complete form factor $f_E + f_M$ vanishes when $m\to 0$ corresponding to the vanishing of the effect of the axion in the static limit. As expected the correction to the pressure has a dependence in $1/d^4$ and corrections in $md$.

\subsection{The Axionic pressure}

\begin{figure}[h]
    \centering
    \includegraphics[width=0.7\textwidth]{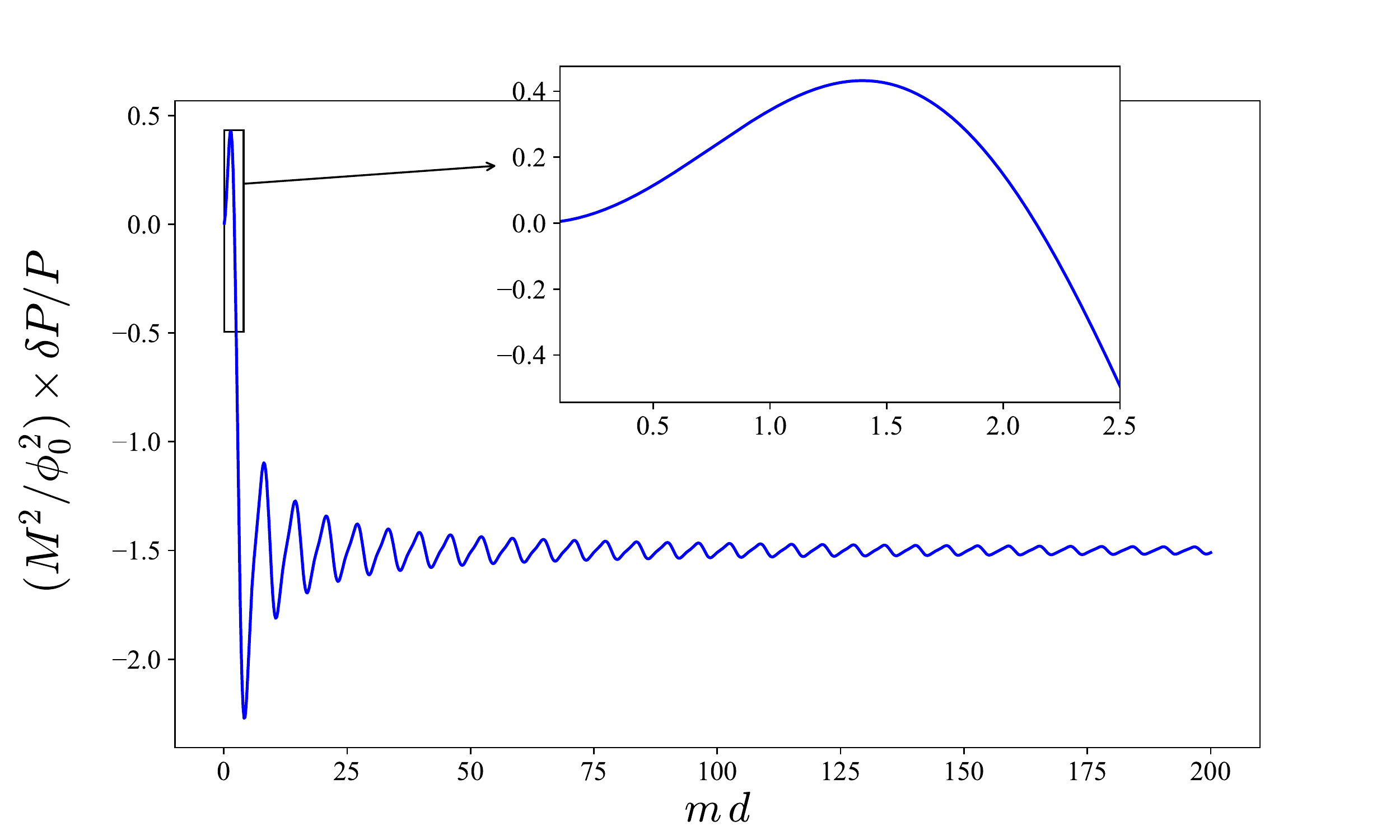}
    \caption{{Axionic quantum correction to the Casimir pressure as a function of the mass-distance product in the ideal case. Notice that at large distance compared to $1/m$ the axion correction is repulsive. }}
    \label{fig:deltaP}
\end{figure}

 The form factors $f_{E,M}$ are ultraviolet (UV) sensitive and dominated by the largest momenta and energies.
In this case and in the limit $md \ll \omega_{\rm Pl} d$, we find that the electric form factor behaves like
\be
f_E\simeq \frac{1}{4}\int \slashed{d}^3 {\hat \Delta}\; \frac{\hat \Delta }{\tanh^3 \hat \Delta}\left (1- e^{imdx/\hat\Delta}\right )^2,
\ee
where we have defined the vector $\vec{\hat \Delta}=(x,\vec y)$. This integral is infrared convergent but diverges quartically in the UV where the cut-off scale is of the order of  $\omega_{\rm Pl}d$ as no pressure is exerted at higher frequencies as the metal becomes transparent. Using this explicit cut-off,  the form factor behaves like
\be
f_E\simeq \frac{\pi}{8} \left (\omega_{\rm Pl}d\right )^4 \left( 2-\frac{4}{md} \sin md + \frac{1}{md}\sin 2md\right).
\ee
This gives a contribution to the pressure in
\be
\delta P_E \simeq  \frac{\omega_{\rm Pl}^4}{128\pi^2} \frac{\phi_0^2}{M^2}\left ( 2-\frac{4}{md} \sin md + \frac{1}{md}\sin 2md \right ).
\ee
This leads to a constant pressure in the limit $d\to \infty$ which must be substracted as before.
Doing so,  the renormalised form factor becomes
\be
f_E\simeq \frac{1}{4}\int \slashed{d}^3 \;{\hat \Delta} \hat \Delta \left (\frac{1}{\tanh^3 \hat \Delta}-1\right )\left(1- e^{imdx/\hat\Delta}\right )^2\simeq \frac{3}{2}\int \slashed{d}^3 {\hat \Delta}  \frac{\hat \Delta}{e^{2\hat \Delta} -1}\left(1- e^{imdx/\hat\Delta}\right )^2.
\ee
This gives explicitly
\be
f_E \simeq 3\left ( 1-\frac{2}{md} \sin md + \frac{1}{2md}\sin 2md\right )\int \slashed{d}^3 {\hat \Delta} \; \frac{\hat \Delta}{e^{2\hat \Delta} -1}.
\ee
The last factor reconstructs the TE casimir pressure in the ideal case.
 As a result we have 
\be
\delta P_E \simeq -\frac{3}{2}\left ( 1-\frac{2}{md} \sin md + \frac{1}{2md}\sin 2md \right )\frac{ \phi_0^2}{M^2} P_E.
\ee
There is also the magnetic contribution implying that the correction to the Casimir pressure due to the axion is given by
\be
\frac{\delta P}{P}\simeq -\frac{3}{2}\left ( 1-\frac{2}{md} \sin md + \frac{1}{2md}\sin 2md\right )\frac{ \phi_0^2}{M^2}.
\ee
At short distance we have
\be
\frac{\delta P}{P}\simeq \frac{ \phi_0^2}{2M^2} (md)^2.
\ee
The dependence of the correction with $md$ is plotted in Fig.~\ref{fig:deltaP}. At short distance the correction is positive and evolves quadratically with $d$ for a fixed mass, as shown in the inset of the figure. Interestingly we find that that for $md={\cal O}(1)$, the pressure changes sign and the plates repulse. This type of behaviour is similar to the one obtained when the axion depends on time or space linearly \cite{Kharlanov:2009pv,Fukushima:2019sjn}. {For larger distances, the correction to the axion-induced Casimir pressure oscillates and becomes constant. The pressure itself vanishes as $1/d^4$ so that for large distances the usual CDM equation of state $p=0$ is retrieved}. The suppression of the quantum effect by $\phi_0^2/M^2$ implies that for dark matter in the galactic halo this contribution is negligible. 
On the other hand, for axion clumps of very large densities, we may hope that the quantum effects could be relevant. The analysis of this possibility is left for future work.

\section{Conclusion}

We have described the effects of the axion oscillations in the environment of a Casimir experiment on the pressure exerted on the metallic plates. We have focused on metallic plates with dissipation and retrieved the Casimir pressure when axions are neglected, i.e. the Lifschitz theory, and then corrected this result at leading order in the axion coupling to the photons. We have developed these calculations in the Schwinger-Keldysh formulation of the $in-in$ formalism as it is particularly suited to evaluate the vacuum expectation value of operators even in the case of dissipative systems. We have also used a {field-}doubling trick in order to describe dissipative systems in a Lagrangian way. For this we have introduced a ``shadow" field whose role is to run history in reverse and bring back energy from infinity. This guarantees that a sound quantum field theory formulation can be defined. This doubling trick makes the connection with classical electrodynamics non-trivial. However we have shown how the conserved energy momentum tensor of the doubled theory can be written as the sum of a well defined  energy momentum tensor and its time-reversed. Identifying the energy momentum tensor of electrodynamics in the presence of dissipation in this manner, we have shown how the pressure on the plates deduces from the spatial components of this tensor exactly reproduce the Lifschitz theory.

We have then calculated the corrections to the Green's function induced by the axion field at second order in the axion coupling constant. As expected, the resulting pressure vanishes for massless axions. Indeed in this case, the axion term disappears from the action for trivial topologies and no axion effect is expected. On the other hand, we find that for small distances compared to the axion's Compton wavelength, the induced pressure varies quadratically with the distance and is attractive. For larger distances {and a given mass}, the axion contribution changes sign and oscillates {with respect to the distance}. This is a breaking of the Kenneth-Klich theorem \cite{Kenneth:2006vr} which happens when parity is violated \cite{Jiang:2018ivv}. Here the role of parity breaking medium is played by the dark matter background with an oscillating axion. 
We have not considered the phenomenology associated to this new effect. In particular, the axion-induced pressure is small due to {the tiny coupling and} the small background dark matter density in the galactic halo unless we take into account the possibility of enhanced effects in axion clumps. This is left for further investigation.
\acknowledgments

We would like to thank Thomas Colas for useful remarks. This project has received funding from the European
Research Council (ERC) under the European Union’s
Horizon 2020 research and innovation programme (Grant
agreement No. 865306).

\bibliography{ref}

\appendix

\section{Dissipation}
\label{app:diss}
\begin{figure}[h]
    \centering
    \includegraphics[width=0.5\linewidth]{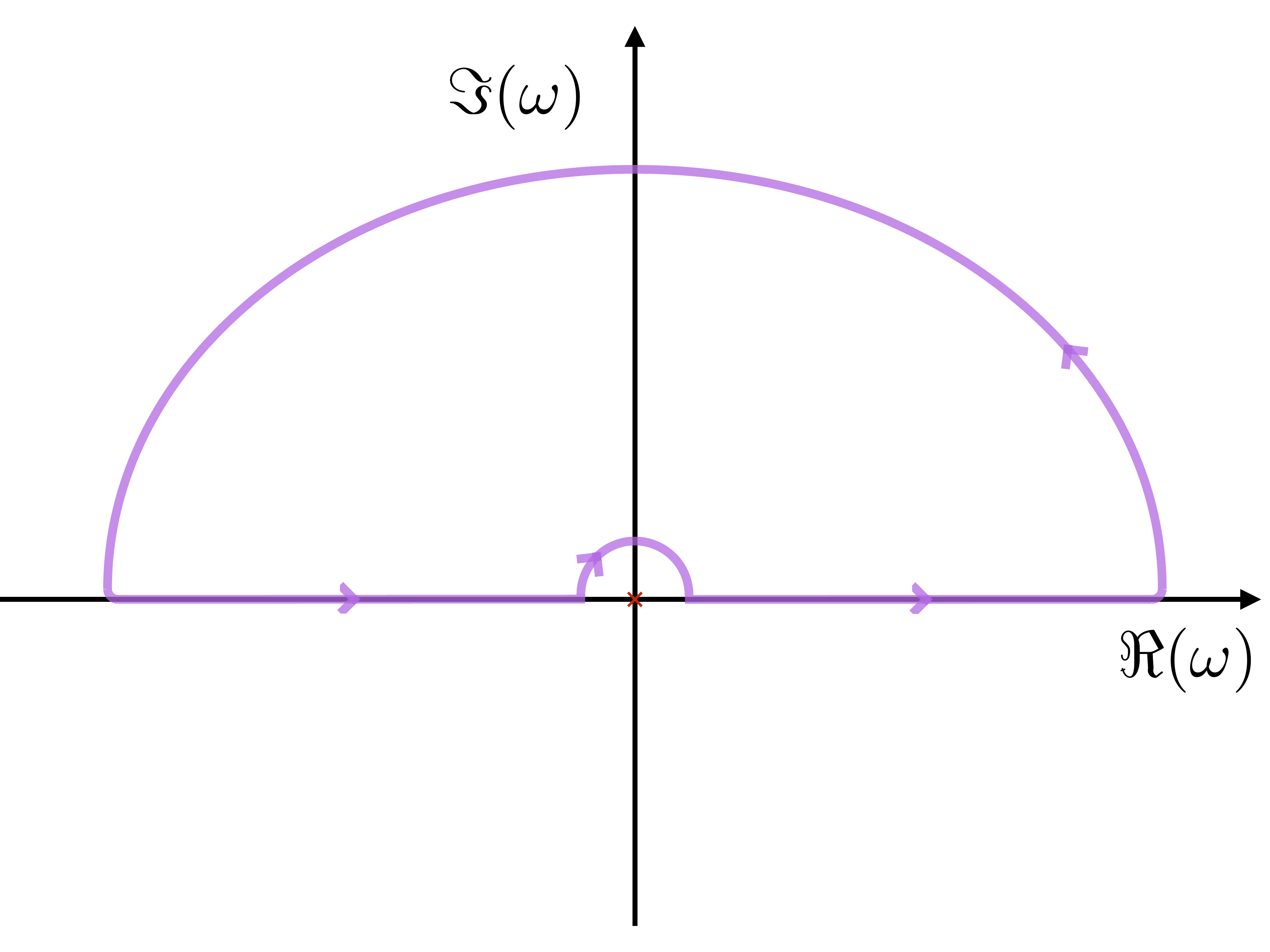}
    \caption{Integration contour used for the evaluation of integral~\ref{eq:dissipation}.}
    \label{fig:dissipation}
\end{figure}

Let us analyse general properties of the permittivity function. We have given the example of the case of metals in the Drude model. More generally, $\epsilon(\omega)-1$ is assumed to be an analytic function in the upper half plane with possible singularities in the lower half plane. As a result we have that
\be
\epsilon (t)= \slashed{\delta} (t) + \int \slashed{d} \omega \; e^{-i\omega t} \left (\epsilon (\omega)-1\right ) \label{eq:dissipation}
\ee
By closing the contour in the upper half-plane, we see that the Cauchy theorem tells us that $\epsilon(t)=\slashed{\delta} (t), \ t<0$.
When $\epsilon(\omega)$ has a pole at the origin as for metals, we define the integral by drawing a small circle around the origin in the complex plane as shown in Fig.~\ref{fig:dissipation}. This guarantees that $\epsilon(t)-\delta (t)$ is only non-zero for $t\ge 0$.

The free Hamiltonian is expressed as
\be
H_{0}=  \int d^3x \;\left ( \partial_0 \tilde a_i \partial_0 a_i  +  \partial_i \tilde a _j \partial_i a_j\right ) =\sum_{\alpha=\pm}\int d^3 x\;\left( \partial _0 \tilde a_\alpha \partial_0 a_\alpha + \partial_i \tilde a _\alpha \partial_i a_\alpha\right)
\ee
corresponding to two systems of independent free scalar fields.
In fact one can define
\be
a_\alpha=\frac{a^+_\alpha + a_\alpha^-}{2}, \ \tilde  a_\alpha= a^+_\alpha- a^-_\alpha
\ee
or equivalently
\be
a_\alpha^+= a_\alpha+ \frac{\tilde a_\alpha}{2}, \ a_\alpha^-= a_\alpha- \frac{\tilde a_\alpha}{2}
\ee
to find that
\be
H_0= H_+-H_-
\ee
where we have the free Hamiltonians
\be
H_{\pm}=  \sum_{\alpha=\pm}\frac{1}{2} \int d^3 x\;\left ( \partial _0  a_\alpha^{\pm} \partial_0 a_\alpha^{\pm} + \partial_i  a _\alpha^{\pm} \partial_i a_\alpha^{\pm}\right )
\ee
The fact that the free Hamiltonian is the difference between two free Hamiltonian of scalar fields in the usual sense is reminiscent of the Keldysh formalism used in the main text.
In fact, going back to the full action including the permittivity we find that
we obtain
\be
S(a_\alpha,\tilde a_\alpha)= S(a^+_\alpha) -S(a^-_\alpha) + K(a^+_\alpha, a^-_\alpha)
\ee
where
\be
S_{}(a^\pm_{\alpha\rm })=\frac{1}{2}\int d^4x \,\left ( \partial_0 a_{\alpha }^\pm \epsilon\star_t \partial_0  a_{\alpha }^\pm- \partial_i  a_{\alpha}^\pm \partial^i  a_{\alpha }^\pm +\frac{\phi}{M} \partial_0  a_{\alpha }^\pm P_{i\alpha\beta}\star_x \partial^i a_{\beta}^\pm+\frac{\tilde \phi}{M}\partial_0 a_{\alpha }^\pm P_{i\alpha\beta}\star_x \partial^i  a_{\beta}^\pm \right ).
\ee
Notice that here the $\pm$ fields with an upper index are not identified with the polarisation basis of fields $\pm$ with a lower index. The $\pm$ fields with an upper index are simply combinations of the gauge fields and their shadows.
This is the type of action functional  corresponding dissipative systems \cite{Galley:2012hxx}. In this case,
the influence functional is given by
\be
K(a^+_\alpha, a^-_\alpha)=\frac{1}{2}\int d^4x \;\partial_0 a_{\alpha }^+ (\epsilon-\hat \epsilon)\star_t \partial_0  a_{\alpha }^-
\ee
and couples the two fields $a_\alpha^\pm$. Notice that it involves the permittivity combination $\epsilon-\hat \epsilon$ whose Fourier transform
$ \epsilon (\omega)- \epsilon(-\omega)= 2i \Im (\epsilon (\omega))$ only depends on the imaginary part of the permittivity function, i.e. the influence functional
is only present when there is dissipation \cite{Galley:2012hxx}.

\section{~$in-in$ vs. $in-out$}
\label{app:in}
In this section we will recall the link between the $in-in$ and $in-out$ formulations of quantum field theory when all the usual assumptions used in particle physics are respected. In particular, we assume that the Hamiltonian is Hermitian implying that the evolution operator is unitary. We also consider the case where Poincar\'e invariance is respected implying that the notion of particle state is well-defined. The ground state of the theory is also assumed to be unique and corresponds to the absence of particles.
We are interested in quantum averages in the Schr\"odinger picture
\be
\left \langle {\cal O}\right \rangle_\psi(t) \equiv \langle \psi(t)\vert  {\cal O}\vert \psi(t)\rangle
\ee
where the ket $\vert \psi \rangle$ evolves with time $\vert \psi \rangle (t)= U(t,-\infty) \vert \psi \rangle$. The system is prepared at time minus infinity and then evolves with the unitary evolution operator.
We focus on vacuum averages where $\vert \psi \rangle= \vert 0\rangle $ is the vacuum of the non-interacting theory at time $t=-\infty$ before the interactions are switched on. This vacuum is the $in$ vacuum of the theory $\vert 0\rangle = \vert {\rm in}\rangle$. We can then use the fact that the free Hamiltonian $H_0\vert 0\rangle =0$ has vanishing energy in vacuum to write
\be
\langle {\cal O}\rangle_{\rm vac}(t) \equiv \left \langle {\rm {in}}\left \vert  U^\dagger_I(t,-\infty){\cal O}_I (t) U_I(t,-\infty)\right \vert {\rm {in}}\right \rangle
\ee
in the terms of the operator ${\cal O}_I(t)$ and the evolution operator in the interaction picture.
We will be interested in the vacuum expectation value (vev) of ${\cal O}$ when all the transient effects have disappeared corresponding to $t=\infty$.
Now we can identify the $out$ vacuum as
\be
\vert {\rm out}\rangle = S \vert {\rm in}\rangle
\ee
where the scattering $S$-matrix is defined as $S=U_I (\infty,-\infty)$
As a result we have
\be
\langle {\cal O}\rangle_{\rm vac}(\infty) = \left \langle \rm {out}\left \vert {\cal O}_I (\infty ) S\right \vert \rm {in}\right \rangle
\ee
This is the origin of the $in-out$ evaluation of vevs using explicit $S$ matrix elements.
Stability of the vacuum and unitarity imply that
\be
\vert {\rm out}\rangle = e^{i\alpha} \vert{\rm in}\rangle
\ee
where
\be
e^{i\alpha}= \langle {\rm in}\vert S \vert {\rm in}\rangle
\ee
leading to
\be
\langle {\cal O}\rangle_{\rm vac}(\infty) = \frac{\langle \rm {int}\vert {\cal O}_I (\infty ) S\vert \rm {in}\rangle }{ \langle {\rm in}\vert S \vert {\rm in}\rangle}.
\ee
This is the Gellmann-Low formula. Now for Lagrangian field theories, we can use the Dyson formula
\be
S= T\left ( e^{i \int d^4x {\cal L}_I}\right )
\ee
where ${\cal L}_I$ is the interaction Lagrangian. This leads to the usual result in terms of path integrals
\be
\langle {\cal O}\rangle_{\rm vac}(\infty) = \frac{\int {\cal D}\phi {\cal O}(\infty)e^{i\int d^4x {\cal L}_I(\phi,\partial_\mu\phi)}}{\int {\cal D}\phi e^{i\int d^4x {\cal L}_I(\phi,\partial_\mu \phi)}}
\ee
where we have assumed that the field theory is defined by the field $\phi$.

None of these results stand in the dissipative case as the evolution operator is not unitarity. Moreover, the notion of particle is ill-defined in the Casimir setup as translation invariance is broken. As a result, the vev's must be evaluated in the Schwinger-Keldysh formalism where such  assumptions are not made.
All this implies that there are memory effects which cannot be cancelled, for instance
\be
\langle 0\vert a_\alpha (t_1) a_\beta (t_2) \vert 0 \rangle = \langle 0\vert U_I^\dagger (t_1, -\infty) a_{\alpha I} (t_1)U_I(t_1,-\infty) U^\dagger (t_2,-\infty) a_{\beta I}(t_2) U_I(t_2,-\infty)  \vert 0 \rangle.
\ee
In the absence of dissipation this reads
\be
\langle 0\vert a_\alpha (t_1) a_\beta (t_2) \vert 0 \rangle = \langle 0\vert U_I ( -\infty, t_1) a_{\alpha I}(t_1) U(t_1,t_2) a_{\beta I}(t_2) U_I(t_2,-\infty) (t_2) \vert 0 \rangle
\ee
which can be represented as
\be
\langle 0\vert a_\alpha (t_1) a_\beta (t_2) \vert 0 \rangle = \langle 0\vert T_C(  a_{\alpha I}(t_1) a_{\beta I}(t_2) U_I(-\infty,-\infty)) \vert 0 \rangle
\ee
where the time ordering is along the path $C$ which goes from $-\infty$ to $\infty$ and back to $-\infty$.
When dissipation is present we obtain
\be
U_I(t_1,-\infty ) U_I^\dagger (t_2,-\infty )\ne U_I(t_1,t_2)
\ee
where $U(t_1,t_2)=  T\left ( e^{-i\int_{t_2}^{t_1} {\cal H}_I (t') dt'}\right )$ due to $ H_I^\dagger\ne H_I$.
As as result the Schwinger-Keldysh representation of the correlators does not
work because of dissipation. This is why in the main text, we only consider the time evolution of single operators.

\section{Quantum dissipation}
We focus on the interior of the cavity of size $d$ seen as an open system when the reflection coefficients are not unity. Its energy is given by the two contributions
\be
\frac{E^+}{A}=  -\frac{d}{\pi} \int \slashed{d}^2 p_\parallel \int_{0}^\infty d\omega\; \frac{\omega^2}{\Delta} \frac{1}{\left (\frac{1+\frac{\xi}{\Delta}}{1-\frac{\xi}{\Delta}}\right )^2 e^{2\Delta d}-1}
\ee
for the TE polarisation
and
\be
\frac{E^-}{A}=  -\frac{d}{\pi} \int \slashed{d}^2 p_\parallel \int_{0}^\infty d\omega\; \frac{\omega^2}{\Delta} \frac{1}{\left (\frac{1+\frac{\xi}{\epsilon(i\omega)\Delta}}{1-\frac{\xi}{\epsilon(i\omega)\Delta}}\right )^2 e^{2\Delta d}-1}
\ee
for the TM case.
When dissipation is present, the first law of thermodynamics gives
\be
dE= -P dV + \delta Q
\ee
where $\delta Q$ is the heat associated with the change of volume. This tells us that the pressure is not simply the derivative of $E$. For the cavity of size $d$
we can define the quantum work $W_{\pm}$ for the two polarisations as
\be
P_\pm= -\partial_d W_\pm
\ee
which reads
\be
W_\pm = \frac{A}{2} \int \slashed{d} \omega \slashed{d}^2 p_\parallel\; \ln \left (1- r_{TE/TM}^{2}e^{-2 \Delta d}\right ).
\ee
With this we can identify the loss of energy from the cavity  as the energy penetrating the plates
\be
Q_\pm = E_R^\pm -W_\pm
\ee
which is a function of the distance $d$.
In the ideal case, the reflection coefficients are unity and the cavity behaves like a closed system. In this case no energy flows from the cavity to the plates, i.e. $Q_\pm=0$. This is non-vanishing for real metals as the reflection coefficients are not equal to unity anymore and the cavity behaves like an open system. Some of the energy will disappear and $ Q_\pm $ should account for some of the energy flowing away to infinity and some being dissipated in the metal when the imaginary part of $\epsilon$ is non-vanishing.

\section{The electric Green's function}
\label{sec:ele}

In this appendix we will give explicit formulae for the Green's function satisfying
\be
-\partial_0( \epsilon\star_t \partial_0 G_E) + \Delta G_E=  \delta^{(4)}(x^\mu-y^\mu).
\ee
In the following we will use time-translation invariance and space-translation invariance in the $(x,y)$ plane along the plaques  to choose $y^\mu=(\vec 0,z_0,0)$.
In terms of Fourier decomposition the Green's function satisfies
\be
\left (\partial_z^2 -p_\parallel^2 + \epsilon(\omega) \omega^2\right ) G_E= \delta(z-z_0).
\ee
We will separate the $z$-axis into three intervals and give the solution in each case using continuity of the Green's function at the boundaries
$z=0$ and $z=d$. The first derivative is continuous there too whilst there is a jump $\left [\frac{dG_E}{dz} \right ]_{z=z_0}=1$ of the first derivative when $G_E$ itself is continuous at $z=z_0$. Moreover we impose that $\lim_{\vert z\vert \to \infty} G_E=0$. This determines a unique solution.

\subsection{$z_0\in [0,d]$}
When $0\le z_0 \le d$ is between the plates
the Green's function is then defined by
\be
G_E=G^E_- e^{\xi z},\ z<0
\ee
and
\be
G_E= G^E_+ e^{-\xi(z-d)}, \ z>d
\ee
where we have now
\be
G^E_-= \Theta_E(\omega,p_\parallel)  \left ( \cos \Delta  (z_0-d) - \frac{\xi}{\Delta} \sin \Delta (z_0-d)\right )
\ee
together with
\be
G^E_+= {\Theta_E (\omega,p_\parallel)} \left ( \cos \Delta z_0 + \frac{\xi}{\Delta} \sin \Delta z_0\right )
\ee
and the Green's function between the plates
\ba
&& 0\le z\le z_0, \ G_E= G^E_- \left ( \cos \Delta z +\frac{\xi}{\Delta} \sin \Delta  z\right )\nonumber \\
&& z_0 \le z \le d, \ G_E= G^E_+\left ( \cos \Delta (z-d) -\frac{\xi}{\Delta} \sin \Delta (z-d)\right ) \nonumber \\.
\ea
where
\be
\Theta_E (\omega,p_\parallel)= \frac{1}{\Delta \left ( (1- \frac{\xi^2}{\Delta^2}) \sin \Delta  d -2 \frac{\xi}{\Delta} \cos \Delta d\right )}.
\ee

\subsection{$z_0<0$}
This case is different from the previous case as
\be
G_E=G^E_- e^{\xi z},\ z\le z_0
\ee
and
\be
G_E= A e^{\xi z} + B e^{-\xi z}, \ z\in [z_0,0]
\ee
whilst
\ba
&& 0\le z\le d, \ G_E= G^E_+ \left [ \left (\sin \Delta d -\frac{\xi}{\Delta} \cos \Delta d\right ) \sin \Delta z+ \left (\cos \Delta d+ \frac{\xi}{\Delta} \sin \Delta d\right ) \cos \Delta z\right ]\nonumber \\
&&  z\ge d, \ G_E= G^E_+ e^{-\xi (z-d)} \nonumber \\.
\ea
We find that
\ba
&&A= \frac{G^E_+}{2}\left [ \cos \Delta d+ \frac{\xi}{\Delta} \sin \Delta d + \frac{\Delta}{\xi}\left (\sin \Delta d -\frac{\xi}{\Delta} \cos \Delta d\right )\right ]\nonumber \\
&&B=\frac{G^E_+}{2}\left [ \cos \Delta d+ \frac{\xi}{\Delta} \sin \Delta d - \frac{\Delta}{\xi}\left (\sin \Delta d -\frac{\xi}{\Delta} \cos \Delta d\right)\right ]\nonumber\\
\ea
and we have
\be
G^E_+= \Theta_E(\omega,p_\parallel) e^{\xi z_0}.
\ee
Finally we find that
\be
G^E_-=\Theta_E(\omega,p_\parallel)\left [\cosh \xi z_0 \left (\cos \Delta d+ \frac{\xi}{\Delta} \sin \Delta d \right )+ \frac{\Delta}{ \xi} \sinh \xi z_0 \left (\sin \Delta d -\frac{\xi}{\Delta} \cos \Delta d\right )\right ]
\ee
\subsection{$z_0>d$}
We have now
\be
z<0,\ G_E= G^E_- e^{\xi z}
\ee
and
\be
z \in [0,d],\ G_E= G^E_-\left ( \cos \Delta z +\frac{\xi}{\Delta} \sin \Delta z\right )
\ee
whilst we have
\be
z\in [d,z_0], \ G_E= \frac{G^E_-}{2}\left [ \sin \Delta d \left ( \frac{\xi}{\Delta} + \frac{\Delta}{\xi}\right)e^{-\xi (z-d)} +\left (2 \cos \Delta d+ \sin \Delta d \left (\frac{\xi}{\Delta}-\frac{\Delta}{\xi}\right)\right )e^{\xi(z-d)}\right ]
\ee
and finally
\be
z>z_0, \ G_E= G^E_+ e^{-\xi( z-d)}
\ee
where we find
\be
G^E_-= \Theta_E (\omega, p_\parallel) e^{-\xi(z_0-d)}
\ee
and
\ba
G^E_+ &&=\Theta_E (\omega, p_\parallel)\Big [\cosh \xi (z_0-d) \left (\cos \Delta d+ \frac{\xi}{\Delta} \sin \Delta d \right)
\nonumber\\
&&+ \frac{\Delta}{ \xi} \sinh \xi (z_0-d) \left (-\sin \Delta d +\frac{\xi}{\Delta} \cos \Delta d\right )\Big ]
\ea
\section{The magnetic Green's functions}
\label{sec:mag}
\subsection{The link to the electric Green's function}

The magnetic Green's function satisfies the identity
\be
-\partial_0^2 G_M + \epsilon^{-1}\star_t \Delta G_M=  \delta^{(4)}(x^\mu-y^\mu)
\ee
Away from the point $x=x_0$, it is a solution of the homogeneous equation and therefore admits the same expansion in exponentials for $z<0$ and $z>d$ whilst it is
a linear combination of $\cos \Delta z$ and $\sin \Delta z$ between $0$ and $d$. The only difference between the two Green's function comes from the boundary conditions involving the continuity of $\partial_z G_E$ for the TE case and $\epsilon^{-1}(\omega) \partial_z G_M$ for the TM polarisation.  This is enough to deduce the expression for the Green's function. Indeed we only have to substitute
\be
\xi \to \frac{\xi}{\epsilon}
\ee
in the coefficients of the electric function to find the magnetic function.
For instance when $0\le z_0 \le d$ is between the plates
the magnetic Green's is function obtained as
\be
G_M=G^M_- e^{\xi z},\ z<0
\ee
and
\be
G_M= G^M_+ e^{-\xi(z-d)}, \ z>d
\ee
where
\be
G^M_-= \Theta_M(\omega,p_\parallel)  \left ( \cos \Delta  (z_0-d) - \frac{\xi}{\epsilon\Delta} \sin \Delta (z_0-d)\right )
\ee
and
\be
G^M_+= {\Theta_M (\omega,p_\parallel)} \left ( \cos \Delta z_0 + \frac{\xi}{\epsilon\Delta} \sin \Delta z_0\right ).
\ee
The magnetic Green's function between the plates is
\ba
&& 0\le z\le z_0, \ G_M= G^M_- \left ( \cos \Delta z +\frac{\xi}{\epsilon\Delta} \sin \Delta  z\right )\nonumber \\
&& z_0 \le z \le d, \ G_M= G^M_+\left ( \cos \Delta (z-d) -\frac{\xi}{\epsilon\Delta} \sin \Delta (z-d)\right ) \nonumber \\.
\ea
where
\be
\Theta_M (\omega,p_\parallel)= \frac{1}{\Delta\left ( (1- \frac{\xi^2}{\epsilon^2 \Delta^2}) \sin \Delta  d -2 \frac{\xi}{\epsilon\Delta} \cos \Delta d\right)}.
\ee
Similar expressions are valid in the other two intervals. All this can be checked by an explicit calculation.

\subsection{Convolution and the magnetic Green's function}

    Some properties of the TM two-point functions are useful. 
The magnetic two-point function are given by
\be
\Delta_M (x;y)= \left \langle a^1_-(x) (\hat \epsilon \star_{t_y}\tilde a^1_-)(y)\right \rangle
\ee
where the permittivity function is convolved with the second field $\tilde a^1_\alpha$, and the convolution is expressed in time leading to
\be
\Delta_M (x;y)= \int dt \;\epsilon (t-t_y)\left \langle a^1_-(\vec x, t_x) \tilde a^1_-(\vec y, t)\right \rangle.
\ee
As the two-point functions respect time translation invariance,,  we have $\Delta_M(x;y)= \Delta_M (\vec x, t_x-t_y;\vec y, 0)$ and therefore
\be
\Delta_M (x;y)= \int dt \;\epsilon (t-t_y) \Delta_M (\vec x, t_x-t;\vec y,0) .
\ee
Changing variable to $\tilde t= t_x+t_y -t$ we find
\be
\Delta_M (x;y)= \int d\tilde t \;\epsilon (t_x-\tilde t) \Delta_M (\vec x, \tilde t -t_y;\vec y,0)= \int d\tilde t \;\epsilon (t_x-\tilde t)\left \langle a^1_- (\vec x, \tilde t) \tilde a^1_- (\vec y, t_y)\right \rangle
\ee
and therefore
\be
\Delta_M (x;y)= (\epsilon\star_{t_x} \Delta^M_-)(x;y)= \left (\hat \epsilon\star_{t_y} \Delta^M_- \right )(x;y)= (\Delta^M_-\star_{t_y} \epsilon)(x;y).
\label{DM}
\ee
where
\be
\Delta^M_- (x;y)=\left \langle a^1_- (\vec x, \tilde t) \tilde a^1_- (\vec y, t_y)\right \rangle.
\ee
In regions of space where $\epsilon(\omega)=1$, we will identify $\Delta_M$ and $\Delta^M_-$.

\subsection{Magnetic resonances}

\label{sec:reso}
The resonance equation for the $TM$ polarisation is
\be
\Theta_M (\omega, p_\parallel)=0
\ee
where
\be
\Theta (\omega,p_\parallel)= \frac{1}{\Delta\left ( (1- \frac{\xi^2}{\epsilon^2\Delta^2}) \sin \Delta  d -2 \frac{\xi}{\epsilon\Delta} \cos \Delta d\right)}.
\ee
This has solutions for
\be
\tan \Delta d= \frac{2\frac{\xi}{\epsilon\Delta}}{1- \frac{\xi^2}{\epsilon^2\Delta^2}}.
\ee
 We consider the cases when $\gamma \ll \vert \omega\vert \ll \omega_{\rm Pl}$ implying that $\Delta \ll \omega_{\rm Pl}$ and
\be
\frac{\xi}{\Delta} \simeq \frac{\omega_{\rm Pl}}{\Delta}\left (1- i\frac{\gamma}{2\omega}\left (1+\frac{\Delta^2}{\omega_{\rm Pl}}\right )\right)
\ee
which is large and therefore
\be
\tan \Delta d=  \frac{2\xi}{\epsilon \Delta}\simeq -2 \frac{\omega^2}{\omega_{\rm Pl}\Delta}\left (1+\frac{i\gamma}{2\omega} - \frac{\Delta^2}{2 \omega_{\rm Pl}^2}\right ).
\ee
The solutions are obtained by iteration and  are
\be
\Delta_n d= n\pi -2 \frac{(\omega_n^{(0)})^2 d}{n\pi \omega_{\rm Pl}}\left (1+i\frac{\gamma}{2\omega_n^{(0)}}\right ), \ n\ne 0
\label{res}
\ee
where to leading order
\be
\omega_n^{(0)}= {\rm sign}(n) \sqrt{ \frac{n^2\pi^2}{d^2}+p_\parallel^2}.
\ee
The imaginary part of the resonance frequency is obtained by squaring (\ref{res1}) implying that
\be
\omega_n^2 =  \frac{n^2\pi^2}{d^2}+p_\parallel^2- 2i\frac{\gamma \omega_n^{(0)}}{\omega_{\rm Pl}d}
\ee
Similarly to the $TE$ case, the resonances are below the real axis. Above the plasma frequency, the resonance condition
is the same as the $TE$ case and there are no solutions.

\end{document}